\documentclass[aps,prx,floatfix,twocolumn,superscriptaddress,longbibliography]{revtex4-1}
\usepackage[english]{babel} 
\usepackage[latin1]{inputenc}
\usepackage[usenames,dvipsnames]{color}

\usepackage{amsmath,amssymb,amsfonts}
\usepackage[colorlinks=true,linkcolor=blue,urlcolor=blue,citecolor=blue]{hyperref}
\usepackage[percent]{overpic}
\usepackage[latin1]{inputenc}
\usepackage{amsmath,amssymb}
\usepackage{graphicx,color}
\usepackage{braket}
\usepackage{bm}
\usepackage[caption = false]{subfig}

\usepackage{titlesec}
\titleformat{\section}[runin]{\normalfont\itshape}{}{3pt}{}[.]

\usepackage[resetlabels,labeled]{multibib}
\newcites{Fig}{References of Figures}

\newcommand{\be}{\begin{eqnarray}}
\newcommand{\ee}{\end{eqnarray}}

\renewcommand{\vec}[1]{\bm{#1}}

\definecolor{green(html/cssgreen)}{rgb}{0.0, 0.5, 0.0}

\begin{document}
	\title{Clock model and parafermions in Rashba nanowires}
	\author{Flavio Ronetti}
	\affiliation{Department of Physics, University of Basel, Klingelbergstrasse 82, CH-4056 Basel, Switzerland}
	\author{Daniel Loss}
	\affiliation{Department of Physics, University of Basel, Klingelbergstrasse 82, CH-4056 Basel, Switzerland}
	\author{Jelena Klinovaja}
	\affiliation{Department of Physics, University of Basel, Klingelbergstrasse 82, CH-4056 Basel, Switzerland}
	
\begin{abstract} 
We consider a  semiconducting nanowire with  Rashba spin-orbit interaction subjected to a magnetic field and
in the presence of strong electron-electron interactions. When the ratio between Fermi and Rashba momenta is tuned to $1/2$, two competing resonant multi-particle scattering processes are present simultaneously and the interplay between them brings the system into a  gapless critical parafermion phase. This critical phase is described by a self-dual sine-Gordon model, which we are able to map explicitly onto the low-energy sector of the $\mathbb{Z}_4$ parafermion clock chain model. Finally, we show that by alternating regions in which only one of these two processes is present one can generate localized zero-energy parafermion bound states. 
\end{abstract}

\maketitle
\textit{Introduction.} 
Spin-orbit interaction (SOI) plays a prominent role in a wide range of spectacular phenomena in condensed matter physics~\cite{LossBook2002,Winkler03}. Extensive investigations in this direction have been carried out, also due to its fundamental role in the implementation of spin-based quantum information platforms~\cite{Loss98,Hanson07,Kloeffel13,Golovach06,Nowack07,Froning20,Froning20b}.
Among the numerous phenomena governed by SOI in condensed matter physics, one of the most fascinating outcomes is the realization of helical liquids at edges of topological insulators~\cite{Wu06,Konig07,Hasan10} or in semiconducting nanowires (NWs)~\cite{Streda03,Pershin04,Meng13,Kammhuber17}. 
 Besides their high potential for spintronics applications, helical liquids attract a lot of attention because, in presence of proximity-induced superconductivity, they can be used to engineer $p$-wave superconductors~\cite{Kitaev01} hosting Majorana bound states at their boundaries~\cite{Braunecker10,Oreg10,Lutchyn10,Potter11,Sticlet12,Klinovaja12c,Halperin12,SanJose12,Rainis13,Mourik12,Das12,Deng12,Scheller14,Deng16,Lutchyn18,Deng18,Prada20}.

If Rashba SOI is combined with strong electron-electron interactions, even more fascinating states of matter can emerge, notably fractional topological insulators~\cite{Levin09,Klinovaja14d,Meng15,Sagi15,Santos15,Stern16,Volpez17,Rachel18,Laubscher19b} 
or fractional helical liquids in Rashba NWs~\cite{Oreg14}. The most striking experimental signature of these systems is a fractional charge conductance, which signals the presence of fractionally charged excitations~\cite{Cheng12,Vaezi13,Meng14,Klinovaja14c,Aseev18,Oreg19}.  
When coupled to superconductors, fractional helical liquids become fully gapped and they host zero-energy parafermion bound states~\cite{Oreg14,Klinovaja14c,Orth15,Sagi17,Thakurathi17,Pedder17,Laubscher19,Fleckenstein19,Klinovaja15}, similar to Majorana bound states but obeying a richer braiding statistics~\cite{Fendley12,Klinovaja14a,Vaezi14,Mong14,Alicea16,Hutter16,Chew18,Rossini19,Groenendijk19,Santos20}. 

\begin{figure}[t]
	\centering
	\includegraphics[width=0.4\linewidth]{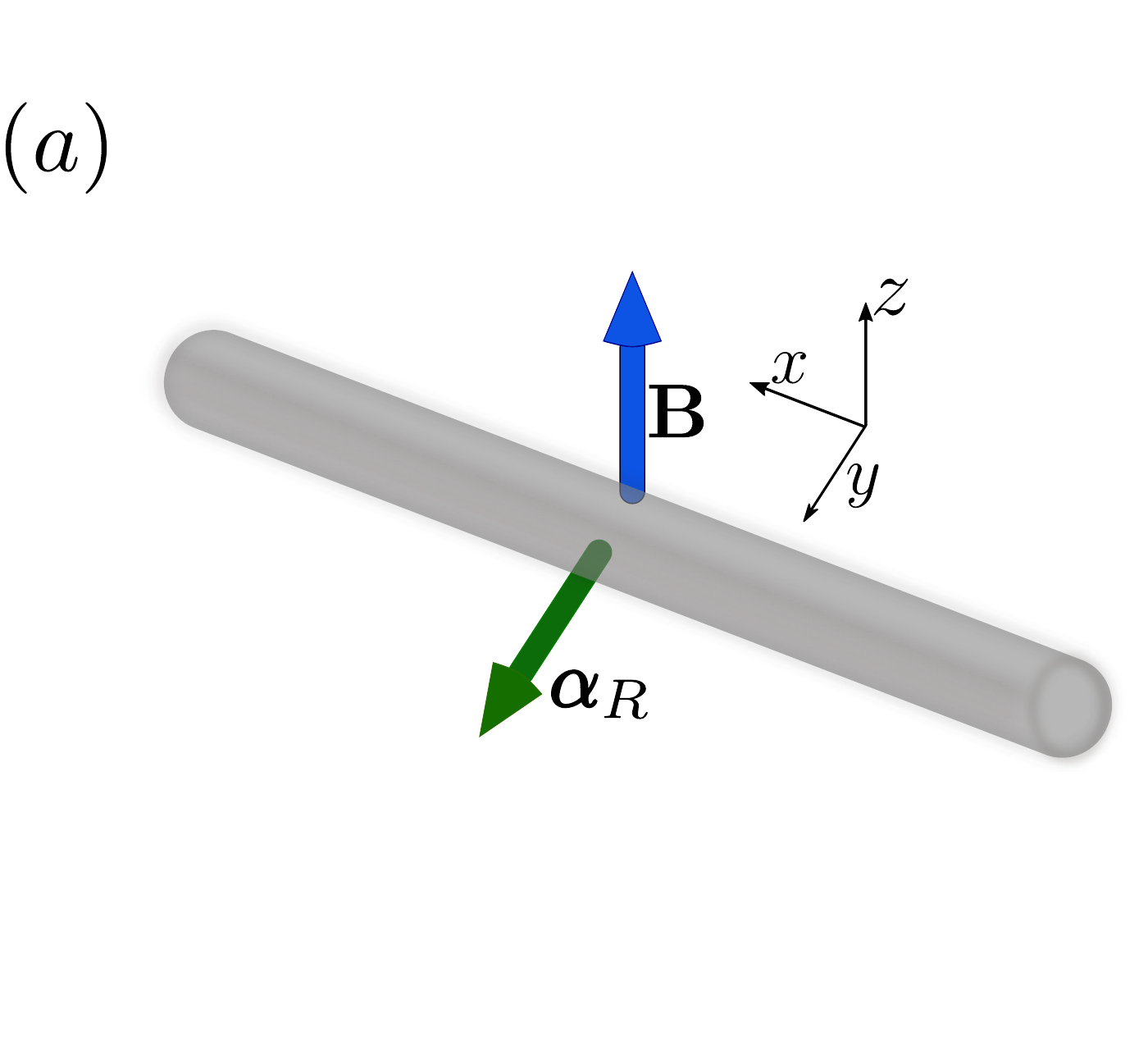} 
	\includegraphics[width=0.58\linewidth]{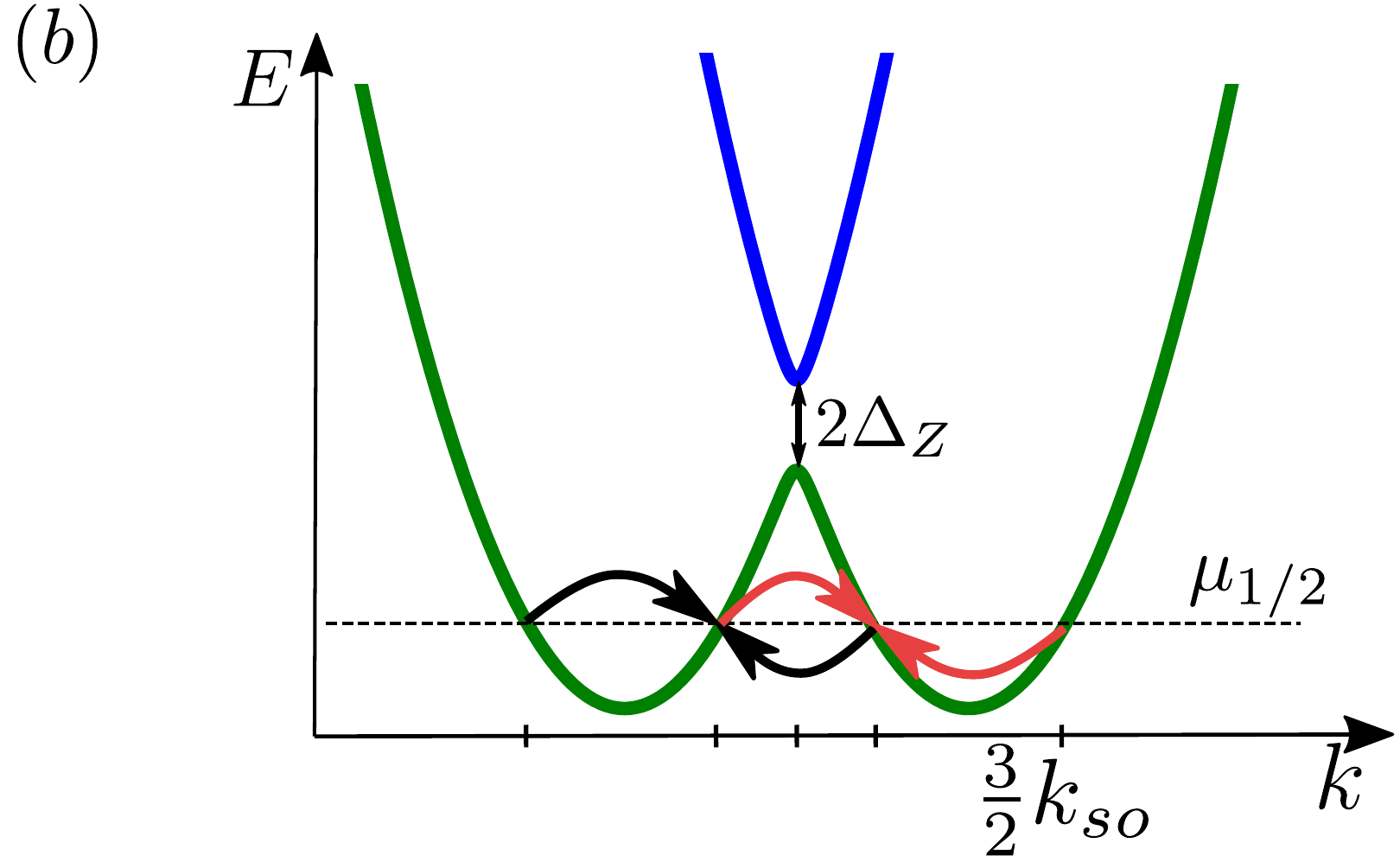}
	\caption{(a) NW aligned along $x$ direction with uniform Rashba SOI vector $\boldsymbol{\alpha}_R$  pointing along  $y$ direction. A magnetic field $\mathbf{B}$ is applied perpendicular to $\boldsymbol{\alpha}_R$, {\it i.e. } along the $z$ axis to open a partial Zeeman gap $2\Delta_Z$ in the spectrum at zero momentum. (b) Spectrum of a 
	one-dimensional NW with strong Rashba SOI and Zeeman gap. The chemical potential is tuned to $\mu_{1/2}$ such that $k_F / k_{so}=1 / 2$, where $k_{so}$ is the Rashba and $k_F$ the Fermi  momentum. Due to electron-electron interactions, there are two competing momentum-conserving  resonant scattering processes (red and black arrows), which lead to a gapless $\mathbb{Z}_4$ parafermion phase.}
	\label{fig:rashba}
\end{figure}

In Rashba nanowires, the main focus  so far has been on odd denominator filling factors. In this work, we uncover a minimal setup of high experimental relevance with parafermion phases which requires only the intrinsic ingredients of a Rashba NW, namely SOI and electron-electron interactions; in particular, no superconductivity and no exotic quantum Hall phases 
are involved~\cite{Barkeshli13,Barkeshli14}.
At the simplest possible even-denominator filling factor $\nu=1/2$, we find the striking result that
two non-commuting processes are simultaneously generated by multi-particle interactions such that, instead of opening a gap, they  leave the system in a gapless phase hosting parafermion excitations, see Fig.~\ref{fig:rashba}. Using bosonization techniques, we analyze the nature of this gapless phase and show that it is described by a $\mathbb{Z}_4$ self-dual sine-Gordon  model. We identify the obtained model with the low-energy limit of the $\mathbb{Z}_4$ parafermion clock chain model~\cite{Fradkin80,Fendley12,Calzona18,Mazza18}.
Remarkably, the $\mathbb{Z}_4$ parafermions emerge in our setup due to purely intrinsic ingredients: spin-orbit and electron-electron interactions of an isolated Rashba NW.
Finally,  an additional magnetic field applied parallel to the SOI vector breaks the balance between two non-commuting processes, leaving only one of them in resonance.  Localized $\mathbb{Z}_4$ zero-energy parafermion  bound states can then emerge at the interfaces between two different dominant processes.  The remaining gapless fermion modes can be easily gapped out  by a spatially oscillating magnetic field, generated e.g. by  nanomagnets~\cite{Tokura06,Pioro08,Desjardins19,Sapkota19}.

\textit{Model.} We consider a one-dimensional Rashba NW, orientied along $x$ direction, in the presence of strong electron-electron interactions, see Fig.~\ref{fig:rashba}.  The Rashba SOI, assumed to be uniform, is characterized by the SOI vector $\boldsymbol{\alpha}_R$ aligned along  $y$ direction. An external magnetic field $\mathbf B$ is applied perpendicular to the SOI vector $\boldsymbol{\alpha}_R$  and opens a partial gap at $k=0$.  The corresponding Hamiltonian is written as $H =\sum_{\sigma,\sigma'} \int dx ~\Psi_{\sigma}^{\dagger}(x)\mathcal{H}_{\sigma\sigma'}\Psi_{\sigma'}(x)$, where $\Psi_{\sigma}(x)$ is the annihilation operator acting on an electron with spin $\sigma/2=\pm 1/2$ at position $x$ of the NW and the Hamiltonian density $\mathcal{H}$ reads (we set $\hbar = 1$):
\begin{equation}
\mathcal{H} = 
-\frac{\partial_x^2}{2 m} - \mu +i \alpha_R \partial_x \sigma_y + \Delta_Z \sigma_z,\label{Eq:ham0}
\end{equation}
where  the Pauli matrices $\sigma_i$ act on the electron spin. Here, $\mu$ is the chemical potential, $m$  the effective mass and $\Delta_Z=g \mu_B B$, where $g$ is the $g$-factor and $\mu_B$  the Bohr magneton. In addition, we define the SOI momentum (energy) $k_{so}=m \alpha_R$  ($E_{so}=k_{so}^2/2m$).  In order to deal with electron-electron interactions, it is convenient to linearize the spectrum around the Fermi points. The corresponding expanded fermion operators are $\Psi_{\sigma}(x) = R_{\sigma}e^{i k_{R\sigma}x} + L_{\sigma}e^{i k_{L\sigma}x}$, where $k_{r\sigma} = r k_F - \sigma k_{so}$ with  the Fermi momentum $k_F$ being calculated from $k_{so}$.

Electron-electron interaction can be divided into two types of terms corresponding to small and large momentum contributions. The first type of interaction can be taken into account in a Hamiltonian that is quadratic in fermion densities and whose form is assumed in accordance with the standard Luttinger liquid description~\cite{Giamarchi03}. Large momentum interaction terms can be built from a product of single-electron operators as $\mathcal{O}_g=R^{s_{R\uparrow}}_{\uparrow}L^{s_{L\uparrow}}_{\uparrow}R^{s_{R\downarrow}}_{\downarrow}L^{s_{L\downarrow}}_{\downarrow}$, where $s_{r \sigma}$ are integers, while $R_{\sigma}^{s_{R \sigma}}=\left(R_{\sigma}^{\dagger}\right)^{|s_{R \sigma}|}$ and $L_{\sigma}^{s_{L \sigma}}=\left(L_{\sigma}^{\dagger}\right)^{|s_{L \sigma}|}$, when $s_{r \sigma}<0$. These multi-particle scattering processes must obey charge and momentum conservation. In general, momentum is conserved only at certain values of filling $\nu=k_F / k_{so}$~\cite{Kane02,Klinovaja14c}. From now on, we fix the chemical potential to $\mu = \mu_{1 / 2}$ such that the filling factor is $\nu = 1 / 2$. In this case, only the two following multi-particle scattering processes conserve momentum (see Fig.~\ref{fig:rashba}):
\begin{align}
H_{\rm int} &= \Lambda_1\int dx~ L^{\dagger}_{\uparrow}(x)R_{\uparrow}(x)L^{\dagger}_{\downarrow}(x)R_{\uparrow}(x) \nonumber\\&+\Lambda_2 \int dx~ L^{\dagger}_{\downarrow}(x)R_{\uparrow}(x)L^{\dagger}_{\downarrow}(x)R_{\downarrow}(x)+\text{h.c.}\label{eq:pert}
\end{align}
These perturbations are generated at first order in interaction strength and the two coefficients are given by $\Lambda_{1,2} \propto \Delta_Z U_{2k_F}/\mu_{1/2}$, where $U_{2k_F}$ is the interaction potential. As a result, the two amplitudes $\Lambda_1$ and $\Lambda_2$ are identical by construction.

Interestingly, one can show that these two perturbations do not commute with each other. The simultaneous presence of two non-commuting back-scattering processes at this even denominator filling factor is quite intriguing, given the fact that, for odd denominator 
fillings, only a single cosine perturbation after bosonization is induced, resulting in the opening of a partial gap~\cite{Oreg14,Meng14}. In contrast, in our case, as long as the degeneracy between these two processes is preserved, two competing cosine terms are present (see below) such that the system remains gapless and critical properties emerge.

\textit{Bosonization and renormalization group (RG) flow.} To analyze  $H_{\rm int}$ defined in Eq.~\eqref{eq:pert}, it is convenient to use  the standard bosonization representation of  fermion fields~\cite{vonDelft98,Giamarchi03}:
\begin{equation}
r_{\sigma}=\frac{1}{\sqrt{2\pi a}} e^{i r \left[\phi_{\rho}(x)+\sigma \phi_{\sigma}(x) - \theta_{\rho}(x)-\sigma \theta_{\sigma}(x)\right]},
\end{equation}
where $a$ is a short-length cut-off, and where we introduced  four boson fields $\phi_{\rho}$, $\phi_{\sigma}$, $\theta_{\rho}$, and $\theta_{\sigma}$.

In this bosonic basis, the Hamiltonian involving kinetic and small-momentum interaction terms becomes diagonal and reads
\begin{equation}
H_0 = \sum_{\mu=\rho,\sigma} \frac{u_{\mu} K_{\mu}}{2\pi}\int dx ~\left[\partial_x\theta_{\mu}(x)\right]^2 +  \frac{u_{\mu}}{2\pi K_{\mu}} \int dx ~\left[\partial_x\phi_{\mu}(x)\right]^2, 
\label{eq:HLL}
\end{equation}
where $u_{\rho}$ and $u_{\sigma}$ are the renormalized charge and spin velocities {\it resp.}, and $K_{\rho,\sigma}$ the  Luttinger liquid (LL) parameters, which characterize the strength of interaction in the NW. In case of repulsive interactions, $K_{\rho}\le1$ and $K_{\sigma}\ge 1$. 
The bosonized form of the interaction term $H_{\rm int}$ turns into two cosine terms, 
\begin{align}
H_{\rm int} &= \frac{\Lambda_1}{4\pi^2a^2} \int dx ~ \cos\left[\sqrt{2}\left(2\phi_{\rho} - \theta_{\sigma} + \phi_{\sigma}\right)\right]+\nonumber\\
+& \frac{\Lambda_2}{4\pi^2a^2} \int dx ~ \cos\left[\sqrt{2}\left(2\phi_{\rho} - \theta_{\sigma} - \phi_{\sigma}\right)\right].
\end{align}
To reproduce the correct commutation relations, obtained in the fermion picture for these two cosine perturbations, we introduce an alternative bosonization procedure with generalized boson commutation relations~\cite{Hsu19}.

\begin{figure}[t]
	\centering
	\includegraphics[scale=0.41]{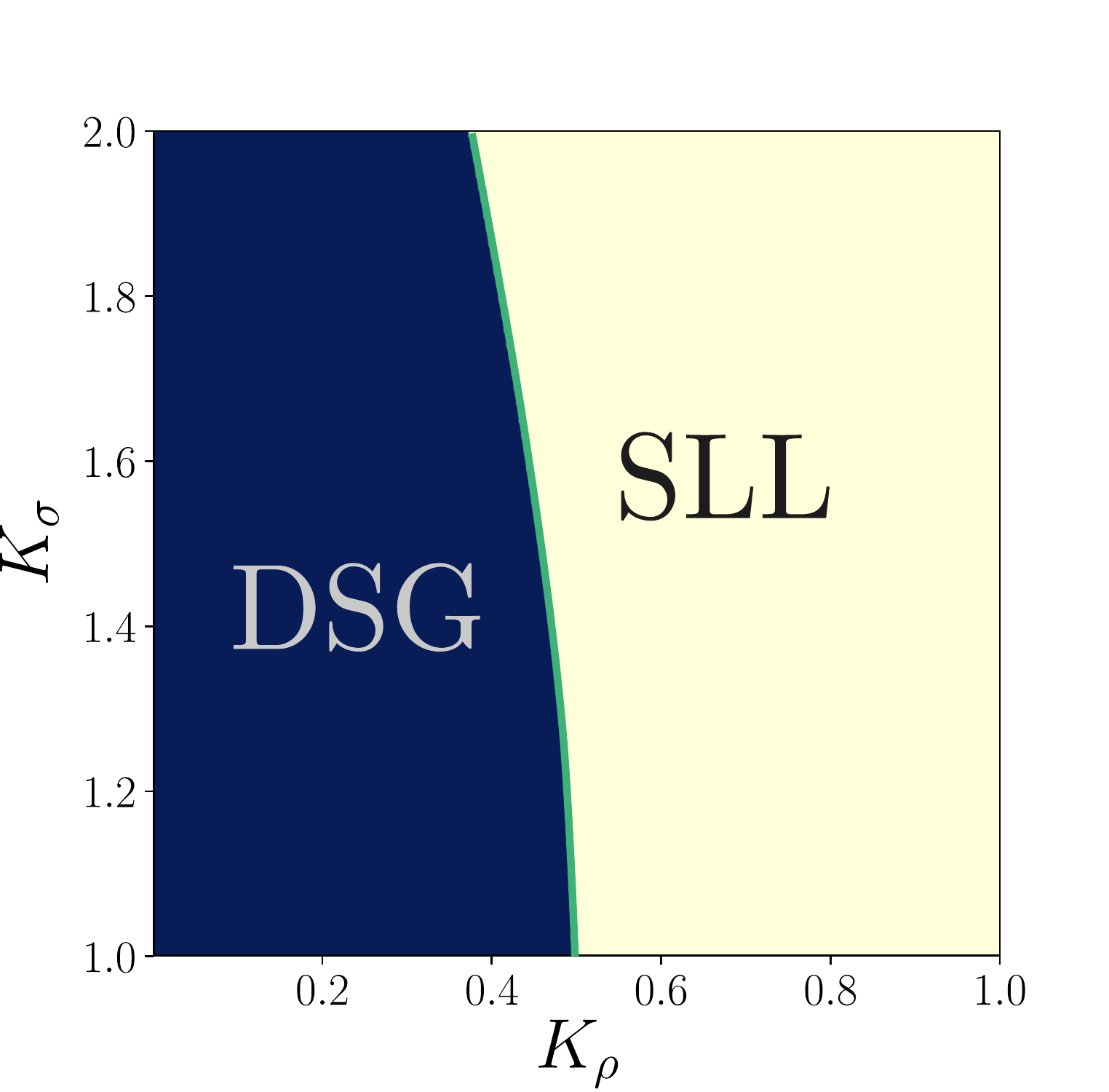}
	\caption{Phase diagram as  function of initial values of LL parameters $K_{\rho}$ and $K_{\sigma}$ 
	for repulsive interactions. In the typical physical situation for which $K_{\sigma}\sim 1$, both perturbations are relevant for $K_{\rho}<1/2$. When  $K_{\sigma}$ increases, the range of $K_{\rho}$ for which the perturbations are relevant is sligthly reduced, meaning that  stronger interactions are required. In this case (blue region), the system is described by the DSG  model, see Eq.~\eqref{eq:DSGM}. When the interaction is not strong enough to make the pertubations relevant (yellow region), the system stays in the spinful LL   phase, see Eq.~\eqref{eq:HLL}.} 
	\label{fig:diagram}
\end{figure}

Next, it is crucial to establish the range of the interaction parameters $K_{\rho}$ and $K_{\sigma}$ for which the cosine perturbations are relevant in the RG sense. The RG equations for the coupling constants $\Lambda_i$ and  $K_{\rho/ \sigma}$ are derived in a standard way, see SM \ref{secSm:action}. The resulting phase diagram is shown in Fig.~\ref{fig:diagram}. Let us start by commenting  on a typical parameter  regime for which $K_{\sigma}\sim 1$: in this case, both perturbations are relevant in the regime of strong interactions for $K_{\rho}<1/2$. When the value of $K_{\sigma}$ increases, the range of $K_{\rho}$ for which the perturbations are relevant is slightly reduced, meaning that  stronger interactions are required. When the interaction is not strong enough to make the pertubations relevant, the system stays in the spinful Luttinger liquid (SLL)  phase described by $H_0$ in Eq.~\eqref{eq:HLL}. When the perturbations are relevant, the initial values of  $K_{\rho}$ and $K_{\sigma}$ flow under the RG to the renormalized values $K^{*}_{\rho} = 0$ and $K^{*}_{\sigma} = 1$. The RG flow of $K_{\rho,\sigma}$
will be crucial to determine the final expression for the Hamiltonian describing the emerging gapless phase.

\textit{Double sine-Gordon (DSG) model.} We focus now on the regime in which both cosine terms are relevant. It is convenient  to rotate the boson fields to a new basis 
\begin{equation}
\left(\begin{matrix}\eta_1\\ \eta_2\\\eta_3\\\eta_4\end{matrix}\right)  =  \frac{1}{2\sqrt{2}}\left(\begin{matrix}
0 & -2  & 1& -1 \\ 0 & -2 & 1 & 1 \\ 0 & 1 & 0 & 0 \\ \frac{1}{2}& 0 &1 & 0 
\end{matrix}\right)\left(\begin{matrix}\theta_{\rho}\\\phi_{\rho}\\ \theta_{\sigma}\\ \phi_{\sigma} \end{matrix}\right).
\end{equation}
This choice is motivated by the fact that the arguments of the cosines now contain only a single boson field $\eta_1$ and $\eta_2$, respectively. In order to focus on the effects induced by the multi-particle processes, it is useful to integrate out the fields $\eta_3$ and $\eta_4$, which do not appear in the cosine arguments. We note that, although velocities and interaction coefficients associated with $\eta_1$ and $\eta_2$ acquire a complicated expression in terms of $K_{\rho}$ and $K_{\sigma}$, the final form of the Hamiltonian can be simplified by taking into account that the Luttinger liquid parameters flow to $K_{\rho}=K^{*}_{\rho}$  and $K_{\sigma}=K^{*}_{\sigma}$. As a result, the effective Hamiltonian becomes
\begin{align}
H_{\rm DSG} &=\sum_{i=1,2} \int dx \left[\frac{v}{2\pi}   \left(\partial_x \eta_i\right)^2 + \frac{\Lambda_i}{4\pi^2a^2} \cos\left(k\eta_i(x)\right)\right] ,
\label{eq:DSGM}
\end{align}
where $k=4$ and $v=4 u_{\sigma}$ (see SM~\ref{secSm:action} for more details). We emphasize that  the boson fields appearing in the cosines  do not commute and obey the commutation relations $\left[\eta_1(x),\eta_2(x')\right]= i \frac{\pi}{4} \text{sign}(x-x')$. Due to the presence of two non-commuting cosines, $H_{\rm DSG}$
DSG model~\cite{Boyanovsky89,Lecheminant02}. We note that the duality transformations $\Lambda_1 \leftrightarrow \Lambda_2$ and $\eta_1 \leftrightarrow \eta_2$ leave $H_{\rm DSG}$ invariant. Thus, the case $\Lambda_1=\Lambda_2$ implies self-duality and, therefore, the emergence of critical properties resulting in exotic gapless modes inside the NW.
In our case, the two  amplitudes $\Lambda_1=\Lambda_2$ are enforced to be the same by symmetry. In addition, it is interesting to note that $H_{\rm DSG} $ [see Eq.~\eqref{eq:DSGM}] satisfies the global symmetry $\mathbb{Z}_4$:$ ~ \eta_1\rightarrow \eta_1 + \frac{2\pi}{4}$,  $\eta_2\rightarrow \eta_2$ and its dual symmetry $\mathbb{Z}^{\rm dual}_4$:$ ~ \eta_1\rightarrow \eta_1$, $ \eta_2\rightarrow \eta_2+ \frac{2\pi}{4}$. 
For the special case $\Lambda_1=\Lambda_2$,  $H_{\rm DSG} $ is known as $\mathbb{Z}_4$ 
self-dual sine-Gordon  model~\cite{Fateev85}.

For later purpose, we point out that the resonance between the two cosine perturbations can be detuned by adding a magnetic field parallel to the SOI direction accompanied by a corresponding readjustment of the chemical potential. As a consequence, only one perturbation would conserve momentum, thus resulting in either $\Lambda_1=0$ or $\Lambda_2=0$.
In this case, a partial gap is opened and the system enters  a fractional helical liquid phase with fractional conductance 
$G=2e^2/3h$, which we obtain by standard methods~\cite{Meng14}.

\textit{Mapping to the $\mathbb{Z}_4$ parafermion clock model.} 
In the following we construct parafermion operators  from the boson fields $\eta_1$ and $\eta_2$ and show that  $H_{\rm DSG}$, Eq.~\eqref{eq:DSGM}, can be identified as the continuum limit of the $\mathbb{Z}_4$  clock  model~\cite{Fradkin80,Fendley12,Calzona18,Mazza18}. We show that this mapping holds for general $k>1$~\cite{footnote1}, starting from the $\mathbb{Z}_k$ 
clock chain model in parafermion representation~\cite{Fendley12,Sagi17}:
\begin{align}
H_{\rm cl}&=-\sum_{j}\sum_{\alpha=1}^{k-1}(-1)^{\alpha}
\Big[J_{1,\alpha}\left(e^{i\pi /k}\chi_{2,j}^{\dagger}\chi_{1,j+1}\right)^{\alpha}+\nonumber\\ &+J_{2,\alpha}\left(e^{-i\pi /k} \chi_{2,j}^{\dagger}\chi_{1,j}\right)^{\alpha}\Big],
\label{eq:para}
\end{align} 
where $J_{i,\alpha}^* = J_{i,k-\alpha}$ 
and $\chi_{p,j}$, $p=1,2$, obey  $\mathbb{Z}_k$ parafermion statistics (see SM~\ref{secSm:para}),  
\begin{align}
&\chi_{p,j}^k = 1, ~\chi_{p,j}^{k-1} = \chi_{p,j}^{\dagger},\,\,\, \chi_{1,i} \chi_{2,i} = \omega \chi_{2,i} \chi_{1,i},\label{eq:para_stat1}\\
&\chi_{p,i} \chi_{p',m} = \omega \chi_{p',m} \chi_{p,i},\,\, i< m\,\,\label{eq:para_stat2}.
\end{align}
Next, we introduce a bosonic representation for the parafermion operators: 
$\chi_{1,j}=\sigma_{1,j} \prod_{l=1}^{j-1} \sigma^{\dagger}_{2,l}\sigma_{2,l+1}$ 
and 
$\chi_{2,j} =e^{i \frac{\pi }{k}(k-1)} \chi_{1,j}\sigma_{2,j}^{\dagger}\sigma_{2,j+1} $, where 
\begin{align}
\sigma_{p,j} &=A_k e^{i \eta_{p}(x_j)}+B_k e^{-i (k-1) \eta_{p}(x_j)},\label{eq:sigma}
\end{align}
with $x_j = j a$. The  $k$-dependent coefficients $A_k$ and $B_k$  are real and obey the relations $A^2_k + B^2_k = 1$ and $k A_k^{k-1}B_k =1$. For $k=4$,  $A_{4}=B_4=1/\sqrt{2}$. The boson fields $\eta_p$ satisfy 
$\left[\eta_p(x),\eta_{p'}(x')\right]= i \frac{\pi}{k} \delta_{p,p'}\text{sign}(x-x')$.
Importantly, due to this non-trivial commutation relation, one can show that this bosonic representation indeed satisfies the $\mathbb{Z}_k$ parafermion statistics in Eqs.~\eqref{eq:para_stat1} and~\eqref{eq:para_stat2} (see SM~\ref{secSm:para}).
\begin{figure}[t]
		\centering
		\includegraphics[scale=0.29]{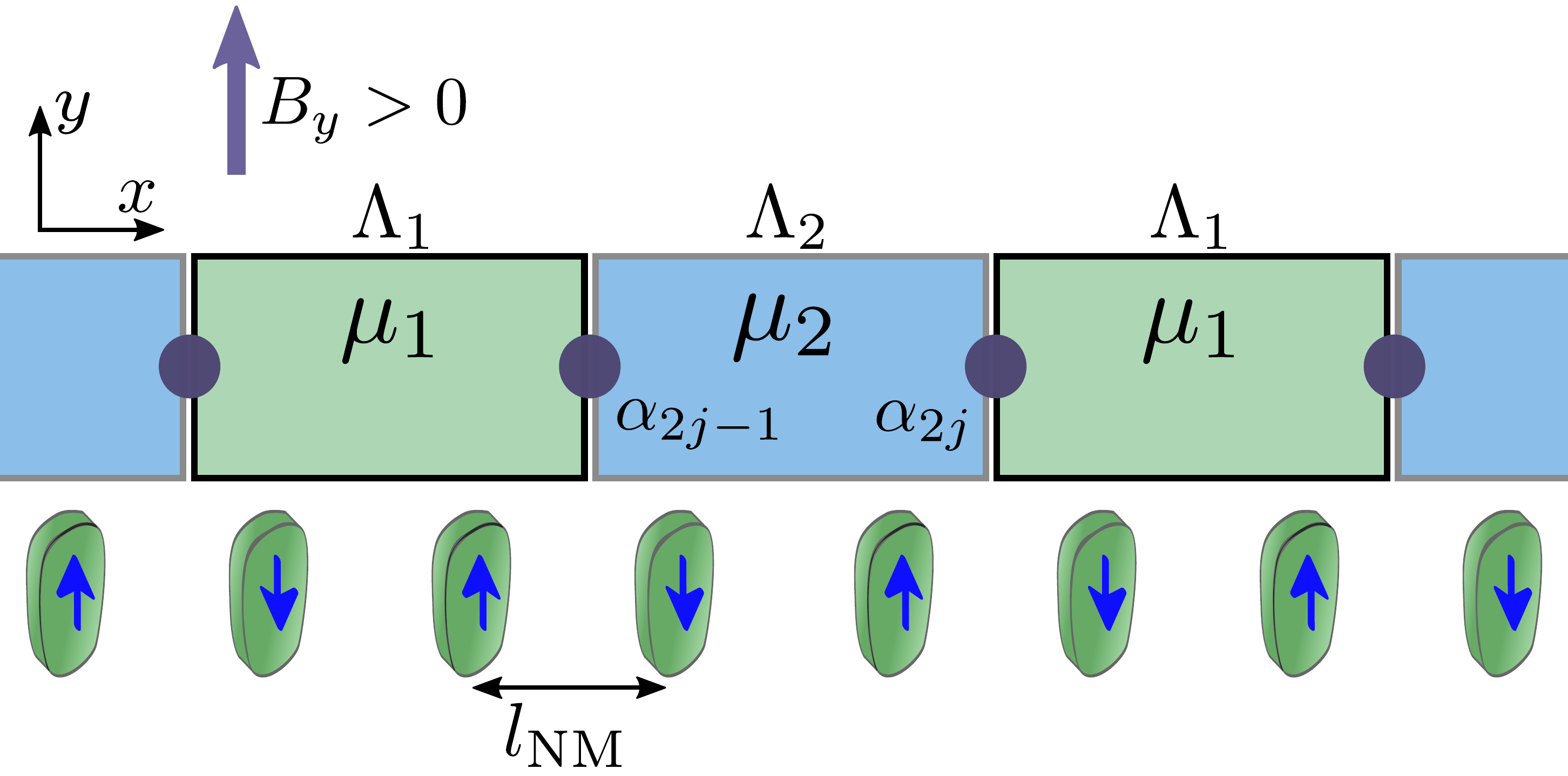}
		\caption{
			Scheme to localize $\mathbb{Z}_4$ parafermions in a single Rashba NW. The alternating values of chemical potential $\mu_i$ generate domains with alternating non-zero amplitudes $\Lambda_i$. 
			At the interface between two such domains, a single  zero-energy parafermion mode $\alpha_{2j}$ (purple disk) emerges.
			The remaining propagating modes can be  gapped out by an additional 
			magnetic field spatially rotating
			in the $xy$-plane with a substantial Fourier component of period $2l_{\rm NM}=\pi/2k_{so}$. 
			Such magnetic textures can be implemented by a row of nanomagnets (green) with alternating magnetizations (blue arrows).}
		\label{fig:para}
\end{figure} 
Using then Eq.~\eqref{eq:sigma}, we  obtain the long-wavelength expansions $(e^{i2\pi (\delta_{p,1}-1/2)/k}\chi_{2,i}^{\dagger}\chi_{1,i+\delta_{p,1}})^{\alpha} \approx (-1)^{\alpha+1}(f_k^{(\alpha)}\cos(k \eta_p(x)) + a^2d_k^{(\alpha)} \left[\partial_x\eta_p(x)\right]^2 )$, where
\begin{align}
        d_k^{(\alpha)} &  =  \alpha \left[A_k^2 + (k-1)^2 B_k^2\right] 
        \nonumber\\&
        + (\alpha/2)(\alpha-1)\left[A_k^2-(k-1) B_k^2\right]^2,\\
	f_k^{(\alpha)} &=-2 \sum_{\substack{n\, \text{odd } >0 }}^{\alpha}\left(A_k B_k\right)^n\left(\begin{matrix}\alpha \\ n\end{matrix}\right)\left(\begin{matrix} n\\ 
	\frac{n+1}{2}\end{matrix}\right),
	\end{align}
	with $d_k^{(\alpha)} = d_k^{(k-\alpha)}$ and $ f_k^{(\alpha)} = f_k^{(k-\alpha)}$. This eventually allows us to map 
	the  clock  model Eq.~\eqref{eq:para} in the continuum limit to the $\mathbb{Z}_k$  DSG model Eq.~\eqref{eq:DSGM},
	with the identification ${\Lambda_i} / {(4\pi^2 a)}=\sum_{\alpha=1}^{k-1} f^{(\alpha)}_{k} J_{i,\alpha}$ and ${v} / {(2\pi a)} = \sum_{\alpha=1}^{k-1} d^{(\alpha)}_{k} J_{1,\alpha} = \sum_{\alpha=1}^{k-1} d^{(\alpha)}_{k} J_{2,\alpha}$. For our special case $k=4$, we find $f^{(\alpha)}_{4} = -\alpha$ and $d^{(\alpha)}_{4}= \alpha \left( \alpha+9\right)/2$.

This demonstrates that a Rashba NW at filling factor $\nu=1/2$ is brought into a gapless phase hosting $\mathbb{Z}_4$ parafermion modes~\cite{Fateev85}. Intruigingly, the emergence of these exotic states is  the result of a competition between intrinsic back-scattering processes induced by strong electron-electron interactions  due to the interplay between SOI and magnetic fields in a single-band NW.

\textit{Parafermion bound states.} Since the self-dual DSG model describes a gapless phase,  parafermion modes are propagating. Nevertheless, a partial gap hosting bound states can emerge if the balance between the two competing back-scattering processes is broken by  a magnetic field $B_y$ applied parallel to the SOI direction. 
If, say, the corresponding Zeeman energy $\Delta_Z^{(y)}>0$, 
the phase with $\Lambda_2=0$ ($\Lambda_1=0$) can only be resonant when the chemical potential is tuned to a value $\mu_1 > \mu_{1/2}$ ($\mu_2 < \mu_{1/2}$), where $\mu_{1(2)}=\mu_{1/2}\pm \Delta_Z^{(y)}/2$. In this way, two distinct partially gapped regions can be engineered. In both phases, the remaining propagating modes can be gapped by applying an additional magnetic field pointing perpendicular to the SOI vector and
having a substantial Fourier component of  period  $2\pi/4 k_{so}$~\cite{Oreg14,Klinovaja12}. We also note that there is no need for SOI if a spatially periodic magnetic field has non-equal magnitudes of Zeeman terms along the, say, $x$ and $y$ axes. In this case, the period is determined by the Fermi wavevector $k_F$ and is given by $2\pi/4k_F$. 
For instance, such magnetic textures can be implemented by arrays of nanomagnets~\cite{Braunecker10,Karmakar11,Klinovaja12,Fatin16,Maurer18,Mohanta19,Desjardins2019} with alternating magnetization and separated by a distance $l_{\rm NM}=\pi/4 k_{so}$.  A possible scheme to localize parafermion bound states  is sketched in Fig.~\ref{fig:para}. If the chemical potential alternates between $\mu_1$ and $\mu_2$  in consecutive domains, the two gapped phases are also alternating. Let us index with $j$ the pairs of neighbouring domains formed by a $\Lambda_1$-dominated and a $\Lambda_2$-dominated phase. The fields are pinned to the values $\eta_1 = \frac{2n_j+1}{4}\pi$, for $\Lambda_2=0$, and $\eta_2 = \frac{2m_j+1}{4}\pi$, for $\Lambda_1=0$, while the field $\eta_3$ is pinned uniformly throughout the system. Here, the integer-valued operators $n_j$ and $m_j$ satisfy $[n_j,m_p] = \frac{i}{\pi}\text{sign}(j - p - \epsilon)$, with $\epsilon$ being a vanishingly small positive quantity. Following standard methods~\cite{Clarke13,Klinovaja14a}, one can introduce the following operators at the interfaces
	\begin{equation}
	\alpha_{2j-1} = e^{i \pi \left(m_j +  n_j\right)/2},~~~ \alpha_{2j} = e^{i \pi \left(m_j +  n_{j+1}\right)/2},
	\end{equation}
	which are zero-energy modes and obey  $\mathbb{Z}_4$ parafermion statistics [see Eqs.~\eqref{eq:para_stat1} and \eqref{eq:para_stat2}].
However, we note that there could be fluctuations in the chemical potential or in the strength of the SOI energy. As a result, parafermion bound states appearing in the middle of the gap at each interface could be at different energies, thus resulting in an additional phase difference between them. Let us also note that the obtained phase can be stabilized at values of the chemical potential close to $\mu_{1/2}$. Large deviations from these values are detrimental, especially, if they become larger than the gap opened by $\Lambda_{1,2}$ terms.

Like other schemes of parafermions in one-dimensional systems~\cite{Klinovaja14a,Klinovaja14c,Oreg14}, our bound states could be sensitive to disorder as described above. However, as was shown numerically in recent studies, the degeneracy still can be stabilized in the regime of strong electron-electron interactions~\cite{Calzona18}. Nevertheless, our setup is very promising for demonstrating the existence of parafermions due to its relative simplicity as it requires only intrinsic ingredients such as spin-orbit and electron-electron interactions and weak external magnetic fields but no superconductivity nor exotic quantum Hall states.
Candidate materials to test our predictions are semiconducting NWs such as InAs or InSb~\cite{Prada20,Sato19,Hsu2019}, 
 or, in particular, Ge/Si~\cite{Scappucci20} as well as ballistic one-dimensional channels in $\rm LaAlO_3/SrTiO_3$~\cite{Annadi18,Briggeman19,Briggeman20} 
 or in GaAs~\cite{Kumar19}.
  A first experimental signature of the new phase would be the fractional conductance $G=2e^2/3h$. The localized parafermion bound states (see Fig.~\ref{fig:para}) would then show up as zero-bias conductance peaks or they could be detected in  Aharonov-Bohm setups~\cite{Rainis14}. 

\textit{Conclusions.} We have investigated an interacting Rashba NW at filling factor $\nu=1/2$. We have shown that 
interactions stabilize two resonant multi-particle processes. The competition between these two processes brings the system into a gapless parafermion phase, 
described by the $\mathbb{Z}_4$ self-dual sine-Gordon model.
We provided a mapping between parafermion operators and bosonic fields and 
showed that the $\mathbb{Z}_k$ DSG is the low-energy limit of the $\mathbb{Z}_k$ parafermion clock chain model. 
Finally, we proposed a scheme to generate zero-energy parafermion bound states.

\textit{Acknowledgments.} This work was supported by the Swiss National Science Foundation and NCCR QSIT. This project received funding from the European Union's Horizon 2020 research and innovation program (ERC Starting Grant, grant agreement No 757725).

\clearpage

\widetext
\begin{center}
	\textbf{\large Supplemental Material: Clock model and parafermions in Rashba nanowires}\\
	\vspace{8pt}
	Flavio Ronetti,$^{1}$ 
	Daniel Loss,$^{1}$ and Jelena Klinovaja$^{1}$ \\ \vspace{4pt}
	$^{1}$ {\it Department of Physics, University of Basel,
		Klingelbergstrasse 82, CH-4056 Basel, Switzerland}
\end{center}

\setcounter{section}{0}
\setcounter{equation}{0}
\setcounter{figure}{0}
\makeatletter
\renewcommand{\thesection}{S\arabic{section}}
\renewcommand{\theequation}{S\arabic{equation}}
\renewcommand{\thefigure}{S\arabic{figure}}
\titleformat{\section}[hang]{\large\bfseries}{\thesection.}{5pt}{}

\section{\label{secSm:action}Renormalization group and effective action}
In this Section, we present renormalization group (RG) equations for the cosine perturbations appearing in the main text and we provide details for the calculation of the effective action for the Hamiltonian $H_{\rm DSG}$ defined in Eq.~(7) of the main text.

In order to derive the RG equations, one has to specify the form of the small-momentum interaction matrix $N$ appearing in the Hamiltonian $H_0$,
\begin{equation}
H_0 = \sum_{\sigma,\sigma',r,r'}\int dx \hspace{1mm}n_{r \sigma}(x)N_{r\sigma r'\sigma'}n_{r' \sigma'}(x),\label{eq:H}
\end{equation}
where $n_{R\sigma}(x)=R_{\sigma}^{\dagger}(x)R_{\sigma}(x)$ and $n_{L\sigma}(x)=L_{\sigma}^{\dagger}(x)L_{\sigma}(x)$ are the electronic densities for each channel. The $4\times 4$-matrix $N$ is given by
\begin{equation}
N = \left(
\begin{matrix}
v + g_{\parallel} & g_{\perp} & g_{\parallel} & g_{\perp}\\ g_{\perp} & v + g_{\parallel} & g_{\perp} & g_{\parallel} \\g_{\parallel} & g_{\perp} & v + g_{\parallel} & g_{\perp}\\ g_{\perp} &  g_{\parallel} & g_{\perp} & v + g_{\parallel} 
\end{matrix}\right),
\end{equation}
where $v=2\mu_{1/2}/k_F$ is the velocity and $g_{\parallel}$ ($g_{\perp}$)  the coupling constants for density-density interaction between electrons with the same (opposite) spins, respectively. It is useful to express the Hamiltonian in the basis in which $H_0$ is diagonal, which is given by the following bosonic fields:
\begin{equation}
\left(\begin{matrix}
\theta_{\rho}(x) \\ \phi_{\rho}(x)\\\theta_{\sigma}(x)\\\phi_{\sigma}(x) 
\end{matrix}\right) = \frac{1}{2\sqrt{2}}\left(\begin{matrix}
1 & 1 & 1 & 1 \\ 1 & -1 & 1 & -1 \\ 1 & 1 & -1 & -1 \\ 1 & -1 & -1 & 1
\end{matrix}\right) \left(\begin{matrix}
\phi_{L\uparrow}(x) \\ \phi_{R\uparrow}(x)\\\phi_{L\downarrow}(x)\\\phi_{R\downarrow}(x) 
\end{matrix}\right).
\end{equation}
In order to reproduce the correct commutation relations among the two perturbations in the original fermionic representation [see Eq.~(2) in the main text], the bosonic fields have to satisfy the following commutation relations:
\begin{align}
&\left[\phi_{\rho}(x),\theta_{\rho}(x')\right]=i \pi\ \text{sign}(x-x'),\\
&\left[\phi_{\sigma}(x),\phi_{\sigma}(x')\right]=-i \pi\ \text{sign}(x-x'),\\
&\left[\theta_{\sigma}(x),\theta_{\sigma}(x')\right]=i \pi\ \text{sign}(x-x').
\end{align} 
The commutators between  right/left mover fields are given by $\left[\phi_{r\sigma}(x),\phi_{r'\sigma'}(x')\right] = i \pi M_{r\sigma ,r' \sigma'} \text{sign}(x-x')$. Here, the $4\times 4$-matrix $M$ can be expressed in the basis $\left(\phi_{R\uparrow},\phi_{R\downarrow},\phi_{L\uparrow},\phi_{L\downarrow}\right)$ as
\begin{equation}
M = \left(\begin{matrix}
1& 1& 1 & -1 \\ 1 & 1 & -1 & 1 \\ 1 & -1 & -1 & -1 \\ -1 & 1 & -1 & -1 
\end{matrix}\right).\label{eq:boso_comm_matrix_sol}
\end{equation}
The Hamiltonian becomes 
\begin{align}
H &= H_0 + \frac{\Lambda_1}{4\pi^2a^2} \int dx ~ \cos\left[\sqrt{2}\left(2\phi_{\rho} - \theta_{\sigma} + \phi_{\sigma}\right)\right] + \frac{\Lambda_2}{4\pi^2a^2} \int dx ~ \cos\left[\sqrt{2}\left(2\phi_{\rho} - \theta_{\sigma} - \phi_{\sigma}\right)\right],
\end{align}
where
\begin{equation}
H_0 = \frac{1}{2\pi} \sum_{\mu=\rho,\sigma}\, \int dx \left[ u_{\mu} K_{\mu} ~\left[\partial_x\theta_{\mu}(x)\right]^2 +  \frac{u_{\mu}}{K_{\mu}}  ~\left[\partial_x\phi_{\mu}(x)\right]^2 \right].
\end{equation}
The RG equations for the two cosine perturbations are given by
\begin{align}
\frac{d}{d l}\tilde{\Lambda}\left(l\right) &= \left[2-2 \tilde{K}_{\rho}(l)-\frac{1}{2}\left(\tilde{K}_{\sigma}(l)+\tilde{K}^{-1}_{\sigma}(l)\right)\right]\tilde{\Lambda}(l)\label{eq:RG1}\\\frac{d}{d l}\tilde{K}_{\rho}\left(l\right) &=- 4\tilde{K}^2_{\rho}(l)\tilde{\Lambda}^2(l),\\\frac{d}{d l}\tilde{K}_{\sigma}\left(l\right) &= \left[1-\tilde{K}^2_{\sigma}(l)\right]\tilde{\Lambda}^2(l).
\end{align}
Here, we introduce $\Lambda\equiv\Lambda_1=\Lambda_2$ and use the rescaled amplitude $\tilde{\Lambda}= \Lambda/v$.
These equations have been used to derive the phase diagram shown in Fig.~2 of the main text. Importantly, we note that the Luttinger liquid parameters
$K_{\rho}$ and $K_{\sigma}$
are flowing under the RG and, according to the above equations, they flow to the values 
\begin{align}
K_{\rho}&\rightarrow 0,\\K_{\sigma}&\rightarrow 1.
\end{align}
In order to derive the effective action, we change the basis as follows:
\begin{equation}
\left(\begin{matrix}\eta_1\\ \eta_2\\\eta_3\\\eta_4\end{matrix}\right)  =  \frac{1}{2\sqrt{2}}\left(\begin{matrix}
0 & -2  & 1& -1 \\ 0 & -2 & 1 & 1 \\ 0 & 1 & 0 & 0 \\ \frac{1}{2}& 0 &1 & 0 
\end{matrix}\right)\left(\begin{matrix}\theta_{\rho}\\\phi_{\rho}\\ \theta_{\sigma}\\ \phi_{\sigma} \end{matrix}\right).
\end{equation}
The Hamiltonian becomes
\begin{equation}
H =\frac{1}{2\pi} \int dx ~\partial_x\vec{\eta}^T(x) \tilde{N}~ \partial_x\vec{\eta}(x)+ \frac{\Lambda_1}{4\pi^2a^2} \int dx ~ \cos\left[4\eta_1(x)\right] + \frac{\Lambda_2}{4\pi^2a^2} \int dx ~ \cos\left[4\eta_2(x)\right],
\end{equation}
where
\begin{equation}
\tilde{N} =\left(
\begin{array}{cccc}
v_1 & g_{12} & g_{13} &g_{14} \\
g_{12} &v_2 & g_{23} & g_{24} \\
g_{13} & g_{14} & v_3 &g_{34} \\
g_{14} & g_{24} & g_{34} & v_4 \\
\end{array}
\right)
\end{equation}
and
\begin{align}
v_1 &= v_2 = \frac{2K_{\sigma }^2 u_{\sigma}+8 u_{\rho}  K_{\rho } K_{\sigma }+2u_{\sigma}}{ K_{\sigma }},\\
g_{12}& =  \frac{2K_{\sigma }^2 u_{\sigma}+8 u_{\rho}  K_{\rho } K_{\sigma }-2u_{\sigma}}{ K_{\sigma }},\\
g_{13} &= g_{23} =8 K_{\sigma}u_{\sigma}+32 K_{\rho
}u_{\rho},\label{eq:param1}\\
g_{14} &= g_{24} =  -16K_{\rho }u_{\rho},\\
v_3 & =128
K_{\rho }u_{\rho}+32K_{\sigma }u_{\sigma}+8\frac{u_{\rho}}{K_{\rho }},\\
v_4 &=32 K_{\rho}u_{\rho},\\
g_{34} &=   -64K_{\rho }u_{\rho}.\label{eq:param2}
\end{align}
The commutators between the bosonic fields are given by
\begin{align}
\left[\eta_1(x),\eta_2(x')\right] &= i \frac{\pi}{4} \text{sign}(x-x'),\\
\left[\eta_3(x),\eta_4(x')\right] &=  i  \frac{\pi}{16}\text{sign}(x-x'),\\
\left[\eta_4(x),\eta_4(x')\right] &=i  \frac{\pi}{8} \text{sign}(x-x'),\\
\left[\eta_1(x),\eta_1(x')\right] &= \left[\eta_2(x),\eta_2(x')\right] = \left[\eta_1(x),\eta_3(x')\right] = \left[\eta_1(x),\eta_4(x')\right] = \left[\eta_2(x),\eta_3(x')\right] = \left[\eta_2(x),\eta_4(x')\right] = 0.
\end{align}

The total action can be divided into four contributions,
\begin{equation}
\mathcal{S}=\mathcal{S}_{12}+\mathcal{S}_{34}+\mathcal{S}_{\rm int}+\mathcal{S}_{\rm cos},
\end{equation}
where 
\begin{align}
\mathcal{S}_{12}&=\frac{1}{2\pi}\int dq\ d\omega ~ \left(\eta_1(q,\omega),\eta_2(q,\omega)\right)\left(\begin{matrix}
-v_1 q^2 & g_{12}q^2 + i \frac{q}{4} \omega\\ g_{12}q^2 + i \frac{q}{4} \omega & -v_2 q^2
\end{matrix}\right)\left(\eta_1(-q,\omega),\eta_2(-q,\omega)\right)^T,\\
\mathcal{S}_{34}&=\frac{1}{2\pi}\int dq\ d\omega ~ \left(\eta_3(q,\omega),\eta_4(q,\omega)\right)\left(\begin{matrix}
-v_3 q^2 & g_{34}q^2 + i \frac{q}{16} \omega\\ g_{34}q^2 + i \frac{q}{16} \omega & -v_4 q^2+ i \frac{q}{4} \omega
\end{matrix}\right)\left(\eta_3(-q,\omega),\eta_4(-q,\omega)\right)^T,\\
\mathcal{S}_{\rm int}&=\frac{1}{2\pi}\int dq\ d\omega ~ q^2\sum_{i,k}\frac{g_{ik}}{2}\eta_i(q,\omega) \eta_k(-q,\omega),\\
\mathcal{S}_{\rm cos}&=\frac{1}{4\pi^2 a^2}\int dx\int d\tau \left[\Lambda_1 \cos(4 \eta_1(x,\tau)) + \Lambda_2 \cos(4 \eta_2(x,\tau))\right].
\end{align}
Then, by using the following relation,
\begin{equation}
\prod_{k}\left(\int~\frac{d u_k d u_k^*}{2\pi i}\right)e^{-\sum_{ij} u_i^*A_{ij}u_j+\sum_{i}h_i^* u_i+\sum_{i}u_i^* h_i}=\frac{e^{\sum_{ij}h_i^* (A^{-1})_{ij}h_j}}{\text{Det}A},
\end{equation}
we can integrate out the bosonic fields $\eta_3$ and $\eta_4$, thus obtaining the following effective action
\begin{equation}
\mathcal{S}_{\rm eff} = \mathcal{S}_{12} + \tilde{\mathcal{S}} + \mathcal{S}_{\rm cos},
\end{equation}
where
\begin{align}
\tilde{\mathcal{S}}&=\frac{1}{2\pi}\int dq\ d\omega ~ \left(g_{13}\eta_1(q,\omega)+g_{23}\eta_2(q,\omega),g_{14}\eta_1(q,\omega)+g_{24}\eta_2(q,\omega)\right)q^2\nonumber\\&\times\left(
\begin{array}{cc}
\frac{64 (4 q v_4-i \omega )}{256 q g_{34}^2+16 i \left(q^2+1\right) \omega 
	g_{34}-q \left(256 v_3 v_4 q^2-64 i v_3 \omega  q+\omega
	^2\right)} &\frac{256 g_{34} q+16 i \omega }{256 q g_{34}^2+16 i \left(q^2+1\right) \omega 
	g_{34}-q \left(256 v_3 v_4 q^2-64 i v_3 \omega  q+\omega
	^2\right)}\\
\frac{256 g_{34} q+16 i \omega }{256 q g_{34}^2+16 i \left(q^2+1\right) \omega 
	g_{34}-q \left(256 v_3 v_4 q^2-64 i v_3 \omega  q+\omega
	^2\right)} & \frac{256 q v_3}{256 q g_{34}^2+16 i \left(q^2+1\right) \omega
	g_{34}-q \left(256 v_3 v_4 q^2-64 i v_3 \omega  q+\omega
	^2\right)} \\
\end{array}
\right) \nonumber\\&\times
\left(g_{13}\eta_1(-q,\omega)+g_{23}\eta_2(-q,\omega),g_{14}\eta_1(-q,\omega)+g_{24}\eta_2(-q,\omega)\right)^T.
\end{align}
We note that, since the Luttinger liquid parameters flow as $K_{\rho}\rightarrow 0$, $K_{\sigma}\rightarrow 1$, the coefficients in Eqs.~\eqref{eq:param1}-\eqref{eq:param2} become such that $\tilde{\mathcal{S}}\rightarrow 0$. Moreover, one also finds that $g_{12}\rightarrow 0$ and $v_1 = v_2  = 4u_{\sigma} \equiv v$. As a result, the effective action becomes
\begin{equation}
\mathcal{S}_{\rm eff} = \mathcal{S}_{12}+\mathcal{S}_{\rm cos},
\end{equation}
with 
\begin{align}
\mathcal{S}_{12}&=\frac{1}{2\pi}\int dq\ d\omega ~ \left(\eta_1(q,\omega),\eta_2(q,\omega)\right)\left(\begin{matrix}
-v q^2 & + i \frac{q}{4} \omega\\ + i \frac{q}{4} \omega & -v q^2 \end{matrix}\right)\left(\eta_1(-q,\omega),\eta_2(-q,\omega)\right)^T,
\end{align}
which corresponds to the Hamiltonian $H_{\rm DSG}$ defined in Eq.~(7) of the main text.

\section{\label{secSm:para}Low-energy limit of the $\mathbb{Z}_k$ clock model }

In this Section, the bosonized forms of the operators $\sigma_{p,i}$ ($p=1,2$), given in Eq~(11)  of the main text, are used to prove that the DSG Hamiltonian $H_{\rm DSG}$, defined  in Eq.~(7) of the main text, is also describing the low-energy limit of the $\mathbb{Z}_4$ parafermion clock chain model. Since this low-energy correspondence is valid for the general case of $\mathbb{Z}_k$ symmetry, we provide the mapping for an arbitary value of $k>1$. The complete mapping proceeds in two  steps. First, we remind the reader of the well-known mapping from the $\mathbb{Z}_k$ clock model to the $\mathbb{Z}_k$ parafermion chain~\cite{Fendley}. We emphasize that, since in this step no assumption is necessary, these two models, the $\mathbb{Z}_k$ parafermion chain and the $\mathbb{Z}_k$ clock model,  are entirely equivalent: for this reason, we denote both of the corresponding Hamiltonians with the same symbol $H_{\rm cl}$. In a second step, we introduce a  representation of $\sigma_{1,i}$ and $\sigma_{2,i}$ in terms of the bosonic fields $\eta_1$ and $\eta_2$ introduced in the main text. We prove that this bosonic representation implements the correct commutation relations for $\sigma_{1,i}$ and $\sigma_{2,i}$. Then, we exploit them to prove that the $\mathbb{Z}_k$ self-dual sine-Gordon model, $H_{\rm DSG}$, is the low-energy limit of the clock model and, therefore, of its parafermion representation.\\
\subsection{From $\mathbb{Z}_k$ clock model to $\mathbb{Z}_k$ parafermion chain}
The one-dimensional lattice Hamiltonian for the $\mathbb{Z}_k$ clock model reads~\cite{Fendley,Fateev} 
\begin{equation}
H_{\rm cl}=- \sum_{\alpha=1}^{k-1}\left[\sum_{i=1}^{n-1}J_{1,\alpha}\left(\sigma_{i}^{\dagger}\sigma_{i+1}\right)^{\alpha} +\sum_{i=1}^{n} J^*_{2,\alpha}\left(\tau_i\right)^{\alpha}\right],\label{eq:pottsSM}
\end{equation}
where $n=L/a$, with $a$ being the lattice constant, and $k>1$. The operators $\sigma_{i}$ and $\tau_i$ satisfy the following set of relations:
\begin{align}
\sigma_{i}^k = 1, \hspace{3mm} &\sigma_{i}^{k-1} = \sigma_{i}^{\dagger},\hspace{3mm}\tau_{i}^k = 1, \hspace{3mm} \tau_{i}^{k-1} = \tau_{i}^{\dagger}\label{eq:commsigma1}\\
&\sigma_{i}\tau_{i} = \omega\tau_i\sigma_{i},\label{eq:commsigma2}
\end{align}
where $\omega= e^{ \frac{2\pi i}{k}}$ and the local operators $\sigma_{i}$ and $\tau_j$ commute at different sites $i\neq j$.
In order for the Hamiltonian $H_{\rm cl}$ in Eq.~(\ref{eq:pottsSM}) to be hermitian, the coefficients must satisfy  the following relations: $J_{1,\alpha}^* = J_{1,k-\alpha}$ and $J_{2,\alpha}^* = J_{2,k-\alpha}$. 
Next, $H_{\rm cl}$ expressed in terms of the clock operators $\sigma_{i}$ and $\tau_{i}$ can be mapped onto a parafermion  representation with the help of the  operators $\chi_{p,i}$ defined as 
\begin{equation}
\chi_{1,i}=\sigma_{i}\prod_{l<i} \tau_l, ~~~ \chi_{2,i} =\omega^{(k-1)/2}  \chi_{1,i}\tau_i .
\end{equation}
The operators $\chi_{p,i}$ (for $p=1,2$) obey  $\mathbb{Z}_k$ parafermion statistics,
\begin{align}
&\chi_{p,i}^k = 1, ~\chi_{p,i}^{k-1} = \chi_{p,i}^{\dagger},\label{eq:para_stat1_SM}\\
&\chi_{p,i} \chi_{p',m} = \omega \chi_{p',m} \chi_{p,i},\,\, i< m,\,\, {\rm and}\,\,\, \chi_{1,i} \chi_{2,i} = \omega \chi_{2,i} \chi_{1,i}
\label{eq:para_stat2_SM}.
\end{align} 
Using these relations we can map the clock model defined in Eq.~(\ref{eq:pottsSM}) onto the parafermion chain~\cite{Fradkin} 
\begin{equation}
H_{\rm cl}=-\sum_{\alpha=1}^{k-1}\left[\sum_{i=1}^{n-1}J_{1,\alpha}\left(\omega^{-(k-1)/2}\chi_{2,i}^{\dagger}\chi_{1,i+1}\right)^{\alpha} +\sum_{i=1}^{n} J^*_{2,\alpha}\left(\omega^{-(k-1)/2}\chi_{1,i}^{\dagger}\chi_{2,i}\right)^{\alpha}\right].\label{seq:parafermions}
\end{equation}
This Hamiltonian can be rewritten as
\begin{align}
H_{\rm cl}=-\sum_{\alpha=1}^{k-1}\left[\sum_{j=1}^{n-1}J_{1,\alpha}\left(\omega^{-(k-1)/2}\chi_{2,j}^{\dagger}\chi_{1,j+1}\right)^{\alpha} +\sum_{j=1}^{n} J_{2,\alpha}\left(\omega^{(k-1)/2}\chi_{2,j}^{\dagger}\chi_{1,j}\right)^{\alpha}\right]\nonumber\\=-\sum_{\alpha=1}^{k-1}(-1)^{\alpha}
\left[\sum_{j=1}^{n-1}J_{1,\alpha}\left(e^{i\pi /k}\chi_{2,j}^{\dagger}\chi_{1,j+1}\right)^{\alpha}+\sum_{j=1}^{n}J_{2,\alpha}\left(e^{-i\pi /k} \chi_{2,j}^{\dagger}\chi_{1,j}\right)^{\alpha}\right],
\label{eq:para2}
\end{align}
where [with respect to Eq.~\eqref{seq:parafermions}] we rewrote the second term  using the hermitian conjugate of the same term in Eq.~\eqref{seq:parafermions}, thus obtaining the form given in the main text [see Eq.~(8)].

\subsection{$\mathbb{Z}_k$ double sine-Gordon model (DSGM)}
As preparation for the mapping of the clock model Eq.~\eqref{eq:pottsSM} onto the DSGM in the  continuum limit $a/L \ll 1$, we 
recall some of  the essential properties of the DSGM given by 
\begin{align}
H_{\rm DSG} &=\int_{0}^{L} dx \left[\frac{u K}{2\pi}   \left(\partial_x \theta(x)\right)^2  + \frac{u}{2\pi K}   \left(\partial_x \phi(x)\right)^2+ \frac{\Lambda_1}{4\pi^2a^{2}} \cos\left(\sqrt{2k}\phi(x)\right) + \frac{\Lambda_2}{4\pi^2a^{2}} \cos\left( \sqrt{2k}\theta(x)\right)\right] ,
\label{eq:DSGphitheta}
\end{align}
where we fix $K\le 1$ and where the dual bosonic fields $\phi$ and $\theta$ satisfy the commutation relation
\begin{equation}
\label{eq:thetacomm}
\left[\phi(x),\theta(x')\right] = \frac{i \pi}{2}\text{sign}(x-x')
\end{equation}
and are related to the $\eta_p$ fields as
\begin{align}
\eta_1(x)&=\sqrt{\frac{2}{k}}\phi(x),\\\eta_2(x)&=\sqrt{\frac{2}{k}}\theta(x).
\end{align}
The $\mathbb{Z}_k$ DSGM in terms of $\eta_p$ reads
\begin{align}
H_{\rm DSG} &= \int_{0}^{L} dx \left[\frac{v K}{2\pi}   \left(\partial_x \eta_2(x)\right)^2  + \frac{v}{2\pi K}   \left(\partial_x \eta_1(x)\right)^2+ \frac{\Lambda_1}{4\pi^2a^{2}} \cos\left(k \eta_1(x)\right) + \frac{\Lambda_2}{4\pi^2a^{2}} \cos\left(k \eta_2(x)\right)\right] ,
\label{eq:DSGeta}
\end{align}
where $v = 2u /k$.

Next, we list some useful relations for bosonic operators needed in the following derivation~\cite{Giamarchi,Manisha}. If  $A$ and $B$ are bosonic operators that are linear functions of bosonic creation and annihilation operators, we have
\begin{equation}
e^Ae^B = :e^{A+B}: e^{\frac{1}{2}\left\langle A^2 + 2AB + B^2\right\rangle }\, , \label{eq:expord}
\end{equation}
where $:\dots:$ stands for normal ordering and $\left\langle...\right\rangle$ denotes the bosonic ground state expectation value~\cite{Giamarchi}. 
Further we will make use of the relations~\cite{Manisha,Giamarchi}:
\begin{align}
\left\langle\left[\phi(x,\tau)- \phi(0,0)\right]^2\right\rangle &= K\ln \frac{\sqrt{x^2 + u^2\tau^2}}{a}, ~~~\left\langle\left[\theta(x,\tau)- \theta(0,0)\right]^2\right\rangle = \frac{1}{K}\ln \frac{\sqrt{x^2 + u^2\tau^2}}{a},\\
\left\langle \phi^2(0,0)\right\rangle &= -\frac{K}{2}\ln \frac{2\pi a}{L}, ~~~\left\langle \theta^2(0,0)\right\rangle = -\frac{1}{2 K}\ln  \frac{2\pi a}{L},
\end{align} 
where, again, 
$a$ is the lattice spacing and the expectation values and time evolution (in imaginary time $\tau$), $\phi(x,\tau)$ and $\theta(x,\tau)$, are governed by the kinetic term of Eq.~\eqref{eq:DSGphitheta}. These relations can be expressed in terms of the bosonic fields $\eta_p$ as
\begin{align}
\left\langle\left[\eta_1(x,\tau)- \eta_1(0,0)\right]^2\right\rangle &= \frac{2K}{k}\ln \frac{\sqrt{x^2 + u^2\tau^2}}{a}, ~~~\left\langle\left[\eta_2(x,\tau)- \eta_2(0,0)\right]^2\right\rangle = \frac{2}{Kk}\ln \frac{\sqrt{x^2 + u^2\tau^2}}{a},\\
\left\langle \eta^2_1(0,0)\right\rangle &= -\frac{K}{k}\ln \frac{2\pi a}{L},~~~\left\langle \eta^2_2(0,0)\right\rangle = -\frac{1}{kK}\ln \frac{2\pi a}{L}.
\end{align} 
In the following, we will be interested in the case $\tau = 0$ and suppress the $\tau$-argument. Note that above relations are valid for
translationally invariant systems, e.g. satisfied for periodic boundary conditions. Thus, our mapping is strictly speaking restricted to this case. However,
it is straightforward to describe boundary effects in the continuum theory by allowing for domain walls [see main text and section \ref{secSm:para}].
By using these relations, we obtain
\begin{align}
\label{eq:loga}
&e^{-\frac{1}{2}c^2\langle \left[\eta_{1}(0,0)\right]^2\rangle} = \left(\frac{2\pi a}{L}\right)^{K c^2 / 2k},\\
&e^{-\frac{1}{2}c^2\langle \left[\eta_{2}(0,0)\right]^2\rangle}= \left(\frac{2\pi a}{L}\right)^ {c^2 / \left(2k K\right)},
\end{align}
where $c$ is some real constant.
Using Eq.~\eqref{eq:expord} for $A=i c_1 \eta_p(x_j)$ and $B=i c_2 \eta_p(x_j)$, we find
\begin{equation}
e^{i c_1 \eta_1(x_j)}e^{i c_2 \eta_1(x_j)} =\, :e^{i (c_1+c_2) \eta_1(x_j)}: \left(\frac{2\pi a}{L}\right)^{K (c_1+c_2)^2/2k}, ~~~e^{i c_1 \eta_2(x_j)}e^{i c_2 \eta_2(x_j)} = \, :e^{i (c_1+c_2) \eta_2(x_j)}: \left(\frac{2\pi a}{L}\right)^{ (c_1+c_2)^2/\left(2k K\right)}, 
\end{equation}
while for $A=-i c \eta_p(x_j)$ and $B=i c \eta_p(x_{j+1})$ we get 
\begin{align}
\label{eq:normalordereta}
e^{-i c \eta_1(x_j)}e^{i c \eta_1(x_{j+1})} &=\, :e^{-i c \left[\eta_1(x_j)-\eta_1(x_{j+1})\right]}: e^{-c^2\frac{K}{k}\ln \frac{\sqrt{a^2}}{a}} = :e^{i a c \partial_j\eta_1(x_j)}: + \dots \\ 
~~~e^{-i c \eta_2(x_j)}e^{i c \eta_2(x_{j+1})} &=\, :e^{-i c \left[\eta_2(x_j)-\eta_2(x_{j+1})\right]}: e^{-c^2\frac{1}{kK}\ln \frac{\sqrt{a^2}}{a}}= :e^{i a c \partial_j\eta_2(x_j)}: + \dots
\end{align}
where $\partial_j \equiv \partial_{x_j}$. The last step, where the derivative is introduced, is valid only in the continuum limit $\frac{a}{L}\rightarrow 0$; the dots $\dots$ stand for the subleading terms that we drop by taking this limit.

For the following calculations, it is important to comment about the $\mathbb{Z}_k$ DSGM when the cosine terms are normal ordered. In this case, the 
Hamiltonian is rewritten as
\begin{align}
H_{\rm DSG} &= \int_{0}^{L} dx \left[\frac{v K}{2\pi}   \left(\partial_x \eta_2(x)\right)^2 + \frac{v}{2\pi K}   \left(\partial_x \eta_1(x)\right)^2  
+ \frac{\Lambda_1}{4\pi^2L^{2}}\left(\frac{2\pi a}{L}\right)^{\frac{k K}{2}-2} :\cos\left(k \eta_1(x)\right): + \frac{\Lambda_2}{4\pi^2L^{2}}\left(\frac{2\pi a}{L}\right)^{\frac{k}{2 K}-2} :\cos\left(k \eta_2(x)\right):\right].
\label{eq:DSGeta_normal}
\end{align}
In this representation, the cosine terms seem to be of higher-order  in $\frac{a}{L}$ compared to the kinetic term and one might argue of dropping them. However, this argument corresponds only to a tree-level RG analysis. For large enough values of $\Lambda_i$, one has to consider also the effect of the higher-order corrections to the RG flow. It has been shown that, for the $\mathbb{Z}_k$ DSGM, according to the third-order RG equations, these cosine terms are relevant and that the system flows to  multicritical fixed points, separating the $\Lambda_1$-dominated and the $\Lambda_2$-dominated phases~\cite{Katharina,Boyanovsky,Chiral_sup}. We also note that, when $K \ne 1$, the system  flows to the phase dominated by that cosine term with the largest scaling dimension (i.e. only a single cosine term remains in the Hamiltonian). 
In conclusion, in order to show consistenly that the continuum limit of the $\mathbb{Z}_k$ clock model is indeed given by the $\mathbb{Z}_k$ DSGM, one has to keep the lowest order cosine terms in $\frac{a}{L}$ in the expansion of the clock model Hamiltonian, even though they might be of higher order compared to other terms in this expansion.

\subsection{Introducing new operators}
For later purposes, it is convenient to rewrite Eq.~\eqref{eq:pottsSM} as
\begin{equation}
H_{\rm cl}=- \sum_{\alpha=1}^{k-1} \left[\sum_{i=1}^{n-1}J_{1,\alpha}\left(\sigma_{1,i}^{\dagger}\sigma_{1,i+1}\right)^{\alpha} +\sum_{i=1}^{n}J_{2,\alpha} \left(\sigma_{2,i}^{\dagger}\sigma_{2,i+1}\right)^{\alpha}\right]\label{eq:pottsSM2},
\end{equation}
where we define $\sigma_{1,i}\equiv \sigma_i$ and $\sigma_{2,i}$ is related to $\tau_i$ as $\tau_i=\sigma_{2,i}^{\dagger}\sigma_{2,i+1}$ and is assumed to possess the following  properties:
\begin{align}
\label{eq:sigma2a}
&\sigma_{2,j}^k = 1, ~~~ \sigma_{2,j}^{k-1} = \sigma_{2,j}^{\dagger}\\
&\sigma_{1,m}\sigma_{2,j}^{\dagger} =e^{i\pi \text{sign}\left(m-j+\epsilon\right)/k} \sigma_{2,j}^{\dagger}\sigma_{1,m}, \label{eq:sigma2b}\\
&\sigma_{1,m}\sigma_{2,j} =e^{-i\pi \text{sign}\left(m-j+\epsilon\right)/k} \sigma_{2,j}\sigma_{1,m},\ \label{eq:sigma2c}
\end{align}
where this expression is assumed in the limit $\epsilon \rightarrow 0^+$. With these properties, the commutation relations between $\sigma_{p,i}$, $p=1,2$, indeed implements the correct commutation relations between $\sigma_{1,m}$ and $\tau_j$, as we can easily verify:
\begin{align}
\label{eq:tausigma}
\sigma_{1,m}\tau_j &= \sigma_{1,m}\sigma_{2,j}^{\dagger}\sigma_{2,j+1}
=e^{i\pi \text{sign}\left(m-j+\epsilon\right)/k} \sigma_{2,j}^{\dagger}\sigma_{1,m}\sigma_{2,j+1} 
\nonumber \\
&= e^{i\pi \text{sign}\left(m-j+\epsilon\right)/k}e^{-i\pi \text{sign}\left(m-j-1+\epsilon\right)/k} \sigma_{2,j}^{\dagger}\sigma_{2,j+1} \sigma_{1,m} \nonumber\\&= e^{i\pi\left[ \text{sign}\left(m-j+\epsilon\right)- \text{sign}\left(m-j-1+\epsilon\right)\right]/k}\tau_j\sigma_{1,m} 
\nonumber \\
&=e^{i2\pi \delta_{m,j}/k} \tau_j\sigma_{1,m},
\end{align}
where we used the fact that, when $m$ and $j$ are integers, the following relation holds  true: 
\begin{equation}
\lim\limits_{\epsilon\rightarrow 0^+}\left[\text{sign}\left(m-j+\epsilon\right)- \text{sign}\left(m-j-1+\epsilon\right)\right]= 2\delta_{m,j}.
\end{equation}
By using Eqs.~\eqref{eq:sigma2a}-\eqref{eq:sigma2c}, we see that the desired properties of $\tau_i$ [see Eqs.~\eqref{eq:commsigma1} and \eqref{eq:commsigma2}] are also satisfied: 
\begin{equation}
\tau_j^k = \left(\sigma^{\dagger}_{2,j}\right)^k\sigma_{2,j+1}^k = 1,
\end{equation}
and
\begin{equation}
\tau_j^{k-1} = \left(\sigma^{\dagger}_{2,j}\right)^{k-1}\sigma_{2,j+1}^{k-1} = \sigma_{2,j}\sigma_{2,j+1}^{\dagger} = \sigma_{2,j+1}^{\dagger}\sigma_{2,j} = \tau^{\dagger}_j.
\end{equation}
It is instructive to note that the parafermion operators $\chi_{p,j}$ can be expressed in a simplified form in terms of the operators $\sigma_{p,j}$ by using the property that $\sigma^{\dagger}_{p,j}\sigma_{p,j}=1$. We find then
\begin{align}
\chi_{1,j} = \sigma_{1,j}\prod_{l<j}\sigma_{2,l}^{\dagger}\sigma_{2,l+1} = \sigma_{1,j}\sigma_{2,1}^{\dagger}\sigma_{2,2}\sigma^{\dagger}_{2,2}\sigma_{2,3}\dots\sigma_{2,j-2}^{\dagger}\sigma_{2,j-1}\sigma_{2,j-1}^{\dagger}\sigma_{2,j} = \sigma_{1,j}\sigma_{2,1}^{\dagger}\sigma_{2,j},
\end{align}
and
\begin{align}
\chi_{2,j} =\omega^{(k-1)/2} \chi_{1,j}\sigma_{2,j}^{\dagger}\sigma_{2,j+1} =\omega^{(k-1)/2} \sigma_{1,j}\sigma_{2,1}^{\dagger}\sigma_{2,j}\sigma_{2,j}^{\dagger}\sigma_{2,j+1} = \omega^{(k-1)/2}  \sigma_{1,j}\sigma_{2,1}^{\dagger}\sigma_{2,j+1}.
\end{align}

\subsection{Bosonic mapping and continuum limit}
Next, we introduce a bosonic representation for $\sigma_{p,j}$ in terms of $\eta_p(x_j)$:
\begin{equation}
\sigma_{p,j} = A_ke^{i \eta_p(x_j)}+B_ke^{-i (k-1) \eta_p(x_j)},\label{eq:sigma_p}\,\,\,\,\,\,  p=1,2, 
\end{equation}
where $A_k$ and $B_k$ are real coefficients which depend on $k$ (but not on $a$); below we will specify the constraints they must satisfy. The fields $\eta_p$ satisfy the following commutation relations [see Eq.~\eqref{eq:thetacomm}]:
\begin{align}
\left[\eta_p(x_m),\eta_p(x_j)\right] &= 0,\\
\left[\eta_1(x_m),\eta_2(x_j)\right] &= i\frac{\pi}{k}\text{sign}(x_m-x_j).\label{eq:commSM}
\end{align} 

We recall that the $\mathbb{Z}_k$ DSGM obeys the global symmetry $\mathbb{Z}_k$:$ ~ \eta_1\rightarrow \eta_1 + \frac{2\pi}{k}$,  $\eta_2\rightarrow \eta_2$ and its dual symmetry $\mathbb{Z}^{\rm dual}_k$:$ ~ \eta_1\rightarrow \eta_1$, $ \eta_2\rightarrow \eta_2+ \frac{2\pi}{k}$. 
Under these transformations, it follows from Eq. \eqref{eq:sigma_p} that the $\sigma_{p,j}$ fields behave as
\begin{align}
&\mathbb{Z}_k: ~ \sigma_{1,j} \rightarrow \omega\sigma_{1,j}, ~ \sigma_{2,i}\rightarrow \sigma_{2,j}\\
&\mathbb{Z}^{\rm dual}_k: ~ \sigma_{1,j}\rightarrow \sigma_{1,j} , ~ \sigma_{2,j} \rightarrow \omega \sigma_{2,j},
\end{align} 
as expected for clock model operators~\cite{Fendley}. 

Next we prove that $\sigma_{p,j}$ in the bosonic representation, Eq. \eqref{eq:sigma_p},
indeed satisfies the commutation relations given in Eqs. \eqref{eq:commsigma1}, \eqref{eq:commsigma2}, \eqref{eq:sigma2a}, \eqref{eq:sigma2b}, and \eqref{eq:sigma2c}. 
First we consider the property
\begin{align}
\sigma_{p,j}^{\dagger}=\sigma_{p,j}^{-1}, 
\end{align}
or equivalently 
\begin{align}
\sigma_{p,j}^{\dagger}\sigma_{p,j}=1.
\end{align}
By using the bosonic representation of these operators and making use of Eqs. \eqref{eq:loga} - \eqref{eq:normalordereta}, we find
\begin{align}
\sigma_{1,j}^{\dagger}\sigma_{1,j} &=\left[A_ke^{-i \eta_1(x_j)}+B_k e^{i (k-1) \eta_1(x_j)}\right]\left[A_k e^{i \eta_1(x_j)}+B_k e^{-i (k-1) \eta_1(x_j)}\right]\nonumber\\&= A_k^2+ B_k^2 + A_k B_k \left(\frac{2\pi a}{L}\right)^{Kk/2} \left[:e^{-i k \eta_1(x_j)}: + :e^{i k \eta_1(x_j)}:\right]= A_k^2+ B_k^2 + \dots,\\
\sigma_{2,j}^{\dagger}\sigma_{2,j} &=\left[A_ke^{-i \eta_2(x_j)}+B_k e^{i (k-1) \eta_2(x_j)}\right]\left[A_k e^{i \eta_2(x_j)}+B_k e^{-i (k-1) \eta_2(x_j)}\right]\nonumber\\&= A_k^2+ B_k^2 + A_k B_k \left(\frac{2\pi a}{L}\right)^{k/2K}\left[:e^{-i k \eta_2(x_j)}: + :e^{i k \eta_2(x_j)}:\right]= A_k^2+ B_k^2 + \dots,
\end{align}
where the dots stand for terms proportional to positive powers of $\frac{a}{L}$, which we will drop in the continuum limit $\frac{a}{L}\to 0$. 
By imposing that $A_k^2+ B_k^2 = 1$, we arrive in the continuum limit at 
\begin{align}
\sigma_{p,j}^{\dagger}\sigma_{p,j} = 1,
\end{align}
which indeed proves $\sigma_{p,j}^{\dagger}=\sigma_{p,j}^{-1}$. As a second relation, we show that
\begin{align}
\sigma_{p,j}^k& = 1.
\end{align}
In this case, we find similarly
\begin{align}
\sigma_{1,j}^k = k A_k^{k-1}B_k +A_k^k \left(\frac{2\pi a}{L}\right)^{kK/2}:e^{i k \eta_1(x_j)}:+ \sum_{q=2}^{k} \left(\begin{matrix} k\\q\end{matrix}\right) A_k^{k-q}B_k^q :e^{i k \left(1-q\right) \eta_1(x_j)}: \left(\frac{2\pi a}{L}\right)^{\left(1-q\right)^2 kK/2}= k A_k^{k-1}B_k +  \dots,\\
\sigma_{2,j}^k = k A_k^{k-1}B_k +A_k^k \left(\frac{2\pi a}{L}\right)^{k/2K}:e^{i k \eta_2(x_j)}:+ \sum_{q=2}^{k} \left(\begin{matrix} k\\q\end{matrix}\right) A_k^{k-q}B_k^q :e^{i k \left(1-q\right) \eta_2(x_j)}: \left(\frac{2\pi a}{L}\right)^{\left(1-q\right)^2 k/2K}= k A_k^{k-1}B_k +  \dots
\end{align}
Imposing a further condition $k A_k^{k-1}B_k = 1$, we see that the relation $\sigma_{p,j}^k = 1$ is also satisfied. By putting these results together, we have shown that
\begin{align}
\sigma_{p,j}^{\dagger}&=\sigma_{p,j}^{-1},\label{eq:comm1}\\
\sigma_{p,j}^k& = 1,\label{eq:comm2}
\end{align}
provided the coefficients $A_k$ and $B_k$ satisfy both conditions $A_k^2+ B_k^2 = 1$ and $k A_k^{k-1}B_k = 1$. This eventually proves the validity of the relations given in
Eqs. \eqref{eq:commsigma1} in the bosonic representation and in the continuum limit.

The two relations in Eqs.~\eqref{eq:comm1} and \eqref{eq:comm2} together give
\begin{equation}
\sigma_{p,j}^{k-1}=\sigma_{p,j}^\dagger.\label{eq:comm3}
\end{equation}

As a final step, we show that, in the bosonic representation, the correct commutation relations between $\sigma_{1,j}$ and $\tau_j$, Eq. \eqref{eq:tausigma}, are satisfied. We have
\begin{align}
\tau_j &=\sigma_{2,j}^{\dagger}\sigma_{2,j+1} =\left[A_ke^{-i \eta_2(x_j)}+B_k e^{i (k-1) \eta_2(x_j)}\right]\left[A_k e^{i \eta_2(x_{j+1})}+B_k e^{-i (k-1) \eta_2(x_{j+1})}\right]\nonumber\\
&= A_k^2 :e^{-i\left[\eta_2(x_j)-\eta_2(x_{j+1})\right]}: +B_k^2 :e^{i\left[(k-1)\eta_2(x_j)-(k-1)\eta_2(x_{j+1})\right]}: \nonumber \\&+A_k B_k e^{-i\left[\eta_2(x_j)-k\eta_2(x_{j})\right]}e^{i\eta_2(x_{j+1})}+A_k B_k e^{i\left[(k-1)\eta_2(x_j)-k\eta_2(x_{j})\right]}e^{-i(k-1)\eta_2(x_{j+1})} \nonumber\\& = 
A_k^2 :e^{ia\partial_j\eta_2(x_j)}: + B_k^2 :e^{-ia(k-1)\partial_j\eta_2(x_j)}: \nonumber\ \\
&+\left(\frac{2\pi a}{L}\right)^{k/2K}\left[A_k B_k :e^{ia\partial_j\eta_2(x_j)}: :e^{ik \eta_2(x_j)}:+A_k B_k :e^{-ia(k-1)\partial_j\eta_2(x_j)}:  :e^{-ik \eta_2(x_{j})}:\right]\nonumber \\&=A_k^2 :e^{ia\partial_j\eta_2(x_j)}: + B_k^2 :e^{-ia(k-1)\partial_j\eta_2(x_j)}: + \dots \label{eq:esp_tau}
\end{align}
Next, we make use of the Baker-Hausdorff-Campbell relation for two operators $A$ and $B$ whose commutator
is a c-number:
\begin{equation}
\label{eq:BCH}
e^A e^B = e^B e^A e^{\left[A,B\right]}.
\end{equation} 
Using the commutator
\begin{equation}
\left[\eta_1(x_m),\partial_j \eta_2(x_j)\right] = -\frac{i 2\pi}{k a}\delta_{m,j},
\end{equation}
which is obtained by taking the derivative with respect to $x_j$ of Eq.~\eqref{eq:commSM},
we then obtain with Eq.~\eqref{eq:BCH}
\begin{equation}
e^{i c_1 \eta_1(x_m)}e^{i c_2 a \partial_j\eta_2(x_j)} = e^{i c_2 a \partial_j\eta_2(x_j)}e^{i c_1 \eta_1(x_m)} e^{i c_1 c_2 2\pi \delta_{m,j}/k}\label{eq:exp_comm}.
\end{equation}
For the expression $\sigma_{1,m} \tau_j$, one has three different possible values of $c_1 c_2$, which gives the same exponential factor
\begin{align}
c_1 c_2 &= 1 \rightarrow e^{i 2\pi \delta_{m,j}/k},\label{eq:c1c2_1}\\
c_1 c_2 &= -(k-1) \rightarrow e^{-i(k-1) 2\pi \delta_{m,j}/k} = e^{-i2\pi \delta_{m,j}} e^{i 2\pi \delta_{m,j}/k} = e^{i 2\pi \delta_{m,j}/k},\label{eq:c1c2_k-1}\\
c_1 c_2 &= (k-1)^2 \rightarrow e^{i(k-1)^2 2\pi \delta_{m,j}/k} = e^{i2\pi(k-2) \delta_{m,j}} e^{i 2\pi \delta_{m,j}/k} = e^{i 2\pi \delta_{m,j}/k}.\label{eq:c1c2_(k-1)^2}
\end{align}
By using the definition of $\sigma_{1,m}$ (see Eq.~\eqref{eq:sigma_p} for $p=1$) and the expansion in Eq.~\eqref{eq:esp_tau} for $\tau_i$, one obtains
\begin{align}
\sigma_{1,m} \tau_j &= \left[A_k e^{i \eta_1(x_m)}+B_k e^{-i (k-1) \eta_1(x_m)}\right]\left[A_k^2 :e^{ia\partial_j\eta_2(x_j)}: + B_k^2 :e^{-ia(k-1)\partial_j\eta_2(x_j)}: + \dots\right] \nonumber\\=&A^3_k e^{i \eta_1(x_m)}:e^{ia\partial_j\eta_2(x_j)}: + A_k B^2_k e^{i \eta_1(x_m)}:e^{-ia(k-1)\partial_j\eta_2(x_j)}: \nonumber\\+& B_k A_k^2 e^{-i (k-1) \eta_1(x_m)}:e^{ia\partial_j\eta_2(x_j)}:  + B^3_k e^{-i (k-1) \eta_1(x_m)}:e^{-ia(k-1)\partial_j\eta_2(x_j)}:+\dots
\end{align}
Then, in the second line of the previous equation, we can apply Eq.~\eqref{eq:c1c2_1} to the first term, Eq.~\eqref{eq:c1c2_k-1} to the second and third terms, and Eq.~\eqref{eq:c1c2_(k-1)^2} to the last term to commute the exponentials with $\eta_1$ past the exponentials with $\eta_2$, thus obtaining
\begin{align}
\sigma_{1,m} \tau_j =&A^3_k e^{i2\pi \delta_{m,j}/k} :e^{ia\partial_j\eta_2(x_j)}:e^{i \eta_1(x_j)} + A_k B^2_k  e^{i2\pi \delta_{m,j}/k} :e^{-ia(k-1)\partial_j\eta_2(x_j)}:e^{i \eta_1(x_j)} \nonumber\\+& B_k A_k^2  e^{i2\pi \delta_{m,j}/k}:e^{ia\partial_j\eta_2(x_j)}: e^{-i (k-1) \eta_1(x_j)} + B^3_k  e^{i2\pi \delta_{m,j}/k}:e^{-ia(k-1)\partial_j\eta_2(x_j)}: e^{-i (k-1) \eta_1(x_j)}+\dots  \nonumber
\end{align}
From this, we find the following commutation relation for the bosonic representations of $\sigma_{1,m}$ and $\tau_j$ in the limit $a/L \rightarrow 0$
\begin{equation}
\sigma_{1,m}\tau_j  = e^{i2\pi \delta_{m,j}/k} \tau_j  \sigma_{1,m}+\dots \label{sigmatau}
\end{equation}
This eventually proves that Eq. \eqref{eq:tausigma} is satisfied in the bosonic representation in the continuum limit.

Next, by using these operators, we derive now the DSGM as the low-energy limit of the clock Hamiltonian. For this, we  expand $\sigma_{p,j}^{\dagger}\sigma_{p,j+1}$  in powers of $\frac{a}{L}$, 
\begin{align}
\sigma_{p,j}^{\dagger}\sigma_{p,j+1} &= \left[A_ke^{-i \eta_p(x_j)}+B_k e^{i (k-1) \eta_p(x_{j})}\right]\left[A_k e^{i \eta_p(x_{j+1})}+B_k e^{-i (k-1) \eta_p(x_{j+1})}\right]\nonumber\\&=A_k^2 :e^{ia\partial_j\eta_p(x_j)}: + B_k^2 :e^{-ia(k-1)\partial_j\eta_p(x_j)}: + 2 A_k B_k \cos\left[k \eta_p(x_j)\right]+ \dots\nonumber\\&= 1 + 2 A_k B_k \cos\left[k \eta_p(x_j)\right] + \left[A_k^2 - (k-1) B_k^2\right]  i a \partial_j \eta_p(x_j) - \left[A_k^2 + (k-1)^2 B_k^2\right]a^2 :\left[\partial_j\eta_p(x_j)\right]^2:  + \dots\nonumber\\&= 1 + 2 A_k B_k \cos\left[k \eta_p(x_j)\right] + \left[A_k^2 - (k-1) B_k^2\right]  i \frac{a}{L} \partial_{\bar{j}} \eta_p(\bar{x}_j) - \left[A_k^2 + (k-1)^2 B_k^2\right]\left(\frac{a}{L}\right)^2 :\left[\partial_{\bar{j}}\eta_p(\bar{x}_j)\right]^2:  + \dots,
\end{align}
where we introduced the dimensionless variable $\bar{x}_j = \frac{x_j}{L}$ and the corresponding derivative $\partial_{\bar{j}} = L \partial_{j}$ and, in contrast with the similar derivation in Eq.~\eqref{eq:esp_tau}, we kept the lowest order cosine term in $\frac{a}{L}$, in accordance with the discussion below Eq.~\eqref{eq:DSGphitheta}. This rescaling is possible because the variable $x_j$ appears in the fields $\eta_p$ as $\frac{x_j}{L}$.
Since for $k>3$, the Hamiltonian in Eq.~\eqref{eq:pottsSM2} includes also powers of $\sigma_{p,j}^{\dagger}\sigma_{p,j+1}$, the low-energy mapping is complete only when also the powers $(\sigma_{p,j}^{\dagger}\sigma_{p,j+1})^\alpha$ are provided. In order to obtain these expansions, we use the following expression 
\begin{align}
&\left\{1 + \frac{a}{L} C_1 + \left(\frac{a}{L}\right)^2 C_2 + 2 A_k B_k \cos\left[k \eta_p(x_j)\right] + \dots \right\}^{\alpha} = \alpha  \frac{a}{L} C_1 +  \frac{\alpha}{2}\left(\frac{a}{L}\right)^2\left[C_1^2\left(\alpha-1\right) + 2 C_2\right] \nonumber\\&+  \left\{1 + 2 A_k B_k \cos\left[k \eta_p(x_j)\right]\right\}^{\alpha} + \dots,
\end{align}
where $C_1$ and $C_2$ stands for the operators contributing at first order and second order in $\frac{a}{L}$ in $\sigma_{p,j}^{\dagger}\sigma_{p,j+1}$ and where we kept the lowest order cosine term in $\frac{a}{L}$, in accordance with the discussion below Eq.~\eqref{eq:DSGphitheta}. Let us consider the cosine term for $p=1$:
\begin{align}
&\left\{1 + 2A_k B_k \cos\left[k \eta_1(x_j)\right]\right\}^{\alpha} = 1+  \sum_{n=1}^{\alpha}\left(\frac{2A_k B_k}{2}\right)^n\left(\begin{matrix}\alpha \\ n\end{matrix}\right)\left(e^{ik \eta_1(x_j)}+e^{-ik \eta_1(x_j)}\right)^{n}\nonumber\\& =1 + \sum_{n=1}^{\alpha}\left(A_k B_k\right)^n\left(\begin{matrix}\alpha \\ n\end{matrix}\right)\sum_{\beta=0}^{n}\left(\begin{matrix} n\\ \beta\end{matrix}\right) :e^{i \left(n-2\beta\right) k \eta_{1}(x_j)}: \left(\frac{2\pi a}{L}\right)^{\frac{K k}{2}\left(n-2\beta\right)^2} \nonumber\\& =1 + \sum_{n=1}^{\alpha}\left(A_k B_k\right)^n\left(\begin{matrix}\alpha \\ n\end{matrix}\right)\sum_{\beta=0}^{n}\left(\begin{matrix} n\\ \beta\end{matrix}\right) \frac{1}{2}\left[:e^{i \left(n-2\beta\right) k \eta_{1}(x_j)}: + :e^{-i \left(n-2\beta\right) k \eta_{1}(x_j)}:\right] \left(\frac{2\pi a}{L}\right)^{\frac{K k}{2}\left(n-2\beta\right)^2} \nonumber\\& =1 + \frac{\left(\frac{2\pi a}{L}\right)^{\frac{K k}{2}} }{2}\left[:e^{i  k \eta_{1}(x_j)}: + :e^{-i  k \eta_{1}(x_j)}:\right]\sum_{n=1, \text{odd } n}^{\alpha}\left(A_k B_k\right)^n\left(\begin{matrix}\alpha \\ n\end{matrix}\right)\left[\left(\begin{matrix} n\\ 
\frac{n-1}{2}\end{matrix}\right) + \left(\begin{matrix} n\\ 
\frac{n+1}{2}\end{matrix}\right)\right]  + \dots \nonumber\\ &= 1 + 2\cos\left[k \eta_1(x_j)\right] \sum_{n=1, \text{odd } n}^{\alpha}\left(A_k B_k\right)^n\left(\begin{matrix}\alpha \\ n\end{matrix}\right)\left(\begin{matrix} n\\ 
\frac{n+1}{2}\end{matrix}\right)  + \dots, 
\end{align}
where the sum over $n$ is performed only for odd integers and we kept the lowest order cosine term in $\frac{a}{L}$. In the case $p=2$, one has to replace $K\rightarrow \frac{1}{K}$ in the intermediate steps, but the final result is unchanged. 

In conclusion, we have 
\begin{align}
\left(\sigma_{p,j}^{\dagger}\sigma_{p,j+1}\right)^{\alpha} & = \left\{1 + 2 A_k B_k \cos\left[k \eta_p(x_j)\right] + \left[A_k^2 - (k-1) B_k^2\right]  i \frac{a}{L} \partial_{\bar{j}} \eta_p(\bar{x}_j) - \left[A_k^2 + (k-1)^2 B_k^2\right]\left(\frac{a}{L}\right)^2 :\left[\partial_{\bar{j}}\eta_p(\bar{x}_j)\right]^2: + \dots \right\}^{\alpha}  \nonumber\\& = 2\cos\left[k \eta_p(x_j)\right] \sum_{n=1, \text{odd } n}^{\alpha}\left(A_k B_k\right)^n\left(\begin{matrix}\alpha \\ n\end{matrix}\right)\left(\begin{matrix} n\\ 
\frac{n+1}{2}\end{matrix}\right)   \nonumber\\  &+\alpha \left[A_k^2 - (k-1) B_k^2\right]  i \frac{a}{L} \partial_{\bar{j}} \eta_p(\bar{x}_j) + \frac{\alpha(\alpha-1)}{2} \left\{\left[A_k^2 - (k-1) B_k^2\right]  i \frac{a}{L} \partial_{\bar{j}} \eta_p(\bar{x}_j)\right\}^2\nonumber\\&-\alpha\left[A_k^2 + (k-1)^2 B_k^2\right]\left(\frac{a}{L}\right)^2 :\left[\partial_{\bar{j}}\eta_p(\bar{x}_j)\right]^2:  + \dots,\label{eq:expansion_sigma}
\end{align}
where the dots $\dots$ stand for higher powers of $\frac{a}{L}$ and constant terms. One can write the expansion in Eq.~\eqref{eq:expansion_sigma} in a more compact form as
\begin{align}
\left(\sigma_{p,j}^{\dagger}\sigma_{p,j+1}\right)^{\alpha} = -f_{k}^{(\alpha)}  \cos\left[k \eta_p(x_j)\right]  +\alpha \left[A_k^2 - (k-1) B_k^2\right]  i \frac{a}{L} \partial_{\bar{j}} \eta_p(\bar{x}_j) -d_{k}^{(\alpha)} \left(\frac{a}{L}\right)^2 :\left[\partial_{\bar{j}}\eta_p(\bar{x}_j)\right]^2:  + \dots,
\end{align}
where the term $ \left\{ i \frac{a}{L} \partial_{\bar{j}} \eta_p(\bar{x}_j)\right\}^2$ has been written in normal ordered form as $-\left(\frac{a}{L}\right)^2 :\left[\partial_{\bar{j}}\eta_p(\bar{x}_j)\right]^2:$ by adding some unimportant constants absorbed in the remainder $\dots$ Here,
\begin{align}
d_{k}^{(\alpha)} &= \frac{\alpha}{2}\left[2\left(A_k^2 + (k-1)^2 B_k^2\right) + \left(A_k^2-(k-1) B_k^2\right)^2(\alpha-1)\right],\\
f_{k}^{(\alpha)} &=-2 \sum_{n=1, \text{odd } n}^{\alpha}\left(A_k B_k\right)^n\left(\begin{matrix}\alpha \\ n\end{matrix}\right)\left(\begin{matrix} n\\ 
\frac{n+1}{2}\end{matrix}\right) ,
\end{align}
with $d_{k}^{(\alpha)}  = d_{k}^{(k-\alpha)} $ and $f_{k}^{(\alpha)}  = f_{k}^{(k-\alpha)}  $, since, due to the properties $\sigma^{k-1}_{p,i} = \sigma^{\dagger}_{p,i}$, one has that   $(\sigma_{p,j}^{\dagger}\sigma_{p,j+1})^\alpha = (\sigma_{p,j}^{\dagger}\sigma_{p,j+1})^{k-\alpha}$. Then,  taking the continuum limit of Eq.~\eqref{eq:pottsSM} (with $\sum_{j} \rightarrow \int \frac{dx}{a}$), we find 
\begin{align}
H_k = &\int_{0}^{L} dx ~ \left\{\frac{v_1}{2\pi}:\left[\partial_x\eta_1(x)\right]^2: + \frac{v_2}{2\pi}:\left[\partial_x\eta_2(x)\right]^2:\right\} + \frac{\Lambda_1}{4\pi^2a^{2}} \int_{0}^{L} dx ~ \cos\left[k\eta_1(x)\right] +  \frac{\Lambda_2}{4\pi^2a^{2}} \int_{0}^{L} dx ~ \cos\left[k\eta_2(x)\right].
\end{align}
We note that the integral $\int_{0}^{L} dx$ gives zero for the term proportional to $\partial_{\bar{j}} \eta_p(\bar{x}_j)$. Here,
\begin{align}
{v_1} / {(2\pi a)}&=\sum_{\alpha=1}^{k-1} d^{(\alpha)}_k J_{1,\alpha},~~~
{v_2} / {(2\pi a)}=\sum_{\alpha=1}^{k-1} d^{(\alpha)}_k J_{2,\alpha},\\
{\Lambda}_1 / {(4\pi^2 a)}&=\sum_{\alpha=1}^{k-1} f^{(\alpha)}_k J_{1,\alpha},~~~
{\Lambda_2} / {(4\pi^2 a)}=\sum_{\alpha=1}^{k-1} f^{(\alpha)}_k J_{2,\alpha}.
\end{align} 
Starting from $v_1$ and $v_2$ one can also find $K$ and $v$ as $K=\sqrt{v_2/v_1}$ and $v=\sqrt{v_1 v_2}$. We note that in our case we have $K=1$ and $\Lambda_1 = \Lambda_2$.

It is instructive to write down a few special cases for $k$. For $k=2$, we find
\begin{align}
v_1 / (2\pi a)&=  J_{1,1}, ~~~ v_2 / (2\pi a)= J_{2,1},\\{\Lambda}_1 / {(4\pi^2 a)} &= - J_{1,1}, ~~~ {\Lambda}_2 / {(4\pi^2 a)} = - J_{2,1}.
\end{align}

For $k=3$ we find two solutions for $A_k, B_k$. First, we get $A_3 = \sqrt{\frac{1}{3} \left(1+2 \cos \left(\frac{2 \pi }{9}\right)\right)}$ and $B_3 = \frac{\sqrt{2 \sin \left(\frac{\pi }{9}\right)}}{\sqrt[4]{3}}$, and
\begin{align}
v_1 / (2\pi a) &=  \frac{2}{3} \left(1+8 \sqrt{3} \sin \left(\frac{\pi }{9}\right)+2 \cos \left(\frac{2 \pi
}{9}\right)\right) \text{Re}\left[J_{1,1}\right], ~~~ v_2 / (2\pi a) = \frac{2}{3} \left(1+8 \sqrt{3} \sin \left(\frac{\pi }{9}\right)+2 \cos \left(\frac{2 \pi
}{9}\right)\right)\text{Re}\left[J_{2,1}\right],\\{\Lambda}_1 / {(4\pi^2 a)} &= - 2 \frac{\sqrt{2 \sin \left(\frac{\pi }{9}\right) \left(1+2 \cos \left(\frac{2 \pi
		}{9}\right)\right)}}{3^{3/4}}\text{Re}\left[J_{1,1}\right], ~~~ {\Lambda}_2 / {(4\pi^2 a)} = - 2\frac{\sqrt{2 \sin \left(\frac{\pi }{9}\right) \left(1+2 \cos \left(\frac{2 \pi
		}{9}\right)\right)}}{3^{3/4}}\text{Re}\left[J_{2,1}\right].
\end{align}
Second, we get
$A_3 =\frac{1}{2} \sqrt{\frac{1}{3} \left(4+4 \sqrt{3} \sin \left(\frac{2 \pi }{9}\right)-4
	\cos \left(\frac{2 \pi }{9}\right)\right)}$ and $B_3 = \sqrt{1+\frac{1}{12} \left(-4-4 \sqrt{3} \sin \left(\frac{2 \pi }{9}\right)+4 \cos
	\left(\frac{2 \pi }{9}\right)\right)}$,
\begin{align}
v_1 / (2\pi a) &=  \left[3-\sqrt{3} \sin \left(\frac{2 \pi }{9}\right)+\cos \left(\frac{2 \pi }{9}\right) \right]\text{Re}\left[J_{1,1}\right], ~~~ v_2 / (2\pi a) = \left[3-\sqrt{3} \sin \left(\frac{2 \pi }{9}\right)+\cos \left(\frac{2 \pi }{9}\right)\right]\text{Re}\left[J_{2,1}\right],\\{\Lambda}_1 / {(4\pi^2 a)} &=- \frac{\sqrt{-\sin \left(\frac{\pi }{9}\right)+\sin \left(\frac{2 \pi }{9}\right)+\cos
		\left(\frac{\pi }{18}\right)}}{3^{3/4}}\text{Re}\left[J_{1,1}\right], ~~~ {\Lambda}_2 / {(4\pi^2 a)} = \frac{\sqrt{-\sin \left(\frac{\pi }{9}\right)+\sin \left(\frac{2 \pi }{9}\right)+\cos
		\left(\frac{\pi }{18}\right)}}{3^{3/4}}\text{Re}\left[J_{2,1}\right].
\end{align}

For $k=4$, we find again two solutions. First, $A_4=B_4 = \frac{1}{\sqrt{2}}$, and
\begin{align}
v_1 / (2\pi a) &=  10\text{Re}\left[J_{1,1}\right] + 11 J_{1,2}, ~~~ v_2 / a = 10\text{Re}\left[J_{2,1}\right] + 11 J_{2,2},\\{\Lambda}_1 / {(4\pi^2 a)} &= - \text{Re}\left[J_{1,1}\right] -2 J_{1,2}, ~~~ {\Lambda}_2 / {(4\pi^2 a)} = - \text{Re}\left[J_{2,1}\right] - 2 J_{2,2}.
\end{align}
Second, we get $A_4 = \frac{1}{\sqrt{\frac{6}{1+\sqrt[3]{19-3 \sqrt{33}}+\sqrt[3]{19+3 \sqrt{33}}}}}$ and $B_4 = \sqrt{1+\frac{1}{6} \left(-1-\sqrt[3]{19-3 \sqrt{33}}-\sqrt[3]{19+3 \sqrt{33}}\right)}$, and
\begin{align}
v_1 / (2\pi a) &=  \left[\frac{1}{3} \left(23-4 \sqrt[3]{19-3 \sqrt{33}}-4 \sqrt[3]{19+3 \sqrt{33}}\right)\right]\text{Re}\left[J_{1,1}\right] + \nonumber\\& + \left[\frac{1}{9} \left(219-52 \sqrt[3]{19-3 \sqrt{33}}+4 \left(19-3 \sqrt{33}\right)^{2/3}-52
\sqrt[3]{19+3 \sqrt{33}}+4 \left(19+3 \sqrt{33}\right)^{2/3}\right)\right]  J_{1,2},\\ v_2 / (2\pi a) &=  \left[\frac{1}{3} \left(23-4 \sqrt[3]{19-3 \sqrt{33}}-4 \sqrt[3]{19+3 \sqrt{33}}\right)\right]\text{Re}\left[J_{2,1}\right]\nonumber\\& + \left[\frac{1}{9} \left(219-52 \sqrt[3]{19-3 \sqrt{33}}+4 \left(19-3 \sqrt{33}\right)^{2/3}-52
\sqrt[3]{19+3 \sqrt{33}}+4 \left(19+3 \sqrt{33}\right)^{2/3}\right)\right] J_{2,2},\\{\Lambda}_1 / {(4\pi^2 a)} &= -\frac{1}{3} \sqrt{-3+4 \sqrt[3]{19-3 \sqrt{33}}-\left(19-3 \sqrt{33}\right)^{2/3}+4
	\sqrt[3]{19+3 \sqrt{33}}-\left(19+3 \sqrt{33}\right)^{2/3}} \left\{\text{Re}\left[J_{1,1}\right] + 2J_{1,2}\right\},\\ {\Lambda}_2 / {(4\pi^2 a)} &= - \frac{1}{3} \sqrt{-3+4 \sqrt[3]{19-3 \sqrt{33}}-\left(19-3 \sqrt{33}\right)^{2/3}+4
	\sqrt[3]{19+3 \sqrt{33}}-\left(19+3 \sqrt{33}\right)^{2/3}}\left\{\text{Re}\left[J_{2,1}\right] + 2J_{2,2}\right\}.
\end{align}


\begin{thebibliography}{99}%
	\makeatletter
	\providecommand \@ifxundefined [1]{%
		\@ifx{#1\undefined}
	}%
	\providecommand \@ifnum [1]{%
		\ifnum #1\expandafter \@firstoftwo
		\else \expandafter \@secondoftwo
		\fi
	}%
	\providecommand \@ifx [1]{%
		\ifx #1\expandafter \@firstoftwo
		\else \expandafter \@secondoftwo
		\fi
	}%
	\providecommand \natexlab [1]{#1}%
	\providecommand \enquote  [1]{``#1''}%
	\providecommand \bibnamefont  [1]{#1}%
	\providecommand \bibfnamefont [1]{#1}%
	\providecommand \citenamefont [1]{#1}%
	\providecommand \href@noop [0]{\@secondoftwo}%
	\providecommand \href [0]{\begingroup \@sanitize@url \@href}%
	\providecommand \@href[1]{\@@startlink{#1}\@@href}%
	\providecommand \@@href[1]{\endgroup#1\@@endlink}%
	\providecommand \@sanitize@url [0]{\catcode `\\12\catcode `\$12\catcode
		`\&12\catcode `\#12\catcode `\^12\catcode `\_12\catcode `\%12\relax}%
	\providecommand \@@startlink[1]{}%
	\providecommand \@@endlink[0]{}%
	\providecommand \url  [0]{\begingroup\@sanitize@url \@url }%
	\providecommand \@url [1]{\endgroup\@href {#1}{\urlprefix }}%
	\providecommand \urlprefix  [0]{URL }%
	\providecommand \Eprint [0]{\href }%
	\providecommand \doibase [0]{http://dx.doi.org/}%
	\providecommand \selectlanguage [0]{\@gobble}%
	\providecommand \bibinfo  [0]{\@secondoftwo}%
	\providecommand \bibfield  [0]{\@secondoftwo}%
	\providecommand \translation [1]{[#1]}%
	\providecommand \BibitemOpen [0]{}%
	\providecommand \bibitemStop [0]{}%
	\providecommand \bibitemNoStop [0]{.\EOS\space}%
	\providecommand \EOS [0]{\spacefactor3000\relax}%
	\providecommand \BibitemShut  [1]{\csname bibitem#1\endcsname}%
	\let\auto@bib@innerbib\@empty
	\bibitem [{\citenamefont {Awschalom}\ \emph {et~al.}(2002)\citenamefont
		{Awschalom}, \citenamefont {Loss},\ and\ \citenamefont
		{Samarth}}]{LossBook2002}%
	\BibitemOpen
	\bibinfo {editor} {\bibfnamefont {D.~D.}\ \bibnamefont {Awschalom}}, \bibinfo
	{editor} {\bibfnamefont {D.}~\bibnamefont {Loss}}, \ and\ \bibinfo {editor}
	{\bibfnamefont {N.}~\bibnamefont {Samarth}},\ eds.,\ \href {\doibase
		10.1007/978-3-662-05003-3} {\emph {\bibinfo {title} {Semiconductor
				Spintronics and Quantum Computation}}}\ (\bibinfo  {publisher} {Springer
		Berlin Heidelberg},\ \bibinfo {year} {2002})\BibitemShut {NoStop}%
	\bibitem [{\citenamefont {Winkler}(2003)}]{Winkler03}%
	\BibitemOpen
	\bibfield  {author} {\bibinfo {author} {\bibfnamefont {R.}~\bibnamefont
			{Winkler}},\ }\href {\doibase 10.1007/b13586} {\emph {\bibinfo {title}
			{Spin--Orbit Coupling Effects in Two-Dimensional Electron and Hole
				Systems}}}\ (\bibinfo  {publisher} {Springer Berlin Heidelberg},\ \bibinfo
	{year} {2003})\BibitemShut {NoStop}%
	\bibitem [{\citenamefont {Loss}\ and\ \citenamefont
		{DiVincenzo}(1998)}]{Loss98}%
	\BibitemOpen
	\bibfield  {author} {\bibinfo {author} {\bibfnamefont {D.}~\bibnamefont
			{Loss}}\ and\ \bibinfo {author} {\bibfnamefont {D.~P.}\ \bibnamefont
			{DiVincenzo}},\ }\href {\doibase 10.1103/PhysRevA.57.120} {\bibfield
		{journal} {\bibinfo  {journal} {Phys. Rev. A}\ }\textbf {\bibinfo {volume}
			{57}},\ \bibinfo {pages} {120} (\bibinfo {year} {1998})}\BibitemShut
	{NoStop}%
	\bibitem [{\citenamefont {Hanson}\ \emph {et~al.}(2007)\citenamefont {Hanson},
		\citenamefont {Kouwenhoven}, \citenamefont {Petta}, \citenamefont {Tarucha},\
		and\ \citenamefont {Vandersypen}}]{Hanson07}%
	\BibitemOpen
	\bibfield  {author} {\bibinfo {author} {\bibfnamefont {R.}~\bibnamefont
			{Hanson}}, \bibinfo {author} {\bibfnamefont {L.~P.}\ \bibnamefont
			{Kouwenhoven}}, \bibinfo {author} {\bibfnamefont {J.~R.}\ \bibnamefont
			{Petta}}, \bibinfo {author} {\bibfnamefont {S.}~\bibnamefont {Tarucha}}, \
		and\ \bibinfo {author} {\bibfnamefont {L.~M.~K.}\ \bibnamefont
			{Vandersypen}},\ }\href {\doibase 10.1103/RevModPhys.79.1217} {\bibfield
		{journal} {\bibinfo  {journal} {Rev. Mod. Phys.}\ }\textbf {\bibinfo {volume}
			{79}},\ \bibinfo {pages} {1217} (\bibinfo {year} {2007})}\BibitemShut
	{NoStop}%
	\bibitem [{\citenamefont {Kloeffel}\ and\ \citenamefont
		{Loss}(2013)}]{Kloeffel13}%
	\BibitemOpen
	\bibfield  {author} {\bibinfo {author} {\bibfnamefont {C.}~\bibnamefont
			{Kloeffel}}\ and\ \bibinfo {author} {\bibfnamefont {D.}~\bibnamefont
			{Loss}},\ }\href {\doibase 10.1146/annurev-conmatphys-030212-184248}
	{\bibfield  {journal} {\bibinfo  {journal} {Annual Review of Condensed Matter
				Physics}\ }\textbf {\bibinfo {volume} {4}},\ \bibinfo {pages} {51} (\bibinfo
		{year} {2013})}\BibitemShut {NoStop}%
	\bibitem [{\citenamefont {Golovach}\ \emph {et~al.}(2006)\citenamefont
		{Golovach}, \citenamefont {Borhani},\ and\ \citenamefont
		{Loss}}]{Golovach06}%
	\BibitemOpen
	\bibfield  {author} {\bibinfo {author} {\bibfnamefont {V.~N.}\ \bibnamefont
			{Golovach}}, \bibinfo {author} {\bibfnamefont {M.}~\bibnamefont {Borhani}}, \
		and\ \bibinfo {author} {\bibfnamefont {D.}~\bibnamefont {Loss}},\ }\href
	{\doibase 10.1103/PhysRevB.74.165319} {\bibfield  {journal} {\bibinfo
			{journal} {Phys. Rev. B}\ }\textbf {\bibinfo {volume} {74}},\ \bibinfo
		{pages} {165319} (\bibinfo {year} {2006})}\BibitemShut {NoStop}%
	\bibitem [{\citenamefont {Nowack}\ \emph {et~al.}(2007)\citenamefont {Nowack},
		\citenamefont {Koppens}, \citenamefont {Nazarov},\ and\ \citenamefont
		{Vandersypen}}]{Nowack07}%
	\BibitemOpen
	\bibfield  {author} {\bibinfo {author} {\bibfnamefont {K.~C.}\ \bibnamefont
			{Nowack}}, \bibinfo {author} {\bibfnamefont {F.~H.~L.}\ \bibnamefont
			{Koppens}}, \bibinfo {author} {\bibfnamefont {Yu.~V.}\ \bibnamefont
			{Nazarov}}, \ and\ \bibinfo {author} {\bibfnamefont {L.~M.~K.}\ \bibnamefont
			{Vandersypen}},\ }\href {\doibase 10.1126/science.1148092} {\bibfield
		{journal} {\bibinfo  {journal} {Science}\ }\textbf {\bibinfo {volume}
			{318}},\ \bibinfo {pages} {1430} (\bibinfo {year} {2007})}\BibitemShut
	{NoStop}%
	\bibitem [{\citenamefont {Froning}\ \emph {et~al.}({\natexlab{a}})\citenamefont
		{Froning}, \citenamefont {Camenzind}, \citenamefont {van~der Molen},
		\citenamefont {Li}, \citenamefont {Bakkers}, \citenamefont {Zumb\"uhl},\ and\
		\citenamefont {Braakman}}]{Froning20}%
	\BibitemOpen
	\bibfield  {author} {\bibinfo {author} {\bibfnamefont {F.~N.~M.}\
			\bibnamefont {Froning}}, \bibinfo {author} {\bibfnamefont {L.~C.}\
			\bibnamefont {Camenzind}}, \bibinfo {author} {\bibfnamefont {O.~A.~H.}\
			\bibnamefont {van~der Molen}}, \bibinfo {author} {\bibfnamefont
			{A.}~\bibnamefont {Li}}, \bibinfo {author} {\bibfnamefont {E.~P. A.~M.}\
			\bibnamefont {Bakkers}}, \bibinfo {author} {\bibfnamefont {D.~M.}\
			\bibnamefont {Zumb\"uhl}}, \ and\ \bibinfo {author} {\bibfnamefont {F.~R.}\
			\bibnamefont {Braakman}},\ }\href {https://arxiv.org/abs/2006.11175} {\emph
		{\bibinfo {title} {\normalfont{arXiv:2006.11175 (2020)}}}}\BibitemShut
	{NoStop}%
	\bibitem [{\citenamefont {Froning}\ \emph {et~al.}({\natexlab{b}})\citenamefont
		{Froning}, \citenamefont {Ran\u{c}i\'c}, \citenamefont {Het\'enyi},
		\citenamefont {Bosco}, \citenamefont {Rehmann}, \citenamefont {Li},
		\citenamefont {Bakkers}, \citenamefont {Zwanenburg}, \citenamefont {Loss},
		\citenamefont {Zumb\"uhl},\ and\ \citenamefont {Braakman}}]{Froning20b}%
	\BibitemOpen
	\bibfield  {author} {\bibinfo {author} {\bibfnamefont {F.~N.~M.}\
			\bibnamefont {Froning}}, \bibinfo {author} {\bibfnamefont {M.~J.}\
			\bibnamefont {Ran\u{c}i\'c}}, \bibinfo {author} {\bibfnamefont
			{B.}~\bibnamefont {Het\'enyi}}, \bibinfo {author} {\bibfnamefont
			{S.}~\bibnamefont {Bosco}}, \bibinfo {author} {\bibfnamefont {M.~K.}\
			\bibnamefont {Rehmann}}, \bibinfo {author} {\bibfnamefont {A.}~\bibnamefont
			{Li}}, \bibinfo {author} {\bibfnamefont {E.~P. A.~M.}\ \bibnamefont
			{Bakkers}}, \bibinfo {author} {\bibfnamefont {F.~A.}\ \bibnamefont
			{Zwanenburg}}, \bibinfo {author} {\bibfnamefont {D.}~\bibnamefont {Loss}},
		\bibinfo {author} {\bibfnamefont {D.~M.}\ \bibnamefont {Zumb\"uhl}}, \ and\
		\bibinfo {author} {\bibfnamefont {F.~R.}\ \bibnamefont {Braakman}},\ }\href
	{https://arxiv.org/abs/2007.04308} {\emph {\bibinfo {title}
			{\normalfont{arXiv:2007.04308} (2020)}}}\BibitemShut {NoStop}%
	\bibitem [{\citenamefont {Wu}\ \emph {et~al.}(2006)\citenamefont {Wu},
		\citenamefont {Bernevig},\ and\ \citenamefont {Zhang}}]{Wu06}%
	\BibitemOpen
	\bibfield  {author} {\bibinfo {author} {\bibfnamefont {C.}~\bibnamefont
			{Wu}}, \bibinfo {author} {\bibfnamefont {B.~A.}\ \bibnamefont {Bernevig}}, \
		and\ \bibinfo {author} {\bibfnamefont {S.-C.}\ \bibnamefont {Zhang}},\ }\href
	{\doibase 10.1103/PhysRevLett.96.106401} {\bibfield  {journal} {\bibinfo
			{journal} {Phys. Rev. Lett.}\ }\textbf {\bibinfo {volume} {96}},\ \bibinfo
		{pages} {106401} (\bibinfo {year} {2006})}\BibitemShut {NoStop}%
	\bibitem [{\citenamefont {K{\"o}nig}\ \emph {et~al.}(2007)\citenamefont
		{K{\"o}nig}, \citenamefont {Wiedmann}, \citenamefont {Br{\"u}ne},
		\citenamefont {Roth}, \citenamefont {Buhmann}, \citenamefont {Molenkamp},
		\citenamefont {Qi},\ and\ \citenamefont {Zhang}}]{Konig07}%
	\BibitemOpen
	\bibfield  {author} {\bibinfo {author} {\bibfnamefont {M.}~\bibnamefont
			{K{\"o}nig}}, \bibinfo {author} {\bibfnamefont {S.}~\bibnamefont {Wiedmann}},
		\bibinfo {author} {\bibfnamefont {C.}~\bibnamefont {Br{\"u}ne}}, \bibinfo
		{author} {\bibfnamefont {A.}~\bibnamefont {Roth}}, \bibinfo {author}
		{\bibfnamefont {H.}~\bibnamefont {Buhmann}}, \bibinfo {author} {\bibfnamefont
			{L.~W.}\ \bibnamefont {Molenkamp}}, \bibinfo {author} {\bibfnamefont {X.-L.}\
			\bibnamefont {Qi}}, \ and\ \bibinfo {author} {\bibfnamefont {S.-C.}\
			\bibnamefont {Zhang}},\ }\href {\doibase 10.1126/science.1148047} {\bibfield
		{journal} {\bibinfo  {journal} {Science}\ }\textbf {\bibinfo {volume}
			{318}},\ \bibinfo {pages} {766} (\bibinfo {year} {2007})}\BibitemShut
	{NoStop}%
	\bibitem [{\citenamefont {Hasan}\ and\ \citenamefont {Kane}(2010)}]{Hasan10}%
	\BibitemOpen
	\bibfield  {author} {\bibinfo {author} {\bibfnamefont {M.~Z.}\ \bibnamefont
			{Hasan}}\ and\ \bibinfo {author} {\bibfnamefont {C.~L.}\ \bibnamefont
			{Kane}},\ }\href {\doibase 10.1103/RevModPhys.82.3045} {\bibfield  {journal}
		{\bibinfo  {journal} {Rev. Mod. Phys.}\ }\textbf {\bibinfo {volume} {82}},\
		\bibinfo {pages} {3045--3067} (\bibinfo {year} {2010})}\BibitemShut {NoStop}%
	\bibitem [{\citenamefont {St\ifmmode~\check{r}\else \v{r}\fi{}eda}\ and\
		\citenamefont {\ifmmode~\check{S}\else \v{S}\fi{}eba}(2003)}]{Streda03}%
	\BibitemOpen
	\bibfield  {author} {\bibinfo {author} {\bibfnamefont {P.}~\bibnamefont
			{St\ifmmode~\check{r}\else \v{r}\fi{}eda}}\ and\ \bibinfo {author}
		{\bibfnamefont {P.}~\bibnamefont {\ifmmode~\check{S}\else \v{S}\fi{}eba}},\
	}\href {\doibase 10.1103/PhysRevLett.90.256601} {\bibfield  {journal}
		{\bibinfo  {journal} {Phys. Rev. Lett.}\ }\textbf {\bibinfo {volume} {90}},\
		\bibinfo {pages} {256601} (\bibinfo {year} {2003})}\BibitemShut {NoStop}%
	\bibitem [{\citenamefont {Pershin}\ \emph {et~al.}(2004)\citenamefont
		{Pershin}, \citenamefont {Nesteroff},\ and\ \citenamefont
		{Privman}}]{Pershin04}%
	\BibitemOpen
	\bibfield  {author} {\bibinfo {author} {\bibfnamefont {Y.~V.}\ \bibnamefont
			{Pershin}}, \bibinfo {author} {\bibfnamefont {J.~A.}\ \bibnamefont
			{Nesteroff}}, \ and\ \bibinfo {author} {\bibfnamefont {V.}~\bibnamefont
			{Privman}},\ }\href {\doibase 10.1103/PhysRevB.69.121306} {\bibfield
		{journal} {\bibinfo  {journal} {Phys. Rev. B}\ }\textbf {\bibinfo {volume}
			{69}},\ \bibinfo {pages} {121306(R)} (\bibinfo {year} {2004})}\BibitemShut
	{NoStop}%
	\bibitem [{\citenamefont {Meng}\ and\ \citenamefont {Loss}(2013)}]{Meng13}%
	\BibitemOpen
	\bibfield  {author} {\bibinfo {author} {\bibfnamefont {T.}~\bibnamefont
			{Meng}}\ and\ \bibinfo {author} {\bibfnamefont {D.}~\bibnamefont {Loss}},\
	}\href {\doibase 10.1103/PhysRevB.88.035437} {\bibfield  {journal} {\bibinfo
			{journal} {Phys. Rev. B}\ }\textbf {\bibinfo {volume} {88}},\ \bibinfo
		{pages} {035437} (\bibinfo {year} {2013})}\BibitemShut {NoStop}%
	\bibitem [{\citenamefont {Kammhuber}\ \emph {et~al.}(2017)\citenamefont
		{Kammhuber}, \citenamefont {Cassidy}, \citenamefont {Pei}, \citenamefont
		{Nowak}, \citenamefont {Vuik}, \citenamefont {G\"{u}l}, \citenamefont {Car},
		\citenamefont {Plissard}, \citenamefont {Bakkers}, \citenamefont {Wimmer},\
		and\ \citenamefont {Kouwenhoven}}]{Kammhuber17}%
	\BibitemOpen
	\bibfield  {author} {\bibinfo {author} {\bibfnamefont {J.}~\bibnamefont
			{Kammhuber}}, \bibinfo {author} {\bibfnamefont {M.~C.}\ \bibnamefont
			{Cassidy}}, \bibinfo {author} {\bibfnamefont {F.}~\bibnamefont {Pei}},
		\bibinfo {author} {\bibfnamefont {M.~P.}\ \bibnamefont {Nowak}}, \bibinfo
		{author} {\bibfnamefont {A.}~\bibnamefont {Vuik}}, \bibinfo {author}
		{\bibfnamefont {\"{O}.}\ \bibnamefont {G\"{u}l}}, \bibinfo {author}
		{\bibfnamefont {D.}~\bibnamefont {Car}}, \bibinfo {author} {\bibfnamefont
			{S.~R.}\ \bibnamefont {Plissard}}, \bibinfo {author} {\bibfnamefont {E.~P.
				A.~M.}\ \bibnamefont {Bakkers}}, \bibinfo {author} {\bibfnamefont
			{M.}~\bibnamefont {Wimmer}}, \ and\ \bibinfo {author} {\bibfnamefont {L.~P.}\
			\bibnamefont {Kouwenhoven}},\ }\href
	{https://doi.org/10.1038/s41467-017-00315-y} {\bibfield  {journal} {\bibinfo
			{journal} {Nature Communications}\ }\textbf {\bibinfo {volume} {8}},\
		\bibinfo {pages} {478} (\bibinfo {year} {2017})}\BibitemShut {NoStop}%
	\bibitem [{\citenamefont {Kitaev}(2001)}]{Kitaev01}%
	\BibitemOpen
	\bibfield  {author} {\bibinfo {author} {\bibfnamefont {A.~Y.}\ \bibnamefont
			{Kitaev}},\ }\href {\doibase 10.1070/1063-7869/44/10s/s29} {\bibfield
		{journal} {\bibinfo  {journal} {Physics-Uspekhi}\ }\textbf {\bibinfo {volume}
			{44}},\ \bibinfo {pages} {131} (\bibinfo {year} {2001})}\BibitemShut
	{NoStop}%
	\bibitem [{\citenamefont {Braunecker}\ \emph {et~al.}(2010)\citenamefont
		{Braunecker}, \citenamefont {Japaridze}, \citenamefont {Klinovaja},\ and\
		\citenamefont {Loss}}]{Braunecker10}%
	\BibitemOpen
	\bibfield  {author} {\bibinfo {author} {\bibfnamefont {B.}~\bibnamefont
			{Braunecker}}, \bibinfo {author} {\bibfnamefont {G.~I.}\ \bibnamefont
			{Japaridze}}, \bibinfo {author} {\bibfnamefont {J.}~\bibnamefont
			{Klinovaja}}, \ and\ \bibinfo {author} {\bibfnamefont {D.}~\bibnamefont
			{Loss}},\ }\href {\doibase 10.1103/PhysRevB.82.045127} {\bibfield  {journal}
		{\bibinfo  {journal} {Phys. Rev. B}\ }\textbf {\bibinfo {volume} {82}},\
		\bibinfo {pages} {045127} (\bibinfo {year} {2010})}\BibitemShut {NoStop}%
	\bibitem [{\citenamefont {Oreg}\ \emph {et~al.}(2010)\citenamefont {Oreg},
		\citenamefont {Refael},\ and\ \citenamefont {von Oppen}}]{Oreg10}%
	\BibitemOpen
	\bibfield  {author} {\bibinfo {author} {\bibfnamefont {Y.}~\bibnamefont
			{Oreg}}, \bibinfo {author} {\bibfnamefont {G.}~\bibnamefont {Refael}}, \ and\
		\bibinfo {author} {\bibfnamefont {F.}~\bibnamefont {von Oppen}},\ }\href
	{\doibase 10.1103/PhysRevLett.105.177002} {\bibfield  {journal} {\bibinfo
			{journal} {Phys. Rev. Lett.}\ }\textbf {\bibinfo {volume} {105}},\ \bibinfo
		{pages} {177002} (\bibinfo {year} {2010})}\BibitemShut {NoStop}%
	\bibitem [{\citenamefont {Lutchyn}\ \emph {et~al.}(2010)\citenamefont
		{Lutchyn}, \citenamefont {Sau},\ and\ \citenamefont {Das~Sarma}}]{Lutchyn10}%
	\BibitemOpen
	\bibfield  {author} {\bibinfo {author} {\bibfnamefont {R.~M.}\ \bibnamefont
			{Lutchyn}}, \bibinfo {author} {\bibfnamefont {J.~D.}\ \bibnamefont {Sau}}, \
		and\ \bibinfo {author} {\bibfnamefont {S.}~\bibnamefont {Das~Sarma}},\ }\href
	{\doibase 10.1103/PhysRevLett.105.077001} {\bibfield  {journal} {\bibinfo
			{journal} {Phys. Rev. Lett.}\ }\textbf {\bibinfo {volume} {105}},\ \bibinfo
		{pages} {077001} (\bibinfo {year} {2010})}\BibitemShut {NoStop}%
	\bibitem [{\citenamefont {Potter}\ and\ \citenamefont {Lee}(2011)}]{Potter11}%
	\BibitemOpen
	\bibfield  {author} {\bibinfo {author} {\bibfnamefont {A.~C.}\ \bibnamefont
			{Potter}}\ and\ \bibinfo {author} {\bibfnamefont {P.~A.}\ \bibnamefont
			{Lee}},\ }\href {\doibase 10.1103/PhysRevB.83.094525} {\bibfield  {journal}
		{\bibinfo  {journal} {Phys. Rev. B}\ }\textbf {\bibinfo {volume} {83}},\
		\bibinfo {pages} {094525} (\bibinfo {year} {2011})}\BibitemShut {NoStop}%
	\bibitem [{\citenamefont {Sticlet}\ \emph {et~al.}(2012)\citenamefont
		{Sticlet}, \citenamefont {Bena},\ and\ \citenamefont {Simon}}]{Sticlet12}%
	\BibitemOpen
	\bibfield  {author} {\bibinfo {author} {\bibfnamefont {D.}~\bibnamefont
			{Sticlet}}, \bibinfo {author} {\bibfnamefont {C.}~\bibnamefont {Bena}}, \
		and\ \bibinfo {author} {\bibfnamefont {P.}~\bibnamefont {Simon}},\ }\href
	{\doibase 10.1103/PhysRevLett.108.096802} {\bibfield  {journal} {\bibinfo
			{journal} {Phys. Rev. Lett.}\ }\textbf {\bibinfo {volume} {108}},\ \bibinfo
		{pages} {096802} (\bibinfo {year} {2012})}\BibitemShut {NoStop}%
	\bibitem [{\citenamefont {Klinovaja}\ \emph
		{et~al.}(2012{\natexlab{a}})\citenamefont {Klinovaja}, \citenamefont
		{Gangadharaiah},\ and\ \citenamefont {Loss}}]{Klinovaja12c}%
	\BibitemOpen
	\bibfield  {author} {\bibinfo {author} {\bibfnamefont {J.}~\bibnamefont
			{Klinovaja}}, \bibinfo {author} {\bibfnamefont {S.}~\bibnamefont
			{Gangadharaiah}}, \ and\ \bibinfo {author} {\bibfnamefont {D.}~\bibnamefont
			{Loss}},\ }\href {\doibase 10.1103/PhysRevLett.108.196804} {\bibfield
		{journal} {\bibinfo  {journal} {Phys. Rev. Lett.}\ }\textbf {\bibinfo
			{volume} {108}},\ \bibinfo {pages} {196804} (\bibinfo {year}
		{2012}{\natexlab{a}})}\BibitemShut {NoStop}%
	\bibitem [{\citenamefont {Halperin}\ \emph {et~al.}(2012)\citenamefont
		{Halperin}, \citenamefont {Oreg}, \citenamefont {Stern}, \citenamefont
		{Refael}, \citenamefont {Alicea},\ and\ \citenamefont {von
			Oppen}}]{Halperin12}%
	\BibitemOpen
	\bibfield  {author} {\bibinfo {author} {\bibfnamefont {B.~I.}\ \bibnamefont
			{Halperin}}, \bibinfo {author} {\bibfnamefont {Y.}~\bibnamefont {Oreg}},
		\bibinfo {author} {\bibfnamefont {A.}~\bibnamefont {Stern}}, \bibinfo
		{author} {\bibfnamefont {G.}~\bibnamefont {Refael}}, \bibinfo {author}
		{\bibfnamefont {J.}~\bibnamefont {Alicea}}, \ and\ \bibinfo {author}
		{\bibfnamefont {F.}~\bibnamefont {von Oppen}},\ }\href {\doibase
		10.1103/PhysRevB.85.144501} {\bibfield  {journal} {\bibinfo  {journal} {Phys.
				Rev. B}\ }\textbf {\bibinfo {volume} {85}},\ \bibinfo {pages} {144501}
		(\bibinfo {year} {2012})}\BibitemShut {NoStop}%
	\bibitem [{\citenamefont {San-Jose}\ \emph {et~al.}(2012)\citenamefont
		{San-Jose}, \citenamefont {Prada},\ and\ \citenamefont {Aguado}}]{SanJose12}%
	\BibitemOpen
	\bibfield  {author} {\bibinfo {author} {\bibfnamefont {P.}~\bibnamefont
			{San-Jose}}, \bibinfo {author} {\bibfnamefont {E.}~\bibnamefont {Prada}}, \
		and\ \bibinfo {author} {\bibfnamefont {R.}~\bibnamefont {Aguado}},\ }\href
	{\doibase 10.1103/PhysRevLett.108.257001} {\bibfield  {journal} {\bibinfo
			{journal} {Phys. Rev. Lett.}\ }\textbf {\bibinfo {volume} {108}},\ \bibinfo
		{pages} {257001} (\bibinfo {year} {2012})}\BibitemShut {NoStop}%
	\bibitem [{\citenamefont {Rainis}\ \emph {et~al.}(2013)\citenamefont {Rainis},
		\citenamefont {Trifunovic}, \citenamefont {Klinovaja},\ and\ \citenamefont
		{Loss}}]{Rainis13}%
	\BibitemOpen
	\bibfield  {author} {\bibinfo {author} {\bibfnamefont {D.}~\bibnamefont
			{Rainis}}, \bibinfo {author} {\bibfnamefont {L.}~\bibnamefont {Trifunovic}},
		\bibinfo {author} {\bibfnamefont {J.}~\bibnamefont {Klinovaja}}, \ and\
		\bibinfo {author} {\bibfnamefont {D.}~\bibnamefont {Loss}},\ }\href {\doibase
		10.1103/PhysRevB.87.024515} {\bibfield  {journal} {\bibinfo  {journal} {Phys.
				Rev. B}\ }\textbf {\bibinfo {volume} {87}},\ \bibinfo {pages} {024515}
		(\bibinfo {year} {2013})}\BibitemShut {NoStop}%
	\bibitem [{\citenamefont {Mourik}\ \emph {et~al.}(2012)\citenamefont {Mourik},
		\citenamefont {Zuo}, \citenamefont {Frolov}, \citenamefont {Plissard},
		\citenamefont {Bakkers},\ and\ \citenamefont {Kouwenhoven}}]{Mourik12}%
	\BibitemOpen
	\bibfield  {author} {\bibinfo {author} {\bibfnamefont {V.}~\bibnamefont
			{Mourik}}, \bibinfo {author} {\bibfnamefont {K.}~\bibnamefont {Zuo}},
		\bibinfo {author} {\bibfnamefont {S.~M.}\ \bibnamefont {Frolov}}, \bibinfo
		{author} {\bibfnamefont {S.~R.}\ \bibnamefont {Plissard}}, \bibinfo {author}
		{\bibfnamefont {E.~P. A.~M.}\ \bibnamefont {Bakkers}}, \ and\ \bibinfo
		{author} {\bibfnamefont {L.~P.}\ \bibnamefont {Kouwenhoven}},\ }\href
	{\doibase 10.1126/science.1222360} {\bibfield  {journal} {\bibinfo  {journal}
			{Science}\ }\textbf {\bibinfo {volume} {336}},\ \bibinfo {pages} {1003}
		(\bibinfo {year} {2012})}\BibitemShut {NoStop}%
	\bibitem [{\citenamefont {Das}\ \emph {et~al.}(2012)\citenamefont {Das},
		\citenamefont {Ronen}, \citenamefont {Most}, \citenamefont {Oreg},
		\citenamefont {Heiblum},\ and\ \citenamefont {Shtrikman}}]{Das12}%
	\BibitemOpen
	\bibfield  {author} {\bibinfo {author} {\bibfnamefont {A.}~\bibnamefont
			{Das}}, \bibinfo {author} {\bibfnamefont {Y.}~\bibnamefont {Ronen}}, \bibinfo
		{author} {\bibfnamefont {Y.}~\bibnamefont {Most}}, \bibinfo {author}
		{\bibfnamefont {Y.}~\bibnamefont {Oreg}}, \bibinfo {author} {\bibfnamefont
			{M.}~\bibnamefont {Heiblum}}, \ and\ \bibinfo {author} {\bibfnamefont
			{H.}~\bibnamefont {Shtrikman}},\ }\href {\doibase 10.1038/nphys2479}
	{\bibfield  {journal} {\bibinfo  {journal} {Nature Physics}\ }\textbf
		{\bibinfo {volume} {8}},\ \bibinfo {pages} {887} (\bibinfo {year}
		{2012})}\BibitemShut {NoStop}%
	\bibitem [{\citenamefont {Deng}\ \emph {et~al.}(2012)\citenamefont {Deng},
		\citenamefont {Yu}, \citenamefont {Huang}, \citenamefont {Larsson},
		\citenamefont {Caroff},\ and\ \citenamefont {Xu}}]{Deng12}%
	\BibitemOpen
	\bibfield  {author} {\bibinfo {author} {\bibfnamefont {M.~T.}\ \bibnamefont
			{Deng}}, \bibinfo {author} {\bibfnamefont {C.~L.}\ \bibnamefont {Yu}},
		\bibinfo {author} {\bibfnamefont {G.~Y.}\ \bibnamefont {Huang}}, \bibinfo
		{author} {\bibfnamefont {M.}~\bibnamefont {Larsson}}, \bibinfo {author}
		{\bibfnamefont {P.}~\bibnamefont {Caroff}}, \ and\ \bibinfo {author}
		{\bibfnamefont {H.~Q.}\ \bibnamefont {Xu}},\ }\href {\doibase
		10.1021/nl303758w} {\bibfield  {journal} {\bibinfo  {journal} {Nano Letters}\
		}\textbf {\bibinfo {volume} {12}},\ \bibinfo {pages} {6414} (\bibinfo {year}
		{2012})}\BibitemShut {NoStop}%
	\bibitem [{\citenamefont {Scheller}\ \emph {et~al.}(2014)\citenamefont
		{Scheller}, \citenamefont {Liu}, \citenamefont {Barak}, \citenamefont
		{Yacoby}, \citenamefont {Pfeiffer}, \citenamefont {West},\ and\ \citenamefont
		{Zumb\"uhl}}]{Scheller14}%
	\BibitemOpen
	\bibfield  {author} {\bibinfo {author} {\bibfnamefont {C.~P.}\ \bibnamefont
			{Scheller}}, \bibinfo {author} {\bibfnamefont {T.-M.}\ \bibnamefont {Liu}},
		\bibinfo {author} {\bibfnamefont {G.}~\bibnamefont {Barak}}, \bibinfo
		{author} {\bibfnamefont {A.}~\bibnamefont {Yacoby}}, \bibinfo {author}
		{\bibfnamefont {L.~N.}\ \bibnamefont {Pfeiffer}}, \bibinfo {author}
		{\bibfnamefont {K.~W.}\ \bibnamefont {West}}, \ and\ \bibinfo {author}
		{\bibfnamefont {D.~M.}\ \bibnamefont {Zumb\"uhl}},\ }\href {\doibase
		10.1103/PhysRevLett.112.066801} {\bibfield  {journal} {\bibinfo  {journal}
			{Phys. Rev. Lett.}\ }\textbf {\bibinfo {volume} {112}},\ \bibinfo {pages}
		{066801} (\bibinfo {year} {2014})}\BibitemShut {NoStop}%
	\bibitem [{\citenamefont {Deng}\ \emph {et~al.}(2016)\citenamefont {Deng},
		\citenamefont {Vaitiekenas}, \citenamefont {Hansen}, \citenamefont {Danon},
		\citenamefont {Leijnse}, \citenamefont {Flensberg}, \citenamefont {Nyg{\r
				a}rd}, \citenamefont {Krogstrup},\ and\ \citenamefont {Marcus}}]{Deng16}%
	\BibitemOpen
	\bibfield  {author} {\bibinfo {author} {\bibfnamefont {M.~T.}\ \bibnamefont
			{Deng}}, \bibinfo {author} {\bibfnamefont {S.}~\bibnamefont {Vaitiekenas}},
		\bibinfo {author} {\bibfnamefont {E.~B.}\ \bibnamefont {Hansen}}, \bibinfo
		{author} {\bibfnamefont {J.}~\bibnamefont {Danon}}, \bibinfo {author}
		{\bibfnamefont {M.}~\bibnamefont {Leijnse}}, \bibinfo {author} {\bibfnamefont
			{K.}~\bibnamefont {Flensberg}}, \bibinfo {author} {\bibfnamefont
			{J.}~\bibnamefont {Nyg{\r a}rd}}, \bibinfo {author} {\bibfnamefont
			{P.}~\bibnamefont {Krogstrup}}, \ and\ \bibinfo {author} {\bibfnamefont
			{C.~M.}\ \bibnamefont {Marcus}},\ }\href {\doibase 10.1126/science.aaf3961}
	{\bibfield  {journal} {\bibinfo  {journal} {Science}\ }\textbf {\bibinfo
			{volume} {354}},\ \bibinfo {pages} {1557} (\bibinfo {year}
		{2016})}\BibitemShut {NoStop}%
	\bibitem [{\citenamefont {Lutchyn}\ \emph {et~al.}(2018)\citenamefont
		{Lutchyn}, \citenamefont {Bakkers}, \citenamefont {Kouwenhoven},
		\citenamefont {Krogstrup}, \citenamefont {Marcus},\ and\ \citenamefont
		{Oreg}}]{Lutchyn18}%
	\BibitemOpen
	\bibfield  {author} {\bibinfo {author} {\bibfnamefont {R.~M.}\ \bibnamefont
			{Lutchyn}}, \bibinfo {author} {\bibfnamefont {E.~P. A.~M.}\ \bibnamefont
			{Bakkers}}, \bibinfo {author} {\bibfnamefont {L.~P.}\ \bibnamefont
			{Kouwenhoven}}, \bibinfo {author} {\bibfnamefont {P.}~\bibnamefont
			{Krogstrup}}, \bibinfo {author} {\bibfnamefont {C.~M.}\ \bibnamefont
			{Marcus}}, \ and\ \bibinfo {author} {\bibfnamefont {Y.}~\bibnamefont
			{Oreg}},\ }\href {\doibase 10.1038/s41578-018-0003-1} {\bibfield  {journal}
		{\bibinfo  {journal} {Nature Reviews Materials}\ }\textbf {\bibinfo {volume}
			{3}},\ \bibinfo {pages} {52} (\bibinfo {year} {2018})}\BibitemShut {NoStop}%
	\bibitem [{\citenamefont {Deng}\ \emph {et~al.}(2018)\citenamefont {Deng},
		\citenamefont {Vaitiek\ifmmode~\dot{e}\else \.{e}\fi{}nas}, \citenamefont
		{Prada}, \citenamefont {San-Jose}, \citenamefont {Nyg\aa{}rd}, \citenamefont
		{Krogstrup}, \citenamefont {Aguado},\ and\ \citenamefont {Marcus}}]{Deng18}%
	\BibitemOpen
	\bibfield  {author} {\bibinfo {author} {\bibfnamefont {M.-T.}\ \bibnamefont
			{Deng}}, \bibinfo {author} {\bibfnamefont {S.}~\bibnamefont
			{Vaitiek\ifmmode~\dot{e}\else \.{e}\fi{}nas}}, \bibinfo {author}
		{\bibfnamefont {E.}~\bibnamefont {Prada}}, \bibinfo {author} {\bibfnamefont
			{P.}~\bibnamefont {San-Jose}}, \bibinfo {author} {\bibfnamefont
			{J.}~\bibnamefont {Nyg\aa{}rd}}, \bibinfo {author} {\bibfnamefont
			{P.}~\bibnamefont {Krogstrup}}, \bibinfo {author} {\bibfnamefont
			{R.}~\bibnamefont {Aguado}}, \ and\ \bibinfo {author} {\bibfnamefont {C.~M.}\
			\bibnamefont {Marcus}},\ }\href {\doibase 10.1103/PhysRevB.98.085125}
	{\bibfield  {journal} {\bibinfo  {journal} {Phys. Rev. B}\ }\textbf {\bibinfo
			{volume} {98}},\ \bibinfo {pages} {085125} (\bibinfo {year}
		{2018})}\BibitemShut {NoStop}%
	\bibitem [{\citenamefont {{Prada}}\ \emph {et~al.}(2020)\citenamefont
		{{Prada}}, \citenamefont {{San-Jose}}, \citenamefont {{de Moor}},
		\citenamefont {{Geresdi}}, \citenamefont {{Lee}}, \citenamefont
		{{Klinovaja}}, \citenamefont {{Loss}}, \citenamefont {{Nyg{\^a}rd}},
		\citenamefont {{Aguado}},\ and\ \citenamefont {{Kouwenhoven}}}]{Prada20}%
	\BibitemOpen
	\bibfield  {author} {\bibinfo {author} {\bibfnamefont {E.}~\bibnamefont
			{{Prada}}}, \bibinfo {author} {\bibfnamefont {P.}~\bibnamefont {{San-Jose}}},
		\bibinfo {author} {\bibfnamefont {M.~W.~A.}\ \bibnamefont {{de Moor}}},
		\bibinfo {author} {\bibfnamefont {A.}~\bibnamefont {{Geresdi}}}, \bibinfo
		{author} {\bibfnamefont {E.~J.~H.}\ \bibnamefont {{Lee}}}, \bibinfo {author}
		{\bibfnamefont {J.}~\bibnamefont {{Klinovaja}}}, \bibinfo {author}
		{\bibfnamefont {D.}~\bibnamefont {{Loss}}}, \bibinfo {author} {\bibfnamefont
			{J.}~\bibnamefont {{Nyg{\^a}rd}}}, \bibinfo {author} {\bibfnamefont
			{R.}~\bibnamefont {{Aguado}}}, \ and\ \bibinfo {author} {\bibfnamefont
			{L.~P.}\ \bibnamefont {{Kouwenhoven}}},\ }\href {\doibase
		10.1038/s42254-020-0228-y} {\bibfield  {journal} {\bibinfo  {journal} {Nature
				Reviews Physics}\ }\textbf {\bibinfo {volume} {2}},\ \bibinfo {pages}
		{575--594} (\bibinfo {year} {2020})}\BibitemShut {NoStop}%
	\bibitem [{\citenamefont {Levin}\ and\ \citenamefont {Stern}(2009)}]{Levin09}%
	\BibitemOpen
	\bibfield  {author} {\bibinfo {author} {\bibfnamefont {M.}~\bibnamefont
			{Levin}}\ and\ \bibinfo {author} {\bibfnamefont {A.}~\bibnamefont {Stern}},\
	}\href {\doibase 10.1103/PhysRevLett.103.196803} {\bibfield  {journal}
		{\bibinfo  {journal} {Phys. Rev. Lett.}\ }\textbf {\bibinfo {volume} {103}},\
		\bibinfo {pages} {196803} (\bibinfo {year} {2009})}\BibitemShut {NoStop}%
	\bibitem [{\citenamefont {Klinovaja}\ and\ \citenamefont
		{Tserkovnyak}(2014)}]{Klinovaja14d}%
	\BibitemOpen
	\bibfield  {author} {\bibinfo {author} {\bibfnamefont {J.}~\bibnamefont
			{Klinovaja}}\ and\ \bibinfo {author} {\bibfnamefont {Y.}~\bibnamefont
			{Tserkovnyak}},\ }\href {\doibase 10.1103/PhysRevB.90.115426} {\bibfield
		{journal} {\bibinfo  {journal} {Phys. Rev. B}\ }\textbf {\bibinfo {volume}
			{90}},\ \bibinfo {pages} {115426} (\bibinfo {year} {2014})}\BibitemShut
	{NoStop}%
	\bibitem [{\citenamefont {Meng}(2015)}]{Meng15}%
	\BibitemOpen
	\bibfield  {author} {\bibinfo {author} {\bibfnamefont {T.}~\bibnamefont
			{Meng}},\ }\href {\doibase 10.1103/PhysRevB.92.115152} {\bibfield  {journal}
		{\bibinfo  {journal} {Phys. Rev. B}\ }\textbf {\bibinfo {volume} {92}},\
		\bibinfo {pages} {115152} (\bibinfo {year} {2015})}\BibitemShut {NoStop}%
	\bibitem [{\citenamefont {Sagi}\ and\ \citenamefont {Oreg}(2015)}]{Sagi15}%
	\BibitemOpen
	\bibfield  {author} {\bibinfo {author} {\bibfnamefont {E.}~\bibnamefont
			{Sagi}}\ and\ \bibinfo {author} {\bibfnamefont {Y.}~\bibnamefont {Oreg}},\
	}\href {\doibase 10.1103/PhysRevB.92.195137} {\bibfield  {journal} {\bibinfo
			{journal} {Phys. Rev. B}\ }\textbf {\bibinfo {volume} {92}},\ \bibinfo
		{pages} {195137} (\bibinfo {year} {2015})}\BibitemShut {NoStop}%
	\bibitem [{\citenamefont {Santos}\ and\ \citenamefont
		{Gutman}(2015)}]{Santos15}%
	\BibitemOpen
	\bibfield  {author} {\bibinfo {author} {\bibfnamefont {R.~A.}\ \bibnamefont
			{Santos}}\ and\ \bibinfo {author} {\bibfnamefont {D.~B.}\ \bibnamefont
			{Gutman}},\ }\href {\doibase 10.1103/PhysRevB.92.075135} {\bibfield
		{journal} {\bibinfo  {journal} {Phys. Rev. B}\ }\textbf {\bibinfo {volume}
			{92}},\ \bibinfo {pages} {075135} (\bibinfo {year} {2015})}\BibitemShut
	{NoStop}%
	\bibitem [{\citenamefont {Stern}(2016)}]{Stern16}%
	\BibitemOpen
	\bibfield  {author} {\bibinfo {author} {\bibfnamefont {A.}~\bibnamefont
			{Stern}},\ }\href {\doibase 10.1146/annurev-conmatphys-031115-011559}
	{\bibfield  {journal} {\bibinfo  {journal} {Annual Review of Condensed Matter
				Physics}\ }\textbf {\bibinfo {volume} {7}},\ \bibinfo {pages} {349--368}
		(\bibinfo {year} {2016})}\BibitemShut {NoStop}%
	\bibitem [{\citenamefont {Volpez}\ \emph {et~al.}(2017)\citenamefont {Volpez},
		\citenamefont {Loss},\ and\ \citenamefont {Klinovaja}}]{Volpez17}%
	\BibitemOpen
	\bibfield  {author} {\bibinfo {author} {\bibfnamefont {Y.}~\bibnamefont
			{Volpez}}, \bibinfo {author} {\bibfnamefont {D.}~\bibnamefont {Loss}}, \ and\
		\bibinfo {author} {\bibfnamefont {J.}~\bibnamefont {Klinovaja}},\ }\href
	{\doibase 10.1103/PhysRevB.96.085422} {\bibfield  {journal} {\bibinfo
			{journal} {Phys. Rev. B}\ }\textbf {\bibinfo {volume} {96}},\ \bibinfo
		{pages} {085422} (\bibinfo {year} {2017})}\BibitemShut {NoStop}%
	\bibitem [{\citenamefont {Rachel}(2018)}]{Rachel18}%
	\BibitemOpen
	\bibfield  {author} {\bibinfo {author} {\bibfnamefont {S.}~\bibnamefont
			{Rachel}},\ }\href {\doibase 10.1088/1361-6633/aad6a6} {\bibfield  {journal}
		{\bibinfo  {journal} {Reports on Progress in Physics}\ }\textbf {\bibinfo
			{volume} {81}},\ \bibinfo {pages} {116501} (\bibinfo {year}
		{2018})}\BibitemShut {NoStop}%
	\bibitem [{\citenamefont {Laubscher}\ \emph
		{et~al.}(2019{\natexlab{a}})\citenamefont {Laubscher}, \citenamefont {Loss},\
		and\ \citenamefont {Klinovaja}}]{Laubscher19b}%
	\BibitemOpen
	\bibfield  {author} {\bibinfo {author} {\bibfnamefont {K.}~\bibnamefont
			{Laubscher}}, \bibinfo {author} {\bibfnamefont {D.}~\bibnamefont {Loss}}, \
		and\ \bibinfo {author} {\bibfnamefont {J.}~\bibnamefont {Klinovaja}},\ }\href
	{\doibase 10.1103/PhysRevResearch.1.032017} {\bibfield  {journal} {\bibinfo
			{journal} {Phys. Rev. Research}\ }\textbf {\bibinfo {volume} {1}},\ \bibinfo
		{pages} {032017(R)} (\bibinfo {year} {2019}{\natexlab{a}})}\BibitemShut
	{NoStop}%
	\bibitem [{\citenamefont {Oreg}\ \emph {et~al.}(2014)\citenamefont {Oreg},
		\citenamefont {Sela},\ and\ \citenamefont {Stern}}]{Oreg14}%
	\BibitemOpen
	\bibfield  {author} {\bibinfo {author} {\bibfnamefont {Y.}~\bibnamefont
			{Oreg}}, \bibinfo {author} {\bibfnamefont {E.}~\bibnamefont {Sela}}, \ and\
		\bibinfo {author} {\bibfnamefont {A.}~\bibnamefont {Stern}},\ }\href
	{\doibase 10.1103/PhysRevB.89.115402} {\bibfield  {journal} {\bibinfo
			{journal} {Phys. Rev. B}\ }\textbf {\bibinfo {volume} {89}},\ \bibinfo
		{pages} {115402} (\bibinfo {year} {2014})}\BibitemShut {NoStop}%
	\bibitem [{\citenamefont {Cheng}(2012)}]{Cheng12}%
	\BibitemOpen
	\bibfield  {author} {\bibinfo {author} {\bibfnamefont {M.}~\bibnamefont
			{Cheng}},\ }\href {\doibase 10.1103/PhysRevB.86.195126} {\bibfield  {journal}
		{\bibinfo  {journal} {Phys. Rev. B}\ }\textbf {\bibinfo {volume} {86}},\
		\bibinfo {pages} {195126} (\bibinfo {year} {2012})}\BibitemShut {NoStop}%
	\bibitem [{\citenamefont {Vaezi}(2013)}]{Vaezi13}%
	\BibitemOpen
	\bibfield  {author} {\bibinfo {author} {\bibfnamefont {A.}~\bibnamefont
			{Vaezi}},\ }\href {\doibase 10.1103/PhysRevB.87.035132} {\bibfield  {journal}
		{\bibinfo  {journal} {Phys. Rev. B}\ }\textbf {\bibinfo {volume} {87}},\
		\bibinfo {pages} {035132} (\bibinfo {year} {2013})}\BibitemShut {NoStop}%
	\bibitem [{\citenamefont {Meng}\ \emph {et~al.}(2014)\citenamefont {Meng},
		\citenamefont {Fritz}, \citenamefont {Schuricht},\ and\ \citenamefont
		{Loss}}]{Meng14}%
	\BibitemOpen
	\bibfield  {author} {\bibinfo {author} {\bibfnamefont {T.}~\bibnamefont
			{Meng}}, \bibinfo {author} {\bibfnamefont {L.}~\bibnamefont {Fritz}},
		\bibinfo {author} {\bibfnamefont {D.}~\bibnamefont {Schuricht}}, \ and\
		\bibinfo {author} {\bibfnamefont {D.}~\bibnamefont {Loss}},\ }\href {\doibase
		10.1103/PhysRevB.89.045111} {\bibfield  {journal} {\bibinfo  {journal} {Phys.
				Rev. B}\ }\textbf {\bibinfo {volume} {89}},\ \bibinfo {pages} {045111}
		(\bibinfo {year} {2014})}\BibitemShut {NoStop}%
	\bibitem [{\citenamefont {Klinovaja}\ and\ \citenamefont
		{Loss}(2014{\natexlab{a}})}]{Klinovaja14c}%
	\BibitemOpen
	\bibfield  {author} {\bibinfo {author} {\bibfnamefont {J.}~\bibnamefont
			{Klinovaja}}\ and\ \bibinfo {author} {\bibfnamefont {D.}~\bibnamefont
			{Loss}},\ }\href {\doibase 10.1103/PhysRevB.90.045118} {\bibfield  {journal}
		{\bibinfo  {journal} {Phys. Rev. B}\ }\textbf {\bibinfo {volume} {90}},\
		\bibinfo {pages} {045118} (\bibinfo {year} {2014}{\natexlab{a}})}\BibitemShut
	{NoStop}%
	\bibitem [{\citenamefont {Aseev}\ \emph {et~al.}(2018)\citenamefont {Aseev},
		\citenamefont {Loss},\ and\ \citenamefont {Klinovaja}}]{Aseev18}%
	\BibitemOpen
	\bibfield  {author} {\bibinfo {author} {\bibfnamefont {P.~P.}\ \bibnamefont
			{Aseev}}, \bibinfo {author} {\bibfnamefont {D.}~\bibnamefont {Loss}}, \ and\
		\bibinfo {author} {\bibfnamefont {J.}~\bibnamefont {Klinovaja}},\ }\href
	{\doibase 10.1103/PhysRevB.98.045416} {\bibfield  {journal} {\bibinfo
			{journal} {Phys. Rev. B}\ }\textbf {\bibinfo {volume} {98}},\ \bibinfo
		{pages} {045416} (\bibinfo {year} {2018})}\BibitemShut {NoStop}%
	\bibitem [{\citenamefont {Shavit}\ and\ \citenamefont {Oreg}(2019)}]{Oreg19}%
	\BibitemOpen
	\bibfield  {author} {\bibinfo {author} {\bibfnamefont {G.}~\bibnamefont
			{Shavit}}\ and\ \bibinfo {author} {\bibfnamefont {Y.}~\bibnamefont {Oreg}},\
	}\href {\doibase 10.1103/PhysRevLett.123.036803} {\bibfield  {journal}
		{\bibinfo  {journal} {Phys. Rev. Lett.}\ }\textbf {\bibinfo {volume} {123}},\
		\bibinfo {pages} {036803} (\bibinfo {year} {2019})}\BibitemShut {NoStop}%
	\bibitem [{\citenamefont {Orth}\ \emph {et~al.}(2015)\citenamefont {Orth},
		\citenamefont {Tiwari}, \citenamefont {Meng},\ and\ \citenamefont
		{Schmidt}}]{Orth15}%
	\BibitemOpen
	\bibfield  {author} {\bibinfo {author} {\bibfnamefont {C.~P.}\ \bibnamefont
			{Orth}}, \bibinfo {author} {\bibfnamefont {R.~P.}\ \bibnamefont {Tiwari}},
		\bibinfo {author} {\bibfnamefont {T.}~\bibnamefont {Meng}}, \ and\ \bibinfo
		{author} {\bibfnamefont {T.~L.}\ \bibnamefont {Schmidt}},\ }\href {\doibase
		10.1103/PhysRevB.91.081406} {\bibfield  {journal} {\bibinfo  {journal} {Phys.
				Rev. B}\ }\textbf {\bibinfo {volume} {91}},\ \bibinfo {pages} {081406(R)}
		(\bibinfo {year} {2015})}\BibitemShut {NoStop}%
	\bibitem [{\citenamefont {Sagi}\ \emph {et~al.}(2017)\citenamefont {Sagi},
		\citenamefont {Haim}, \citenamefont {Berg}, \citenamefont {von Oppen},\ and\
		\citenamefont {Oreg}}]{Sagi17}%
	\BibitemOpen
	\bibfield  {author} {\bibinfo {author} {\bibfnamefont {E.}~\bibnamefont
			{Sagi}}, \bibinfo {author} {\bibfnamefont {A.}~\bibnamefont {Haim}}, \bibinfo
		{author} {\bibfnamefont {E.}~\bibnamefont {Berg}}, \bibinfo {author}
		{\bibfnamefont {F.}~\bibnamefont {von Oppen}}, \ and\ \bibinfo {author}
		{\bibfnamefont {Y.}~\bibnamefont {Oreg}},\ }\href {\doibase
		10.1103/PhysRevB.96.235144} {\bibfield  {journal} {\bibinfo  {journal} {Phys.
				Rev. B}\ }\textbf {\bibinfo {volume} {96}},\ \bibinfo {pages} {235144}
		(\bibinfo {year} {2017})}\BibitemShut {NoStop}%
	\bibitem [{\citenamefont {Thakurathi}\ \emph {et~al.}(2017)\citenamefont
		{Thakurathi}, \citenamefont {Loss},\ and\ \citenamefont
		{Klinovaja}}]{Thakurathi17}%
	\BibitemOpen
	\bibfield  {author} {\bibinfo {author} {\bibfnamefont {M.}~\bibnamefont
			{Thakurathi}}, \bibinfo {author} {\bibfnamefont {D.}~\bibnamefont {Loss}}, \
		and\ \bibinfo {author} {\bibfnamefont {J.}~\bibnamefont {Klinovaja}},\ }\href
	{\doibase 10.1103/PhysRevB.95.155407} {\bibfield  {journal} {\bibinfo
			{journal} {Phys. Rev. B}\ }\textbf {\bibinfo {volume} {95}},\ \bibinfo
		{pages} {155407} (\bibinfo {year} {2017})}\BibitemShut {NoStop}%
	\bibitem [{\citenamefont {Pedder}\ \emph {et~al.}(2017)\citenamefont {Pedder},
		\citenamefont {Meng}, \citenamefont {Tiwari},\ and\ \citenamefont
		{Schmidt}}]{Pedder17}%
	\BibitemOpen
	\bibfield  {author} {\bibinfo {author} {\bibfnamefont {C.~J.}\ \bibnamefont
			{Pedder}}, \bibinfo {author} {\bibfnamefont {T.}~\bibnamefont {Meng}},
		\bibinfo {author} {\bibfnamefont {R.~P.}\ \bibnamefont {Tiwari}}, \ and\
		\bibinfo {author} {\bibfnamefont {T.~L.}\ \bibnamefont {Schmidt}},\ }\href
	{\doibase 10.1103/PhysRevB.96.165429} {\bibfield  {journal} {\bibinfo
			{journal} {Phys. Rev. B}\ }\textbf {\bibinfo {volume} {96}},\ \bibinfo
		{pages} {165429} (\bibinfo {year} {2017})}\BibitemShut {NoStop}%
	\bibitem [{\citenamefont {Laubscher}\ \emph
		{et~al.}(2019{\natexlab{b}})\citenamefont {Laubscher}, \citenamefont {Loss},\
		and\ \citenamefont {Klinovaja}}]{Laubscher19}%
	\BibitemOpen
	\bibfield  {author} {\bibinfo {author} {\bibfnamefont {K.}~\bibnamefont
			{Laubscher}}, \bibinfo {author} {\bibfnamefont {D.}~\bibnamefont {Loss}}, \
		and\ \bibinfo {author} {\bibfnamefont {J.}~\bibnamefont {Klinovaja}},\ }\href
	{\doibase 10.1103/PhysRevResearch.1.032017} {\bibfield  {journal} {\bibinfo
			{journal} {Phys. Rev. Research}\ }\textbf {\bibinfo {volume} {1}},\ \bibinfo
		{pages} {032017} (\bibinfo {year} {2019}{\natexlab{b}})}\BibitemShut
	{NoStop}%
	\bibitem [{\citenamefont {Fleckenstein}\ \emph {et~al.}(2019)\citenamefont
		{Fleckenstein}, \citenamefont {Ziani},\ and\ \citenamefont
		{Trauzettel}}]{Fleckenstein19}%
	\BibitemOpen
	\bibfield  {author} {\bibinfo {author} {\bibfnamefont {C.}~\bibnamefont
			{Fleckenstein}}, \bibinfo {author} {\bibfnamefont {N.~T.}\ \bibnamefont
			{Ziani}}, \ and\ \bibinfo {author} {\bibfnamefont {B.}~\bibnamefont
			{Trauzettel}},\ }\href {\doibase 10.1103/PhysRevLett.122.066801} {\bibfield
		{journal} {\bibinfo  {journal} {Phys. Rev. Lett.}\ }\textbf {\bibinfo
			{volume} {122}},\ \bibinfo {pages} {066801} (\bibinfo {year}
		{2019})}\BibitemShut {NoStop}%
	\bibitem [{\citenamefont {Klinovaja}\ and\ \citenamefont
		{Loss}(2015)}]{Klinovaja15}%
	\BibitemOpen
	\bibfield  {author} {\bibinfo {author} {\bibfnamefont {J.}~\bibnamefont
			{Klinovaja}}\ and\ \bibinfo {author} {\bibfnamefont {D.}~\bibnamefont
			{Loss}},\ }\href {\doibase 10.1103/PhysRevB.92.121410} {\bibfield  {journal}
		{\bibinfo  {journal} {Phys. Rev. B}\ }\textbf {\bibinfo {volume} {92}},\
		\bibinfo {pages} {121410(R)} (\bibinfo {year} {2015})}\BibitemShut {NoStop}%
	\bibitem [{\citenamefont {Fendley}(2012)}]{Fendley12}%
	\BibitemOpen
	\bibfield  {author} {\bibinfo {author} {\bibfnamefont {P.}~\bibnamefont
			{Fendley}},\ }\href {\doibase 10.1088/1742-5468/2012/11/p11020} {\bibfield
		{journal} {\bibinfo  {journal} {Journal of Statistical Mechanics: Theory and
				Experiment}\ }\textbf {\bibinfo {volume} {2012}},\ \bibinfo {pages} {P11020}
		(\bibinfo {year} {2012})}\BibitemShut {NoStop}%
	\bibitem [{\citenamefont {Klinovaja}\ and\ \citenamefont
		{Loss}(2014{\natexlab{b}})}]{Klinovaja14a}%
	\BibitemOpen
	\bibfield  {author} {\bibinfo {author} {\bibfnamefont {J.}~\bibnamefont
			{Klinovaja}}\ and\ \bibinfo {author} {\bibfnamefont {D.}~\bibnamefont
			{Loss}},\ }\href {\doibase 10.1103/PhysRevLett.112.246403} {\bibfield
		{journal} {\bibinfo  {journal} {Phys. Rev. Lett.}\ }\textbf {\bibinfo
			{volume} {112}},\ \bibinfo {pages} {246403} (\bibinfo {year}
		{2014}{\natexlab{b}})}\BibitemShut {NoStop}%
	\bibitem [{\citenamefont {Vaezi}(2014)}]{Vaezi14}%
	\BibitemOpen
	\bibfield  {author} {\bibinfo {author} {\bibfnamefont {A.}~\bibnamefont
			{Vaezi}},\ }\href {\doibase 10.1103/PhysRevX.4.031009} {\bibfield  {journal}
		{\bibinfo  {journal} {Phys. Rev. X}\ }\textbf {\bibinfo {volume} {4}},\
		\bibinfo {pages} {031009} (\bibinfo {year} {2014})}\BibitemShut {NoStop}%
	\bibitem [{\citenamefont {Mong}\ \emph {et~al.}(2014)\citenamefont {Mong},
		\citenamefont {Clarke}, \citenamefont {Alicea}, \citenamefont {Lindner},
		\citenamefont {Fendley}, \citenamefont {Nayak}, \citenamefont {Oreg},
		\citenamefont {Stern}, \citenamefont {Berg}, \citenamefont {Shtengel},\ and\
		\citenamefont {Fisher}}]{Mong14}%
	\BibitemOpen
	\bibfield  {author} {\bibinfo {author} {\bibfnamefont {R.~S.~K.}\
			\bibnamefont {Mong}}, \bibinfo {author} {\bibfnamefont {D.~J.}\ \bibnamefont
			{Clarke}}, \bibinfo {author} {\bibfnamefont {J.}~\bibnamefont {Alicea}},
		\bibinfo {author} {\bibfnamefont {N.~H.}\ \bibnamefont {Lindner}}, \bibinfo
		{author} {\bibfnamefont {P.}~\bibnamefont {Fendley}}, \bibinfo {author}
		{\bibfnamefont {C.}~\bibnamefont {Nayak}}, \bibinfo {author} {\bibfnamefont
			{Y.}~\bibnamefont {Oreg}}, \bibinfo {author} {\bibfnamefont {A.}~\bibnamefont
			{Stern}}, \bibinfo {author} {\bibfnamefont {E.}~\bibnamefont {Berg}},
		\bibinfo {author} {\bibfnamefont {K.}~\bibnamefont {Shtengel}}, \ and\
		\bibinfo {author} {\bibfnamefont {M.~P.~A.}\ \bibnamefont {Fisher}},\ }\href
	{\doibase 10.1103/PhysRevX.4.011036} {\bibfield  {journal} {\bibinfo
			{journal} {Phys. Rev. X}\ }\textbf {\bibinfo {volume} {4}},\ \bibinfo {pages}
		{011036} (\bibinfo {year} {2014})}\BibitemShut {NoStop}%
	\bibitem [{\citenamefont {Alicea}\ and\ \citenamefont
		{Fendley}(2016)}]{Alicea16}%
	\BibitemOpen
	\bibfield  {author} {\bibinfo {author} {\bibfnamefont {J.}~\bibnamefont
			{Alicea}}\ and\ \bibinfo {author} {\bibfnamefont {P.}~\bibnamefont
			{Fendley}},\ }\href {\doibase 10.1146/annurev-conmatphys-031115-011336}
	{\bibfield  {journal} {\bibinfo  {journal} {Annual Review of Condensed Matter
				Physics}\ }\textbf {\bibinfo {volume} {7}},\ \bibinfo {pages} {119--139}
		(\bibinfo {year} {2016})}\BibitemShut {NoStop}%
	\bibitem [{\citenamefont {Hutter}\ and\ \citenamefont {Loss}(2016)}]{Hutter16}%
	\BibitemOpen
	\bibfield  {author} {\bibinfo {author} {\bibfnamefont {A.}~\bibnamefont
			{Hutter}}\ and\ \bibinfo {author} {\bibfnamefont {D.}~\bibnamefont {Loss}},\
	}\href {\doibase 10.1103/PhysRevB.93.125105} {\bibfield  {journal} {\bibinfo
			{journal} {Phys. Rev. B}\ }\textbf {\bibinfo {volume} {93}},\ \bibinfo
		{pages} {125105} (\bibinfo {year} {2016})}\BibitemShut {NoStop}%
	\bibitem [{\citenamefont {Chew}\ \emph {et~al.}(2018)\citenamefont {Chew},
		\citenamefont {Mross},\ and\ \citenamefont {Alicea}}]{Chew18}%
	\BibitemOpen
	\bibfield  {author} {\bibinfo {author} {\bibfnamefont {A.}~\bibnamefont
			{Chew}}, \bibinfo {author} {\bibfnamefont {D.~F.}\ \bibnamefont {Mross}}, \
		and\ \bibinfo {author} {\bibfnamefont {J.}~\bibnamefont {Alicea}},\ }\href
	{\doibase 10.1103/PhysRevB.98.085143} {\bibfield  {journal} {\bibinfo
			{journal} {Phys. Rev. B}\ }\textbf {\bibinfo {volume} {98}},\ \bibinfo
		{pages} {085143} (\bibinfo {year} {2018})}\BibitemShut {NoStop}%
	\bibitem [{\citenamefont {Rossini}\ \emph {et~al.}(2019)\citenamefont
		{Rossini}, \citenamefont {Carrega}, \citenamefont {Calvanese~Strinati},\ and\
		\citenamefont {Mazza}}]{Rossini19}%
	\BibitemOpen
	\bibfield  {author} {\bibinfo {author} {\bibfnamefont {D.}~\bibnamefont
			{Rossini}}, \bibinfo {author} {\bibfnamefont {M.}~\bibnamefont {Carrega}},
		\bibinfo {author} {\bibfnamefont {M.}~\bibnamefont {Calvanese~Strinati}}, \
		and\ \bibinfo {author} {\bibfnamefont {L.}~\bibnamefont {Mazza}},\ }\href
	{\doibase 10.1103/PhysRevB.99.085113} {\bibfield  {journal} {\bibinfo
			{journal} {Phys. Rev. B}\ }\textbf {\bibinfo {volume} {99}},\ \bibinfo
		{pages} {085113} (\bibinfo {year} {2019})}\BibitemShut {NoStop}%
	\bibitem [{\citenamefont {Groenendijk}\ \emph {et~al.}(2019)\citenamefont
		{Groenendijk}, \citenamefont {Calzona}, \citenamefont {Tschirhart},
		\citenamefont {Idrisov},\ and\ \citenamefont {Schmidt}}]{Groenendijk19}%
	\BibitemOpen
	\bibfield  {author} {\bibinfo {author} {\bibfnamefont {S.}~\bibnamefont
			{Groenendijk}}, \bibinfo {author} {\bibfnamefont {A.}~\bibnamefont
			{Calzona}}, \bibinfo {author} {\bibfnamefont {H.}~\bibnamefont {Tschirhart}},
		\bibinfo {author} {\bibfnamefont {E.~G.}\ \bibnamefont {Idrisov}}, \ and\
		\bibinfo {author} {\bibfnamefont {T.~L.}\ \bibnamefont {Schmidt}},\ }\href
	{\doibase 10.1103/PhysRevB.100.205424} {\bibfield  {journal} {\bibinfo
			{journal} {Phys. Rev. B}\ }\textbf {\bibinfo {volume} {100}},\ \bibinfo
		{pages} {205424} (\bibinfo {year} {2019})}\BibitemShut {NoStop}%
	\bibitem [{\citenamefont {Santos}\ and\ \citenamefont
		{B\'eri}(2020)}]{Santos20}%
	\BibitemOpen
	\bibfield  {author} {\bibinfo {author} {\bibfnamefont {R.~A.}\ \bibnamefont
			{Santos}}\ and\ \bibinfo {author} {\bibfnamefont {B.}~\bibnamefont
			{B\'eri}},\ }\href {\doibase 10.1103/PhysRevLett.125.207201} {\bibfield
		{journal} {\bibinfo  {journal} {Phys. Rev. Lett.}\ }\textbf {\bibinfo
			{volume} {125}},\ \bibinfo {pages} {207201} (\bibinfo {year}
		{2020})}\BibitemShut {NoStop}%
	\bibitem [{\citenamefont {Barkeshli}\ \emph {et~al.}(2013)\citenamefont
		{Barkeshli}, \citenamefont {Jian},\ and\ \citenamefont {Qi}}]{Barkeshli13}%
	\BibitemOpen
	\bibfield  {author} {\bibinfo {author} {\bibfnamefont {M.}~\bibnamefont
			{Barkeshli}}, \bibinfo {author} {\bibfnamefont {C.-M.}\ \bibnamefont {Jian}},
		\ and\ \bibinfo {author} {\bibfnamefont {X.-L.}\ \bibnamefont {Qi}},\ }\href
	{\doibase 10.1103/PhysRevB.87.045130} {\bibfield  {journal} {\bibinfo
			{journal} {Phys. Rev. B}\ }\textbf {\bibinfo {volume} {87}},\ \bibinfo
		{pages} {045130} (\bibinfo {year} {2013})}\BibitemShut {NoStop}%
	\bibitem [{\citenamefont {Barkeshli}\ and\ \citenamefont
		{Qi}(2014)}]{Barkeshli14}%
	\BibitemOpen
	\bibfield  {author} {\bibinfo {author} {\bibfnamefont {M.}~\bibnamefont
			{Barkeshli}}\ and\ \bibinfo {author} {\bibfnamefont {X.-L.}\ \bibnamefont
			{Qi}},\ }\href {\doibase 10.1103/PhysRevX.4.041035} {\bibfield  {journal}
		{\bibinfo  {journal} {Phys. Rev. X}\ }\textbf {\bibinfo {volume} {4}},\
		\bibinfo {pages} {041035} (\bibinfo {year} {2014})}\BibitemShut {NoStop}%
	\bibitem [{\citenamefont {Fradkin}\ and\ \citenamefont
		{Kadanoff}(1980)}]{Fradkin80}%
	\BibitemOpen
	\bibfield  {author} {\bibinfo {author} {\bibfnamefont {E.~H.}\ \bibnamefont
			{Fradkin}}\ and\ \bibinfo {author} {\bibfnamefont {L.P.}\ \bibnamefont
			{Kadanoff}},\ }\href {\doibase 10.1016/0550-3213(80)90472-1} {\bibfield
		{journal} {\bibinfo  {journal} {Nucl. Phys. B}\ }\textbf {\bibinfo {volume}
			{170}},\ \bibinfo {pages} {1} (\bibinfo {year} {1980})}\BibitemShut {NoStop}%
	\bibitem [{\citenamefont {Calzona}\ \emph {et~al.}(2018)\citenamefont
		{Calzona}, \citenamefont {Meng}, \citenamefont {Sassetti},\ and\
		\citenamefont {Schmidt}}]{Calzona18}%
	\BibitemOpen
	\bibfield  {author} {\bibinfo {author} {\bibfnamefont {A.}~\bibnamefont
			{Calzona}}, \bibinfo {author} {\bibfnamefont {T.}~\bibnamefont {Meng}},
		\bibinfo {author} {\bibfnamefont {M.}~\bibnamefont {Sassetti}}, \ and\
		\bibinfo {author} {\bibfnamefont {T.~L.}\ \bibnamefont {Schmidt}},\ }\href
	{\doibase 10.1103/PhysRevB.98.201110} {\bibfield  {journal} {\bibinfo
			{journal} {Phys. Rev. B}\ }\textbf {\bibinfo {volume} {98}},\ \bibinfo
		{pages} {201110(R)} (\bibinfo {year} {2018})}\BibitemShut {NoStop}%
	\bibitem [{\citenamefont {Mazza}\ \emph {et~al.}(2018)\citenamefont {Mazza},
		\citenamefont {Iemini}, \citenamefont {Dalmonte},\ and\ \citenamefont
		{Mora}}]{Mazza18}%
	\BibitemOpen
	\bibfield  {author} {\bibinfo {author} {\bibfnamefont {L.}~\bibnamefont
			{Mazza}}, \bibinfo {author} {\bibfnamefont {F.}~\bibnamefont {Iemini}},
		\bibinfo {author} {\bibfnamefont {M.}~\bibnamefont {Dalmonte}}, \ and\
		\bibinfo {author} {\bibfnamefont {C.}~\bibnamefont {Mora}},\ }\href {\doibase
		10.1103/PhysRevB.98.201109} {\bibfield  {journal} {\bibinfo  {journal} {Phys.
				Rev. B}\ }\textbf {\bibinfo {volume} {98}},\ \bibinfo {pages} {201109(R)}
		(\bibinfo {year} {2018})}\BibitemShut {NoStop}%
	\bibitem [{\citenamefont {Tokura}\ \emph {et~al.}(2006)\citenamefont {Tokura},
		\citenamefont {van~der Wiel}, \citenamefont {Obata},\ and\ \citenamefont
		{Tarucha}}]{Tokura06}%
	\BibitemOpen
	\bibfield  {author} {\bibinfo {author} {\bibfnamefont {Y.}~\bibnamefont
			{Tokura}}, \bibinfo {author} {\bibfnamefont {W.~G.}\ \bibnamefont {van~der
				Wiel}}, \bibinfo {author} {\bibfnamefont {T.}~\bibnamefont {Obata}}, \ and\
		\bibinfo {author} {\bibfnamefont {S.}~\bibnamefont {Tarucha}},\ }\href
	{\doibase 10.1103/PhysRevLett.96.047202} {\bibfield  {journal} {\bibinfo
			{journal} {Phys. Rev. Lett.}\ }\textbf {\bibinfo {volume} {96}},\ \bibinfo
		{pages} {047202} (\bibinfo {year} {2006})}\BibitemShut {NoStop}%
	\bibitem [{\citenamefont {{Pioro-Ladri{\`e}re}}\ \emph
		{et~al.}(2008)\citenamefont {{Pioro-Ladri{\`e}re}}, \citenamefont {{Obata}},
		\citenamefont {{Tokura}}, \citenamefont {{Shin}}, \citenamefont {{Kubo}},
		\citenamefont {{Yoshida}}, \citenamefont {{Taniyama}},\ and\ \citenamefont
		{{Tarucha}}}]{Pioro08}%
	\BibitemOpen
	\bibfield  {author} {\bibinfo {author} {\bibfnamefont {M.}~\bibnamefont
			{{Pioro-Ladri{\`e}re}}}, \bibinfo {author} {\bibfnamefont {T.}~\bibnamefont
			{{Obata}}}, \bibinfo {author} {\bibfnamefont {Y.}~\bibnamefont {{Tokura}}},
		\bibinfo {author} {\bibfnamefont {Y.~S.}\ \bibnamefont {{Shin}}}, \bibinfo
		{author} {\bibfnamefont {T.}~\bibnamefont {{Kubo}}}, \bibinfo {author}
		{\bibfnamefont {K.}~\bibnamefont {{Yoshida}}}, \bibinfo {author}
		{\bibfnamefont {T.}~\bibnamefont {{Taniyama}}}, \ and\ \bibinfo {author}
		{\bibfnamefont {S.}~\bibnamefont {{Tarucha}}},\ }\href {\doibase
		10.1038/nphys1053} {\bibfield  {journal} {\bibinfo  {journal} {Nature
				Physics}\ }\textbf {\bibinfo {volume} {4}},\ \bibinfo {pages} {776--779}
		(\bibinfo {year} {2008})}\BibitemShut {NoStop}%
	\bibitem [{\citenamefont {{Desjardins}}\ \emph {et~al.}(2019)\citenamefont
		{{Desjardins}}, \citenamefont {{Contamin}}, \citenamefont {{Delbecq}},
		\citenamefont {{Dartiailh}}, \citenamefont {{Bruhat}}, \citenamefont
		{{Cubaynes}}, \citenamefont {{Viennot}}, \citenamefont {{Mallet}},
		\citenamefont {{Rohart}}, \citenamefont {{Thiaville}}, \citenamefont
		{{Cottet}},\ and\ \citenamefont {{Kontos}}}]{Desjardins19}%
	\BibitemOpen
	\bibfield  {author} {\bibinfo {author} {\bibfnamefont {M.~M.}\ \bibnamefont
			{{Desjardins}}}, \bibinfo {author} {\bibfnamefont {L.~C.}\ \bibnamefont
			{{Contamin}}}, \bibinfo {author} {\bibfnamefont {M.~R.}\ \bibnamefont
			{{Delbecq}}}, \bibinfo {author} {\bibfnamefont {M.~C.}\ \bibnamefont
			{{Dartiailh}}}, \bibinfo {author} {\bibfnamefont {L.~E.}\ \bibnamefont
			{{Bruhat}}}, \bibinfo {author} {\bibfnamefont {T.}~\bibnamefont
			{{Cubaynes}}}, \bibinfo {author} {\bibfnamefont {J.~J.}\ \bibnamefont
			{{Viennot}}}, \bibinfo {author} {\bibfnamefont {F.}~\bibnamefont {{Mallet}}},
		\bibinfo {author} {\bibfnamefont {S.}~\bibnamefont {{Rohart}}}, \bibinfo
		{author} {\bibfnamefont {A.}~\bibnamefont {{Thiaville}}}, \bibinfo {author}
		{\bibfnamefont {A.}~\bibnamefont {{Cottet}}}, \ and\ \bibinfo {author}
		{\bibfnamefont {T.}~\bibnamefont {{Kontos}}},\ }\href {\doibase
		10.1038/s41563-019-0457-6} {\bibfield  {journal} {\bibinfo  {journal} {Nature
				Materials}\ }\textbf {\bibinfo {volume} {18}},\ \bibinfo {pages} {1060--1064}
		(\bibinfo {year} {2019})}\BibitemShut {NoStop}%
	\bibitem [{\citenamefont {Sapkota}\ \emph {et~al.}(2019)\citenamefont
		{Sapkota}, \citenamefont {Eley}, \citenamefont {Bussmann}, \citenamefont
		{Harris}, \citenamefont {Maurer},\ and\ \citenamefont {Lu}}]{Sapkota19}%
	\BibitemOpen
	\bibfield  {author} {\bibinfo {author} {\bibfnamefont {K.~R.}\ \bibnamefont
			{Sapkota}}, \bibinfo {author} {\bibfnamefont {S.}~\bibnamefont {Eley}},
		\bibinfo {author} {\bibfnamefont {E.}~\bibnamefont {Bussmann}}, \bibinfo
		{author} {\bibfnamefont {C.~T.}\ \bibnamefont {Harris}}, \bibinfo {author}
		{\bibfnamefont {L.~N.}\ \bibnamefont {Maurer}}, \ and\ \bibinfo {author}
		{\bibfnamefont {T.~M.}\ \bibnamefont {Lu}},\ }\href {\doibase
		10.1063/1.5098768} {\bibfield  {journal} {\bibinfo  {journal} {AIP Advances}\
		}\textbf {\bibinfo {volume} {9}},\ \bibinfo {pages} {075203} (\bibinfo {year}
		{2019})}\BibitemShut {NoStop}%
	\bibitem [{\citenamefont {Giamarchi}(2003)}]{Giamarchi03}%
	\BibitemOpen
	\bibfield  {author} {\bibinfo {author} {\bibfnamefont {T.}~\bibnamefont
			{Giamarchi}},\ }\href {\doibase 10.1093/acprof:oso/9780198525004.001.0001}
	{\emph {\bibinfo {title} {Quantum Physics in One Dimension}}}\ (\bibinfo
	{publisher} {Oxford University Press},\ \bibinfo {year} {2003})\BibitemShut
	{NoStop}%
	\bibitem [{\citenamefont {Kane}\ \emph {et~al.}(2002)\citenamefont {Kane},
		\citenamefont {Mukhopadhyay},\ and\ \citenamefont {Lubensky}}]{Kane02}%
	\BibitemOpen
	\bibfield  {author} {\bibinfo {author} {\bibfnamefont {C.~L.}\ \bibnamefont
			{Kane}}, \bibinfo {author} {\bibfnamefont {Ranjan}\ \bibnamefont
			{Mukhopadhyay}}, \ and\ \bibinfo {author} {\bibfnamefont {T.~C.}\
			\bibnamefont {Lubensky}},\ }\href {\doibase 10.1103/PhysRevLett.88.036401}
	{\bibfield  {journal} {\bibinfo  {journal} {Phys. Rev. Lett.}\ }\textbf
		{\bibinfo {volume} {88}},\ \bibinfo {pages} {036401} (\bibinfo {year}
		{2002})}\BibitemShut {NoStop}%
	\bibitem [{\citenamefont {von Delft}\ and\ \citenamefont
		{Schoeller}(1998)}]{vonDelft98}%
	\BibitemOpen
	\bibfield  {author} {\bibinfo {author} {\bibfnamefont {J.}~\bibnamefont {von
				Delft}}\ and\ \bibinfo {author} {\bibfnamefont {H.}~\bibnamefont
			{Schoeller}},\ }\href {\doibase
		10.1002/(SICI)1521-3889(199811)7:4<225::AID-ANDP225>3.0.CO;2-L} {\bibfield
		{journal} {\bibinfo  {journal} {Annalen der Physik}\ }\textbf {\bibinfo
			{volume} {7}},\ \bibinfo {pages} {225} (\bibinfo {year} {1998})}\BibitemShut
	{NoStop}%
	\bibitem [{\citenamefont {Hsu}\ \emph {et~al.}(2020)\citenamefont {Hsu},
		\citenamefont {Ronetti}, \citenamefont {Stano}, \citenamefont {Klinovaja},\
		and\ \citenamefont {Loss}}]{Hsu19}%
	\BibitemOpen
	\bibfield  {author} {\bibinfo {author} {\bibfnamefont {C.-H.}\ \bibnamefont
			{Hsu}}, \bibinfo {author} {\bibfnamefont {F.}~\bibnamefont {Ronetti}},
		\bibinfo {author} {\bibfnamefont {P.}~\bibnamefont {Stano}}, \bibinfo
		{author} {\bibfnamefont {J.}~\bibnamefont {Klinovaja}}, \ and\ \bibinfo
		{author} {\bibfnamefont {D.}~\bibnamefont {Loss}},\ }\href {\doibase
		10.1103/PhysRevResearch.2.043208} {\bibfield  {journal} {\bibinfo  {journal}
			{Phys. Rev. Research}\ }\textbf {\bibinfo {volume} {2}},\ \bibinfo {pages}
		{043208} (\bibinfo {year} {2020})}\BibitemShut {NoStop}%
	\bibitem [{\citenamefont {Boyanovsky}(1989)}]{Boyanovsky89}%
	\BibitemOpen
	\bibfield  {author} {\bibinfo {author} {\bibfnamefont {D}~\bibnamefont
			{Boyanovsky}},\ }\href {https://doi.org/10.1088%2F0305-4470%2F22%2F13%2F051}
		{\bibfield  {journal} {\bibinfo  {journal} {Journal of Physics A:
					Mathematical and General}\ }\textbf {\bibinfo {volume} {22}},\ \bibinfo
			{pages} {2601} (\bibinfo {year} {1989})}\BibitemShut {NoStop}%
		\bibitem [{\citenamefont {Lecheminant}\ \emph {et~al.}(2002)\citenamefont
			{Lecheminant}, \citenamefont {Gogolin},\ and\ \citenamefont
			{Nersesyan}}]{Lecheminant02}%
		\BibitemOpen
		\bibfield  {author} {\bibinfo {author} {\bibfnamefont {P.}~\bibnamefont
				{Lecheminant}}, \bibinfo {author} {\bibfnamefont {A.~O.}\ \bibnamefont
				{Gogolin}}, \ and\ \bibinfo {author} {\bibfnamefont {A.~A.}\ \bibnamefont
				{Nersesyan}},\ }\href
		{http://www.sciencedirect.com/science/article/pii/S0550321302004741}
		{\bibfield  {journal} {\bibinfo  {journal} {Nuclear Physics B}\ }\textbf
			{\bibinfo {volume} {639}},\ \bibinfo {pages} {502} (\bibinfo {year}
			{2002})}\BibitemShut {NoStop}%
		\bibitem [{\citenamefont {Fateev}\ and\ \citenamefont
			{Zamolodchikov}(1985)}]{Fateev85}%
		\BibitemOpen
		\bibfield  {author} {\bibinfo {author} {\bibfnamefont {V.A.}\ \bibnamefont
				{Fateev}}\ and\ \bibinfo {author} {\bibfnamefont {A.B.}\ \bibnamefont
				{Zamolodchikov}},\ }\href
		{https://pdfs.semanticscholar.org/9860/8a6624fa46c94276c1915b5245607e83b87a.pdf?_ga=2.166568382.1084256838.1594905273-1867481097.1556538471}
		{\bibfield  {journal} {\bibinfo  {journal} {Sov. Phys. JETP}\ }\textbf
			{\bibinfo {volume} {62}},\ \bibinfo {pages} {215} (\bibinfo {year}
			{1985})}\BibitemShut {NoStop}%
		\bibitem [{foo()}]{footnote1}%
		\BibitemOpen
		\href@noop {} {\bibinfo  {journal} {In the case $k=2$ (Majorana fermions),
				the mapping requires in addition that velocity $v$ is equal to amplitude
				$\Lambda$}\ }\BibitemShut {NoStop}%
		\bibitem [{\citenamefont {Klinovaja}\ \emph
			{et~al.}(2012{\natexlab{b}})\citenamefont {Klinovaja}, \citenamefont
			{Stano},\ and\ \citenamefont {Loss}}]{Klinovaja12}%
		\BibitemOpen
		\bibfield  {journal} {  }\bibfield  {author} {\bibinfo {author} {\bibfnamefont
				{J.}~\bibnamefont {Klinovaja}}, \bibinfo {author} {\bibfnamefont
				{P.}~\bibnamefont {Stano}}, \ and\ \bibinfo {author} {\bibfnamefont
				{D.}~\bibnamefont {Loss}},\ }\href {\doibase 10.1103/PhysRevLett.109.236801}
		{\bibfield  {journal} {\bibinfo  {journal} {Phys. Rev. Lett.}\ }\textbf
			{\bibinfo {volume} {109}},\ \bibinfo {pages} {236801} (\bibinfo {year}
			{2012}{\natexlab{b}})}\BibitemShut {NoStop}%
		\bibitem [{\citenamefont {Karmakar}\ \emph {et~al.}(2011)\citenamefont
			{Karmakar}, \citenamefont {Venturelli}, \citenamefont {Chirolli},
			\citenamefont {Taddei}, \citenamefont {Giovannetti}, \citenamefont {Fazio},
			\citenamefont {Roddaro}, \citenamefont {Biasiol}, \citenamefont {Sorba},
			\citenamefont {Pellegrini},\ and\ \citenamefont {Beltram}}]{Karmakar11}%
		\BibitemOpen
		\bibfield  {author} {\bibinfo {author} {\bibfnamefont {B.}~\bibnamefont
				{Karmakar}}, \bibinfo {author} {\bibfnamefont {D.}~\bibnamefont
				{Venturelli}}, \bibinfo {author} {\bibfnamefont {L.}~\bibnamefont
				{Chirolli}}, \bibinfo {author} {\bibfnamefont {F.}~\bibnamefont {Taddei}},
			\bibinfo {author} {\bibfnamefont {V.}~\bibnamefont {Giovannetti}}, \bibinfo
			{author} {\bibfnamefont {R.}~\bibnamefont {Fazio}}, \bibinfo {author}
			{\bibfnamefont {S.}~\bibnamefont {Roddaro}}, \bibinfo {author} {\bibfnamefont
				{G.}~\bibnamefont {Biasiol}}, \bibinfo {author} {\bibfnamefont
				{L.}~\bibnamefont {Sorba}}, \bibinfo {author} {\bibfnamefont
				{V.}~\bibnamefont {Pellegrini}}, \ and\ \bibinfo {author} {\bibfnamefont
				{F.}~\bibnamefont {Beltram}},\ }\href {\doibase 10.1103/PhysRevLett.107.236804} {\bibfield  {journal} {\bibinfo  {journal}
				{Phys. Rev. Lett.}\ }\textbf {\bibinfo {volume} {107}},\ \bibinfo {pages}
			{236804} (\bibinfo {year} {2011})}\BibitemShut {NoStop}%
		\bibitem [{\citenamefont {Fatin}\ \emph {et~al.}(2016)\citenamefont {Fatin},
			\citenamefont {Matos-Abiague}, \citenamefont {Scharf},\ and\ \citenamefont
			{\ifmmode \check{Z}\else \v{Z}\fi{}uti\ifmmode~\acute{c}\else
				\'{c}\fi{}}}]{Fatin16}%
		\BibitemOpen
		\bibfield  {author} {\bibinfo {author} {\bibfnamefont {G.~L.}\ \bibnamefont
				{Fatin}}, \bibinfo {author} {\bibfnamefont {A.}~\bibnamefont
				{Matos-Abiague}}, \bibinfo {author} {\bibfnamefont {B.}~\bibnamefont
				{Scharf}}, \ and\ \bibinfo {author} {\bibfnamefont {I.}~\bibnamefont
				{\ifmmode \check{Z}\else \v{Z}\fi{}uti\ifmmode~\acute{c}\else \'{c}\fi{}}},\
		}\href {\doibase 10.1103/PhysRevLett.117.077002} {\bibfield  {journal}
			{\bibinfo  {journal} {Phys. Rev. Lett.}\ }\textbf {\bibinfo {volume} {117}},\
			\bibinfo {pages} {077002} (\bibinfo {year} {2016})}\BibitemShut {NoStop}%
		\bibitem [{\citenamefont {Maurer}\ \emph {et~al.}(2018)\citenamefont {Maurer},
			\citenamefont {Gamble}, \citenamefont {Tracy}, \citenamefont {Eley},\ and\
			\citenamefont {Lu}}]{Maurer18}%
		\BibitemOpen
		\bibfield  {author} {\bibinfo {author} {\bibfnamefont {L.N.}\ \bibnamefont
				{Maurer}}, \bibinfo {author} {\bibfnamefont {J.K.}\ \bibnamefont {Gamble}},
			\bibinfo {author} {\bibfnamefont {L.}~\bibnamefont {Tracy}}, \bibinfo
			{author} {\bibfnamefont {S.}~\bibnamefont {Eley}}, \ and\ \bibinfo {author}
			{\bibfnamefont {T.M.}\ \bibnamefont {Lu}},\ }\href {\doibase
			10.1103/PhysRevApplied.10.054071} {\bibfield  {journal} {\bibinfo  {journal}
				{Phys. Rev. Applied}\ }\textbf {\bibinfo {volume} {10}},\ \bibinfo {pages}
			{054071} (\bibinfo {year} {2018})}\BibitemShut {NoStop}%
		\bibitem [{\citenamefont {Mohanta}\ \emph {et~al.}(2019)\citenamefont
			{Mohanta}, \citenamefont {Zhou}, \citenamefont {Xu}, \citenamefont {Han},
			\citenamefont {Kent}, \citenamefont {Shabani}, \citenamefont {\ifmmode
				\check{Z}\else \v{Z}\fi{}uti\ifmmode~\acute{c}\else \'{c}\fi{}},\ and\
			\citenamefont {Matos-Abiague}}]{Mohanta19}%
		\BibitemOpen
		\bibfield  {author} {\bibinfo {author} {\bibfnamefont {N.}~\bibnamefont
				{Mohanta}}, \bibinfo {author} {\bibfnamefont {T.}~\bibnamefont {Zhou}},
			\bibinfo {author} {\bibfnamefont {J.-W.}\ \bibnamefont {Xu}}, \bibinfo
			{author} {\bibfnamefont {J.~E.}\ \bibnamefont {Han}}, \bibinfo {author}
			{\bibfnamefont {A.~D.}\ \bibnamefont {Kent}}, \bibinfo {author}
			{\bibfnamefont {J.}~\bibnamefont {Shabani}}, \bibinfo {author} {\bibfnamefont
				{I.}~\bibnamefont {\ifmmode \check{Z}\else
					\v{Z}\fi{}uti\ifmmode~\acute{c}\else \'{c}\fi{}}}, \ and\ \bibinfo {author}
			{\bibfnamefont {A.}~\bibnamefont {Matos-Abiague}},\ }\href {\doibase
			10.1103/PhysRevApplied.12.034048} {\bibfield  {journal} {\bibinfo  {journal}
				{Phys. Rev. Applied}\ }\textbf {\bibinfo {volume} {12}},\ \bibinfo {pages}
			{034048} (\bibinfo {year} {2019})}\BibitemShut {NoStop}%
		\bibitem [{\citenamefont {Desjardins}\ \emph {et~al.}(2019)\citenamefont
			{Desjardins}, \citenamefont {Contamin}, \citenamefont {Delbecq},
			\citenamefont {Dartiailh}, \citenamefont {Bruhat}, \citenamefont {Cubaynes},
			\citenamefont {Viennot}, \citenamefont {Mallet}, \citenamefont {Rohart},
			\citenamefont {Thiaville}, \citenamefont {Cottet},\ and\ \citenamefont
			{Kontos}}]{Desjardins2019}%
		\BibitemOpen
		\bibfield  {author} {\bibinfo {author} {\bibfnamefont {M.~M.}\ \bibnamefont
				{Desjardins}}, \bibinfo {author} {\bibfnamefont {L.~C.}\ \bibnamefont
				{Contamin}}, \bibinfo {author} {\bibfnamefont {M.~R.}\ \bibnamefont
				{Delbecq}}, \bibinfo {author} {\bibfnamefont {M.~C.}\ \bibnamefont
				{Dartiailh}}, \bibinfo {author} {\bibfnamefont {L.~E.}\ \bibnamefont
				{Bruhat}}, \bibinfo {author} {\bibfnamefont {T.}~\bibnamefont {Cubaynes}},
			\bibinfo {author} {\bibfnamefont {J.~J.}\ \bibnamefont {Viennot}}, \bibinfo
			{author} {\bibfnamefont {F.}~\bibnamefont {Mallet}}, \bibinfo {author}
			{\bibfnamefont {S.}~\bibnamefont {Rohart}}, \bibinfo {author} {\bibfnamefont
				{A.}~\bibnamefont {Thiaville}}, \bibinfo {author} {\bibfnamefont
				{A.}~\bibnamefont {Cottet}}, \ and\ \bibinfo {author} {\bibfnamefont
				{T.}~\bibnamefont {Kontos}},\ }\href
		{https://doi.org/10.1038/s41563-019-0457-6} {\bibfield  {journal} {\bibinfo
				{journal} {Nature Materials}\ }\textbf {\bibinfo {volume} {18}},\ \bibinfo
			{pages} {1060} (\bibinfo {year} {2019})}\BibitemShut {NoStop}%
		\bibitem [{\citenamefont {{Clarke}}\ \emph {et~al.}()\citenamefont {{Clarke}},
			\citenamefont {{Alicea}},\ and\ \citenamefont {{Shtengel}}}]{Clarke13}%
		\BibitemOpen
		\bibfield  {author} {\bibinfo {author} {\bibfnamefont {D.~J.}\ \bibnamefont
				{{Clarke}}}, \bibinfo {author} {\bibfnamefont {J.}~\bibnamefont {{Alicea}}},
			\ and\ \bibinfo {author} {\bibfnamefont {K.}~\bibnamefont {{Shtengel}}},\
		}\href {\doibase 10.1038/ncomms2340} {\bibfield  {journal} {\bibinfo
				{journal} {Nature Communications}\ }\textbf {\bibinfo {volume} {4}},\
			\bibinfo {eid} {1348 (2013).}}\BibitemShut {Stop}%
		\bibitem [{\citenamefont {Sato}\ \emph {et~al.}(2019)\citenamefont {Sato},
			\citenamefont {Matsuo}, \citenamefont {Hsu}, \citenamefont {Stano},
			\citenamefont {Ueda}, \citenamefont {Takeshige}, \citenamefont {Kamata},
			\citenamefont {Lee}, \citenamefont {Shojaei}, \citenamefont {Wickramasinghe},
			\citenamefont {Shabani}, \citenamefont {Palmstr\o{}m}, \citenamefont
			{Tokura}, \citenamefont {Loss},\ and\ \citenamefont {Tarucha}}]{Sato19}%
		\BibitemOpen
		\bibfield  {author} {\bibinfo {author} {\bibfnamefont {Y.}~\bibnamefont
				{Sato}}, \bibinfo {author} {\bibfnamefont {S.}~\bibnamefont {Matsuo}},
			\bibinfo {author} {\bibfnamefont {C.-H.}\ \bibnamefont {Hsu}}, \bibinfo
			{author} {\bibfnamefont {P.}~\bibnamefont {Stano}}, \bibinfo {author}
			{\bibfnamefont {K.}~\bibnamefont {Ueda}}, \bibinfo {author} {\bibfnamefont
				{Y.}~\bibnamefont {Takeshige}}, \bibinfo {author} {\bibfnamefont
				{H.}~\bibnamefont {Kamata}}, \bibinfo {author} {\bibfnamefont {J.~S.}\
				\bibnamefont {Lee}}, \bibinfo {author} {\bibfnamefont {B.}~\bibnamefont
				{Shojaei}}, \bibinfo {author} {\bibfnamefont {K.}~\bibnamefont
				{Wickramasinghe}}, \bibinfo {author} {\bibfnamefont {J.}~\bibnamefont
				{Shabani}}, \bibinfo {author} {\bibfnamefont {C.}~\bibnamefont
				{Palmstr\o{}m}}, \bibinfo {author} {\bibfnamefont {Y.}~\bibnamefont
				{Tokura}}, \bibinfo {author} {\bibfnamefont {D.}~\bibnamefont {Loss}}, \ and\
			\bibinfo {author} {\bibfnamefont {S.}~\bibnamefont {Tarucha}},\ }\href
		{\doibase 10.1103/PhysRevB.99.155304} {\bibfield  {journal} {\bibinfo
				{journal} {Phys. Rev. B}\ }\textbf {\bibinfo {volume} {99}},\ \bibinfo
			{pages} {155304} (\bibinfo {year} {2019})}\BibitemShut {NoStop}%
		\bibitem [{\citenamefont {Hsu}\ \emph {et~al.}(2019)\citenamefont {Hsu},
			\citenamefont {Stano}, \citenamefont {Sato}, \citenamefont {Matsuo},
			\citenamefont {Tarucha},\ and\ \citenamefont {Loss}}]{Hsu2019}%
		\BibitemOpen
		\bibfield  {author} {\bibinfo {author} {\bibfnamefont {C.-H.}\ \bibnamefont
				{Hsu}}, \bibinfo {author} {\bibfnamefont {P.}~\bibnamefont {Stano}}, \bibinfo
			{author} {\bibfnamefont {Y.}~\bibnamefont {Sato}}, \bibinfo {author}
			{\bibfnamefont {S.}~\bibnamefont {Matsuo}}, \bibinfo {author} {\bibfnamefont
				{S.}~\bibnamefont {Tarucha}}, \ and\ \bibinfo {author} {\bibfnamefont
				{D.}~\bibnamefont {Loss}},\ }\href {\doibase 10.1103/PhysRevB.100.195423}
		{\bibfield  {journal} {\bibinfo  {journal} {Phys. Rev. B}\ }\textbf {\bibinfo
				{volume} {100}},\ \bibinfo {pages} {195423} (\bibinfo {year}
			{2019})}\BibitemShut {NoStop}%
		\bibitem [{\citenamefont {Scappucci}\ \emph {et~al.}()\citenamefont
			{Scappucci}, \citenamefont {Kloeffel}, \citenamefont {Zwanenburg},
			\citenamefont {Loss}, \citenamefont {Myronov}, \citenamefont {Zhang},
			\citenamefont {Franceschi}, \citenamefont {Katsaros},\ and\ \citenamefont
			{Veldhorst}}]{Scappucci20}%
		\BibitemOpen
		\bibfield  {author} {\bibinfo {author} {\bibfnamefont {G.}~\bibnamefont
				{Scappucci}}, \bibinfo {author} {\bibfnamefont {C.}~\bibnamefont {Kloeffel}},
			\bibinfo {author} {\bibfnamefont {F.~A.}\ \bibnamefont {Zwanenburg}},
			\bibinfo {author} {\bibfnamefont {D.}~\bibnamefont {Loss}}, \bibinfo {author}
			{\bibfnamefont {M.}~\bibnamefont {Myronov}}, \bibinfo {author} {\bibfnamefont
				{J.-J.}\ \bibnamefont {Zhang}}, \bibinfo {author} {\bibfnamefont {S.~De}\
				\bibnamefont {Franceschi}}, \bibinfo {author} {\bibfnamefont
				{G.}~\bibnamefont {Katsaros}}, \ and\ \bibinfo {author} {\bibfnamefont
				{M.}~\bibnamefont {Veldhorst}},\ }\href {https://arxiv.org/abs/2004.08133}
		{\emph {\bibinfo {title} {\normalfont{arXiv:2004.08133}
					(2020).}}}\BibitemShut {Stop}%
		\bibitem [{\citenamefont {{Annadi}}\ \emph {et~al.}(2018)\citenamefont
			{{Annadi}}, \citenamefont {{Cheng}}, \citenamefont {{Lee}}, \citenamefont
			{{Lee}}, \citenamefont {{Lu}}, \citenamefont {{Tylan-Tyler}}, \citenamefont
			{{Briggeman}}, \citenamefont {{Tomczyk}}, \citenamefont {{Huang}},
			\citenamefont {{Pekker}}, \citenamefont {{Eom}}, \citenamefont {{Irvin}},\
			and\ \citenamefont {{Levy}}}]{Annadi18}%
		\BibitemOpen
		\bibfield  {author} {\bibinfo {author} {\bibfnamefont {A.}~\bibnamefont
				{{Annadi}}}, \bibinfo {author} {\bibfnamefont {G.}~\bibnamefont {{Cheng}}},
			\bibinfo {author} {\bibfnamefont {H.}~\bibnamefont {{Lee}}}, \bibinfo
			{author} {\bibfnamefont {J.-W.}\ \bibnamefont {{Lee}}}, \bibinfo {author}
			{\bibfnamefont {S.}~\bibnamefont {{Lu}}}, \bibinfo {author} {\bibfnamefont
				{A.}~\bibnamefont {{Tylan-Tyler}}}, \bibinfo {author} {\bibfnamefont
				{M.}~\bibnamefont {{Briggeman}}}, \bibinfo {author} {\bibfnamefont
				{M.}~\bibnamefont {{Tomczyk}}}, \bibinfo {author} {\bibfnamefont
				{M.}~\bibnamefont {{Huang}}}, \bibinfo {author} {\bibfnamefont
				{D.}~\bibnamefont {{Pekker}}}, \bibinfo {author} {\bibfnamefont {C.-B.}\
				\bibnamefont {{Eom}}}, \bibinfo {author} {\bibfnamefont {P.}~\bibnamefont
				{{Irvin}}}, \ and\ \bibinfo {author} {\bibfnamefont {J.}~\bibnamefont
				{{Levy}}},\ }\href {\doibase 10.1021/acs.nanolett.8b01614} {\bibfield
			{journal} {\bibinfo  {journal} {Nano Letters}\ }\textbf {\bibinfo {volume}
				{18}},\ \bibinfo {pages} {4473} (\bibinfo {year} {2018})}\BibitemShut
		{NoStop}%
		\bibitem [{\citenamefont {{Briggeman}}\ \emph {et~al.}()\citenamefont
			{{Briggeman}}, \citenamefont {{Lee}}, \citenamefont {{Lee}}, \citenamefont
			{{Eom}}, \citenamefont {{Damanet}}, \citenamefont {{Mansfield}},
			\citenamefont {{Li}}, \citenamefont {{Huang}}, \citenamefont {{Daley}},
			\citenamefont {{Eom}}, \citenamefont {{Irvin}},\ and\ \citenamefont
			{{Levy}}}]{Briggeman19}%
		\BibitemOpen
		\bibfield  {author} {\bibinfo {author} {\bibfnamefont {M.}~\bibnamefont
				{{Briggeman}}}, \bibinfo {author} {\bibfnamefont {H.}~\bibnamefont {{Lee}}},
			\bibinfo {author} {\bibfnamefont {J.-W.}\ \bibnamefont {{Lee}}}, \bibinfo
			{author} {\bibfnamefont {K.}~\bibnamefont {{Eom}}}, \bibinfo {author}
			{\bibfnamefont {F.}~\bibnamefont {{Damanet}}}, \bibinfo {author}
			{\bibfnamefont {E.}~\bibnamefont {{Mansfield}}}, \bibinfo {author}
			{\bibfnamefont {J.}~\bibnamefont {{Li}}}, \bibinfo {author} {\bibfnamefont
				{M.}~\bibnamefont {{Huang}}}, \bibinfo {author} {\bibfnamefont {A.~J.}\
				\bibnamefont {{Daley}}}, \bibinfo {author} {\bibfnamefont {C.-B.}\
				\bibnamefont {{Eom}}}, \bibinfo {author} {\bibfnamefont {P.}~\bibnamefont
				{{Irvin}}}, \ and\ \bibinfo {author} {\bibfnamefont {J.}~\bibnamefont
				{{Levy}}},\ }\href {https://arxiv.org/abs/1912.07164} {\emph {\bibinfo
				{title} {\normalfont{arXiv:1912.07164} (2019).}}}\BibitemShut {Stop}%
		\bibitem [{\citenamefont {{Briggeman}}\ \emph {et~al.}(2020)\citenamefont
			{{Briggeman}}, \citenamefont {{Tomczyk}}, \citenamefont {{Tian}},
			\citenamefont {{Lee}}, \citenamefont {{Lee}}, \citenamefont {{He}},
			\citenamefont {{Tylan-Tyler}}, \citenamefont {{Huang}}, \citenamefont
			{{Eom}}, \citenamefont {{Pekker}}, \citenamefont {{Mong}}, \citenamefont
			{{Irvin}},\ and\ \citenamefont {{Levy}}}]{Briggeman20}%
		\BibitemOpen
		\bibfield  {author} {\bibinfo {author} {\bibfnamefont {M.}~\bibnamefont
				{{Briggeman}}}, \bibinfo {author} {\bibfnamefont {M.}~\bibnamefont
				{{Tomczyk}}}, \bibinfo {author} {\bibfnamefont {B.}~\bibnamefont {{Tian}}},
			\bibinfo {author} {\bibfnamefont {H.}~\bibnamefont {{Lee}}}, \bibinfo
			{author} {\bibfnamefont {J.-W.}\ \bibnamefont {{Lee}}}, \bibinfo {author}
			{\bibfnamefont {Y.}~\bibnamefont {{He}}}, \bibinfo {author} {\bibfnamefont
				{A.}~\bibnamefont {{Tylan-Tyler}}}, \bibinfo {author} {\bibfnamefont
				{M.}~\bibnamefont {{Huang}}}, \bibinfo {author} {\bibfnamefont {C.-B.}\
				\bibnamefont {{Eom}}}, \bibinfo {author} {\bibfnamefont {D.}~\bibnamefont
				{{Pekker}}}, \bibinfo {author} {\bibfnamefont {R.~S.~K.}\ \bibnamefont
				{{Mong}}}, \bibinfo {author} {\bibfnamefont {P.}~\bibnamefont {{Irvin}}}, \
			and\ \bibinfo {author} {\bibfnamefont {J.}~\bibnamefont {{Levy}}},\ }\href
		{\doibase 10.1126/science.aat6467} {\bibfield  {journal} {\bibinfo  {journal}
				{Science}\ }\textbf {\bibinfo {volume} {367}},\ \bibinfo {pages} {769}
			(\bibinfo {year} {2020})}\BibitemShut {NoStop}%
		\bibitem [{\citenamefont {Kumar}\ \emph {et~al.}(2019)\citenamefont {Kumar},
			\citenamefont {Pepper}, \citenamefont {Holmes}, \citenamefont {Montagu},
			\citenamefont {Gul}, \citenamefont {Ritchie},\ and\ \citenamefont
			{Farrer}}]{Kumar19}%
		\BibitemOpen
		\bibfield  {author} {\bibinfo {author} {\bibfnamefont {S.}~\bibnamefont
				{Kumar}}, \bibinfo {author} {\bibfnamefont {M.}~\bibnamefont {Pepper}},
			\bibinfo {author} {\bibfnamefont {S.~N.}\ \bibnamefont {Holmes}}, \bibinfo
			{author} {\bibfnamefont {H.}~\bibnamefont {Montagu}}, \bibinfo {author}
			{\bibfnamefont {Y.}~\bibnamefont {Gul}}, \bibinfo {author} {\bibfnamefont
				{D.~A.}\ \bibnamefont {Ritchie}}, \ and\ \bibinfo {author} {\bibfnamefont
				{I.}~\bibnamefont {Farrer}},\ }\href {\doibase
			10.1103/PhysRevLett.122.086803} {\bibfield  {journal} {\bibinfo  {journal}
				{Phys. Rev. Lett.}\ }\textbf {\bibinfo {volume} {122}},\ \bibinfo {pages}
			{086803} (\bibinfo {year} {2019})}\BibitemShut {NoStop}%
		\bibitem [{\citenamefont {Rainis}\ \emph {et~al.}(2014)\citenamefont {Rainis},
			\citenamefont {Saha}, \citenamefont {Klinovaja}, \citenamefont {Trifunovic},\
			and\ \citenamefont {Loss}}]{Rainis14}%
		\BibitemOpen
		\bibfield  {author} {\bibinfo {author} {\bibfnamefont {D.}~\bibnamefont
				{Rainis}}, \bibinfo {author} {\bibfnamefont {A.}~\bibnamefont {Saha}},
			\bibinfo {author} {\bibfnamefont {J.}~\bibnamefont {Klinovaja}}, \bibinfo
			{author} {\bibfnamefont {L.}~\bibnamefont {Trifunovic}}, \ and\ \bibinfo
			{author} {\bibfnamefont {D.}~\bibnamefont {Loss}},\ }\href {\doibase
			10.1103/PhysRevLett.112.196803} {\bibfield  {journal} {\bibinfo  {journal}
				{Phys. Rev. Lett.}\ }\textbf {\bibinfo {volume} {112}},\ \bibinfo {pages}
			{196803} (\bibinfo {year} {2014})}\BibitemShut {NoStop}%
	\end{thebibliography}

\begin{thebibliography}{1}
	\bibitem{Fendley} P. Fendley, Journal of Statistical Mechanics: Theory and Experiment, P11020 (2012).
	\bibitem{Fateev} V.A. Fateev and A.B. Zamolodchikov, Sov. Phys. JETP \textbf{62}, 215 (1985).
	\bibitem{Fradkin} E. H. Fradkin and L.P. Kadanoff, Nucl. Phys. B \textbf{170}, 1 (1980).
	\bibitem{Giamarchi} T. Giamarchi, \textit{Quantum Physics in One Dimension} (Oxford University Press, 2003).
	\bibitem{Manisha} M. Thakurathi, P. Simon, I. Mandal, J. Klinovaja, and D. Loss, Phys. Rev. B {\bf 97}, 045415 (2017).
	\bibitem{Katharina} K. Laubscher, D. Loss, and J. Klinovaja, Phys. Rev. Research {\bf 1}, 032017(R) (2019).
	\bibitem{Boyanovsky} D. Boyanovsky, J. Phys. A: Math. Gen. {\bf 22}, 2601 (1989).
	\bibitem{Chiral_sup} E. Sagi, A. Haim, E. Berg, F. von Oppen, and Y. Oreg, Phys. Rev. B {\bf 96}, 235144 (2017).
\end{thebibliography}
\end{document}